\documentclass[12pt,preprint,numberedappendix,appendixfloats]{emulateapj}
\usepackage{natbib}
\input{colordvi}

\newcommand{\mic}{~$\mu$m}
\newcommand{\micJy}{~$\mu$Jy}
\newcommand{\spitzer}{{\it Spitzer}}

\slugcomment{Last edited: \today}

\shortauthors{P\'erez-Gonz\'alez et al.}

\begin{document}

\title{The stellar mass assembly of galaxies from z$=$0 to z$=$4.\\ 
Analysis of a sample selected in the rest-frame near-infrared with
Spitzer}

\author{Pablo G. P\'erez-Gonz\'alez\altaffilmark{1,2}, George H. Rieke\altaffilmark{3}, Victor Villar\altaffilmark{1}, Guillermo Barro\altaffilmark{1}, Myra Blaylock\altaffilmark{3}, Eiichi Egami\altaffilmark{3}, Jes\'us Gallego\altaffilmark{1}, Armando Gil de Paz\altaffilmark{1}, Sergio Pascual\altaffilmark{1}, Jaime Zamorano\altaffilmark{1}, Jennifer L. Donley\altaffilmark{3}
}

\altaffiltext{1}{Departamento de Astrof\'{\i}sica, Facultad de CC. F\'{\i}sicas,
Universidad Complutense de Madrid, E-28040 Madrid, Spain}
\altaffiltext{2}{Associate Astronomer at Steward Observatory, The University of Arizona}
\altaffiltext{3}{The University of Arizona, Steward Observatory, 933 N Cherry Avenue, Tucson, AZ 85721}

\begin{abstract} 
Using a sample of $\sim$28,000 sources selected at 3.6--4.5 microns
with \spitzer\, observations of the HDF-N, the CDF-S, and the Lockman
Hole (surveyed area: $\sim$664~arcmin$^2$), we study the evolution of
the stellar mass content of the Universe at 0$<$z$<$4. We calculate
stellar masses and photometric redshifts, based on $\sim$2,000
templates built with stellar population and dust emission models
fitting the UV-to-MIR spectral energy distributions of galaxies with
spectroscopic redshifts. We estimate stellar mass functions for
different redshift intervals. We find that 50\% of the local stellar
mass density was assembled at 0$<$z$<$1 (average SFR:
0.048~$\mathcal{M}_\sun$yr$^{-1}$Mpc$^{-3}$), and at least another
40\% at 1$<$z$<$4 (average SFR:
0.074~$\mathcal{M}_\sun$yr$^{-1}$Mpc$^{-3}$). Our results confirm and
quantify the ``downsizing'' scenario of galaxy formation. The most
massive galaxies ($\mathcal{M}$$>$$10^{12.0}$~$\mathcal{M}_\sun$)
assembled the bulk of their stellar content rapidly (in 1-2~Gyr)
beyond z$\sim$3 in very intense star formation events (producing high
specific SFRs). Galaxies with
$10^{11.5}$$<$$\mathcal{M}$$<$$10^{12.0}$~$\mathcal{M}_\sun$ assembled
half of their stellar mass before z$\sim$1.5, and more than 90\% of
their mass was already in place at z$\sim$0.6. Galaxies with
$\mathcal{M}$$<$$10^{11.5}$~$\mathcal{M}_\sun$ evolved more slowly
(presenting smaller specific SFRs), assembling half of their stellar
mass below z$\sim$1. About 40\% of the local stellar mass density of
$10^{9.0}$$<$$\mathcal{M}$$<$$10^{11.0}$~$\mathcal{M}_\sun$ galaxies
was assembled below z$\sim$0.4, most probably through accretion of
small satellites producing little star formation. The cosmic stellar
mass density at z$>$2.5 is dominated by optically faint
($R$$\gtrsim$25) red galaxies (Distant Red Galaxies or $BzK$ sources)
which account for $\sim$30\% of the global population of galaxies, but
contribute at least 60\% to the cosmic stellar mass density.  Bluer
galaxies (e.g., Lyman Break Galaxies) are more numerous but less
massive, contributing less than 50\% to the global stellar mass
density at high redshift.
\end{abstract}

\section{INTRODUCTION}
\label{intro}

In the last decade, our knowledge about the formation and evolution of
galaxies has increased significantly with the advent of deep and/or
wide photometric and spectroscopic galaxy surveys carried out at
different wavelengths. This advance in our understanding of the
evolution of the Universe is succinctly represented in the so-called
Lilly-Madau plot \citep{1996ApJ...460L...1L,1996MNRAS.283.1388M}, a
diagram showing the evolution of the Star Formation Rate (SFR) density
of the Universe as a function of look-back time (or
redshift). Originally, with only a few points in the diagram, it was
clearly visible that in the last $\sim$8~Gyr (i.e., about 55\% of its
age) the Universe experienced a significant decrease (of about a
factor of 10) in the rate at which new stars were created. Nowadays,
there are more than 80 data points in the Lilly-Madau diagram (see
\citealt{2004ApJ...615..209H} for a nice compilation of results on
this topic; see also
\citealt{2005ApJ...619L..47S}, \citealt{2005ApJ...630...82P}, and 
\citealt{2006ApJ...651..142H}), and the picture is clearer at z$\lesssim$1, 
where there is just a factor of 2 scatter among the estimations coming
from different surveys, and using different selection techniques and
SFR tracers. At z$\gtrsim$1, the uncertainties are larger, up to a
factor of $\sim$5, but there is increasing evidence that the SFR
density remained approximately constant for 4--5~Gyr (from z$\sim$1 to
z$\sim$4).  


Although the Lilly-Madau plot concentrates a large amount of
information about the formation of structures in the Universe, the
(recent) SFR is not the best parameter to characterize the evolution
of a galaxy, as it is an instantaneous parameter. Indeed, the
stellar mass or the metallicity, which are closely linked to the star
formation history, are more appropriate parameters to follow the
evolution of galaxies. Thus, an increasing number of studies explore
the evolution of the cosmic comoving stellar mass density, showing
that it has steadily increased in the last 12~Gyr (see, e.g.,
\citealt{2000ApJ...536L..77B},
\citealt{2003ApJ...587...25D}, \citealt{2004Natur.430..181G}, 
\citealt{2005ApJ...619L.131D}, \citealt{2006A&A...459..745F}; 
see also the references given in Figure~\ref{massdensity}).

Because of the increasingly large scale of cosmological surveys, the
problem of the evolution of galaxies is now being addressed by
dividing the samples into ranges in stellar mass. In this context, the
evolution of galaxies seems to follow a 'downsizing' scenario
\citep{1996AJ....112..839C}, where the most massive galaxies are
formed first and the star formation continues in less massive systems
until more recent epochs
\citep{2004Natur.428..625H,2005ApJ...619L.135J,2005ApJ...621L..89B,
2005ApJ...630...82P,2006ApJ...651..120B,2007A&A...472..403T}. Although the
'downsizing' picture is being confirmed by an increasing number of
works, the quantification of the process is still very limited, given
the necessity of large samples of high redshift galaxies with
multi-wavelength data to explore it (covering from the rest-frame
ultraviolet to the near-infrared and beyond).

In contrast with these observational results, classical models of
galaxy evolution assuming a Cold Dark Matter (CDM) Universe usually
predict that the most massive galaxies assembled late via the
coalescence of small halos that form larger ones
\citep[e.g.,][]{1993MNRAS.264..201K,1998ApJ...498..504B,
2001MNRAS.320..504S}. This contradicts the observational evidence
of the existence of large galaxies at high redshifts (some of them
already harboring old stellar populations at those early epochs, some
with significant recent star formation), detected by their unusually
red colors \citep[among
others,][]{1988ApJ...331L..77E,1999ApJ...519..610D,
2000ApJ...531..624D, 2002ApJ...578L..19I, 2003ApJ...587L..79F} or
their bright emission at sub-millimeter wavelengths \citep[e.g.,][see
also \citealt{2002PhR...369..111B} for a
review]{1997ApJ...490L...5S,1998Natur.394..241H}. More recent models
based on a $\Lambda$CDM cosmology succeed in predicting the early
formation of massive galaxies by introducing very large dust
extinctions, non-standard Initial Mass Functions, and/or suppression
of the star formation due to the quenching of cooling flows due to
supernovae or Active Galactic Nuclei
\citep[e.g.,][]{2000MNRAS.319..168C,2004ApJ...600..580G,
2005MNRAS.356.1191B, 2005ApJ...627..608N,
2006MNRAS.365...11C,2006MNRAS.370..645B}.

In this paper, we observationally characterize the build-up of the
stellar mass of galaxies in the last $\sim$12~Gyr (almost 90\% of the
age of the Universe) as a function of the stellar mass of each
object. This is done by estimating stellar mass functions at different
redshifts. Given that we are interested in the stellar mass assembly
of galaxies, it would be convenient to analyze a sample whose
selection is based precisely on that parameter, the stellar mass. From
studies at low and intermediate redshift, we know that the rest-frame
near-infrared (NIR) emission of galaxies arises mainly from relatively
old stars that usually dominate the total stellar mass of galaxies, in
contrast to younger stellar populations that may contribute little to
the rest-frame NIR emission and stellar mass, but emit strongly at
bluer wavelengths. Indeed, stellar mass estimations based (only) on
photometry at rest-frame wavelengths bluer than $\sim$600~nm are
particularly troublesome because of the ability of a small population
of young stars to dominate the output of a galaxy. In the red and NIR,
the light is dominated by similar stellar populations, but the
rest-frame NIR is preferred for estimating stellar masses because of
its relative immunity to extinction. In addition, data at red
wavelengths is crucial to detect galaxies that are very faint in the
optical (too faint for optical surveys) but may contribute
significantly or even dominate the stellar mass density of the
Universe at high-z (e.g., Extremely Red Objects, EROs,
\citealt{1988ApJ...331L..77E},
\citealt{2000AJ....120..575Y}; or Distant Red Galaxies, DRGs,
\citealt{2003ApJ...587L..79F},
\citealt{2003ApJ...587L..83V}). These galaxies are usually missed 
by selection techniques based on rest-frame ultraviolet colors
\citep[e.g., Lyman Break Galaxies, LBGs;][]{2003ApJ...592..728S}. 
Therefore, a sample selected in the rest-frame NIR is the most
adequate to attempt a stellar mass function analysis. Still,
mass-to-light ratios in the rest-frame NIR from galaxy to galaxy may
still vary by factor of 6--15 (depending on the mean age of the
stellar population, the presence of recent bursts, etc...; see, e.g.,
\citealt{2001ApJ...550..212B}, \citealt{2005ApJ...626..698S}, or 
\citealt{2005ApJ...624L..81L}).  This means that a complete study 
of the optical-to-NIR spectral energy distribution of galaxies in a
galaxy-by-galaxy basis should be performed to obtain robust stellar
mass estimates.

This paper is based on the analysis of a sample of galaxies at
0$<$z$<$4 selected in 3 different fields (to minimize cosmic variance
problems) at 3.6-4.5\mic\, with the Infrared Array Camera
\citep[IRAC, ][]{2004ApJS..154...10F} on-board of the \spitzer\, Space Telescope
\citep{2004ApJS..154....1W}. Even at the highest redshift in the sample, 
the sources are still selected in the rest-frame NIR (approximately
the $J$-band), so an IRAC selected sample uniquely constitutes a
statistically complete sample in stellar mass at all redshifts up to
z$\sim$4 (to a certain lower limit based on the flux cut of the
sample). In addition, the estimations of the stellar masses of our
galaxies always count with a NIR band, which significantly reduces the
uncertainties in the derived stellar masses \citep[see, e.g.,
][]{2006A&A...459..745F}, since the relatively old stellar population
contributing the most to the total stellar mass of galaxies usually
dominates the emission at NIR wavelengths, and also because the NIR is
relatively free of extinction effects and hence is better for
estimating stellar masses than shorter wavelengths. Our sample
selection constitutes an extension (in area, depth, and consequently,
in the number of galaxies detected) of those used by other groups
based on ground-based $K$-band data (e.g.,
\citealt{2004ApJ...608..742D} and
\citealt{2004A&A...424...23F}).

This paper is organized as follows: Section~\ref{data} presents the
dataset and samples of galaxies used in this
work. Section~\ref{models} describes the stellar population and dust
emission models used to estimate photometric redshifts, stellar
masses, and SFRs for all galaxies in our sample. Here, we also discuss
the uncertainties in these parameters. Sections~\ref{photozsect} and
\ref{stellarmass} discuss the main results about photometric
redshifts and stellar masses. More precisely, we present stellar mass
functions and densities, discussing their evolution with
redshift. Section~\ref{comparison} divides our sample into several
sub-types (such as DRGs or LBGs), and discusses the evolution of
galaxies of different natures and their role on the evolution of the
stellar mass density of the Universe as a whole. Section~\ref{sfrs}
analyzes the SFRs of the galaxies in our sample and the evolution of
the cosmic SFR density. Finally, Section~\ref{conclusions} summarizes
the conclusions of this paper.

Throughout this paper, we use a cosmology with $\mathrm
H_{0}=70$~km\,s$^{-1}$\,Mpc$^{-1}$, $\Omega_{\mathrm M}=0.3$ and
$\Lambda=0.7$. All magnitudes refer to the AB system. The results
about stellar masses assume a \citet{1955ApJ...121..161S} universal
(i.e., constant through time) Initial Mass Function (IMF) with
0.1$<$$\mathcal{M}$$<$100~$\mathcal{M}_\sun$ and a single power-law
slope in this range.

\section{SAMPLE SELECTION}
\label{data}

This paper analyzes the main properties of the galaxies selected by
IRAC (hereafter, the IRAC selected sample), which should be close to a
stellar mass selected sample up to the highest redshifts in our
survey. We complemented this dataset with a sample of galaxies
selected in a ground-based optical image (the $I$-band selected
sample\footnote{For this selection, we chose the deepest ground-based
images in a band common (or similar) to the 3 fields, namely, the
Subaru $I$-band images in the LHF and the HDF-N, and the Subaru
$NB816$ image (close to an $I$-band image, and also very deep) in the
CDF-S.}, hereafter), in order to check the effect on our results of
the galaxies missed by IRAC, i.e., galaxies which are relatively faint
in the rest-frame NIR but can be detected in deep optical
imaging. This sample of NIR-faint galaxies should allow us to probe
the stellar mass functions at small masses below the IRAC detection
limits (and at higher masses, where the galaxies should also be
detected by IRAC).

The IRAC sample is drawn from the \spitzer\, GTO \citep[see,
e.g.,][]{2005ApJ...630...82P} and GOODS
\citep{2003mglh.conf..324D} IRAC and MIPS observations of the 
Hubble Deep Field North (HDF-N) and the Chandra Deep Field South
(CDF-S), and the \spitzer\, GTO data in the Lockman Hole Field
(LHF). In each field, we concentrated on a relatively reduced sky area
with the deepest coverage by \spitzer, and also observed by other
X-ray, ultraviolet (UV), optical, near-infrared (NIR), and
mid-infrared (MIR) surveys. In the HDF-N, we focused our analysis in
257~arcmin$^2$ centered at $\alpha=12^h38^m56^s$,
$\delta=+62\arcdeg14\arcmin06\arcsec$, J2000, and including the entire
GOODS ACS footprint; in the CDF-S, we focused on a rectangle of
225~arcmin$^2$ centered at $\alpha=03^h30^m28^s$,
$\delta=-27\arcdeg48\arcmin18\arcsec$, J2000, also including the
entire GOODS ACS footprint; and in the LHF, we used a square area of
183~arcmin$^2$ centered at $\alpha=10^h52^m47^s$,
$\delta=+57\arcdeg29\arcmin06\arcsec$, J2000. This adds up a total
surveyed area of 664~arcmin$^2$.

The reduction, source extraction, and photometry of the IRAC and MIPS
images were performed in the same way explained in
\citet{2005ApJ...630...82P}. We describe the procedure with more details
in Appendix~\ref{catalog}. The IRAC sample was built by detecting
sources separately in the 2 bluer IRAC bands (at 3.6\mic\, and
4.5\mic), and then merging the catalogs, and removing repeated
sources. Aperture photometry was measured in the 4 IRAC images (fixing
the positions and forcing the detection in all bands), obtaining the
final integrated magnitude after applying an aperture correction based
on empirical Point Spread Functions (PSFs). All the sources in the
IRAC sample have measured fluxes at both 3.6\mic\, and 4.5\mic. For
the MIPS 24\mic\, images, we measured integrated fluxes using PSF fits
and aperture corrections. The $I$-band selected sample was built by
detecting sources with {\sc sextractor} \citep{1996A&AS..117..393B} in
the optical images.

Our IRAC selected sample is composed of 9,074 sources in the HDF-N,
9,676 in the CDF-S, and 9,149 in the LHF, for a total of 27,899
sources (i.e., 42~sources$/\mathrm{arcmin}^2$). Out of these, less
than 3\% (700 sources) are identified as stars (see the star-galaxy
separation method in Section~\ref{stargalaxy}). Based on simulations
carried out by adding artificial sources to the IRAC images and trying
to recover their detection and input flux, we estimate that our IRAC
catalogs in the HDF-N and the CDF-S are 75\% (90\%) complete down to
1.6\micJy\, (5.0\micJy) at 3.6\mic, and 1.4\micJy\, (4.0\micJy) at
4.5\mic. For the LHF, where deep GOODS IRAC data are not available,
the 75\% (90\%) completeness levels are 2.2\micJy\, (5.8\micJy) at
3.6\mic, and 2.0\micJy\, (4.8\micJy) at 4.5\mic. Above the 75\%
completeness flux limits, our sample has 7,512 galaxies (after removal
of stars) in the HDF-N, 6,546 galaxies in the CDF-S, and 5,341
galaxies in the LHF, adding a total of 19,399 galaxies
(29.2~sources$/\mathrm{arcmin}^2$). Out of these, 6,686 (35\%)
galaxies are detected by MIPS at 24\mic, 3,483 (18\%) above our 75\%
24\mic\, completeness level [$F(24)$$=$80\micJy].

We concentrated our analysis of the $I$-band selected sample on the
region covered by the other UV-to-MIR surveys, and enclosing a similar
number of sources as those detected with the IRAC selection
(therefore, we considered smaller regions in each field). We used an
area of 101~$\mathrm{arcmin}^2$ centered at $\alpha=12^h37^m00^s$,
$\delta=+62\arcdeg13\arcmin30\arcsec$ (J2000) in the HDF-N,
103~arcmin$^2$ at $\alpha=03^h32^m28^s$
$\delta=-27\arcdeg48\arcmin54\arcsec$ (J2000) in the CDF-S, and
70~arcmin$^2$ at $\alpha=10^h52^m48^s$
$\delta=+57\arcdeg29\arcmin24\arcsec$ in the LHF. The samples are
formed by 7,326 sources (112 of them identified as stars) in the
HDF-N, 6,680 (87 stars) in the CDF-S, and 6,797 (99 stars) in the LHF,
for a total of 20,505 galaxies with $I$$\lesssim$25.5
(75~sources$/\mathrm{arcmin}^2$).

The \spitzer\, data were complemented with other publicly available
and proprietary photometric and spectroscopic data in the 3
fields. These data cover the electromagnetic spectrum from UV to MIR
wavelengths. The description of the different datasets and the
procedure used to get merged UV-to-MIR photometry for each source is
described in detail in Appendix~\ref{catalog}. The spectral energy
distributions (SEDs) of each source were used to remove stars from our
sample, detect candidates to harbor an Active Galactic Nuclei (AGN),
and to estimate photometric redshifts, stellar masses, and SFRs for
the entire sample in a galaxy-by-galaxy basis, as explained in
Section~\ref{models} and Appendix~\ref{models_app}.

\section{ESTIMATION OF PHOTOMETRIC REDSHIFTS, STELLAR MASSES, AND STAR FORMATION RATES}
\label{models}

The estimation of the photometric redshift, stellar mass, and SFR of
each galaxy in our IRAC and $I$-band selected samples was carried out
in a two step process. Given the significant degeneracies inherent to
any stellar population modelling, and in order to get the best
estimations of the interesting parameters, we decided to first build a
reference set of stellar population and dust emission templates. This
trained template set was used in the second step to obtain photometric
redshifts, stellar masses, and SFRs for the entire sample. The
reference template set was built with the $\sim$2,000 galaxies in our
spectroscopic sample with highly reliable redshifts and well-covered
SEDs (with enough data points from the rest-frame UV to NIR/MIR
wavelengths). This is the same approach we chose in
\citet{2005ApJ...630...82P}. As a major improvement of our photometric 
redshift technique described in that paper, we (significantly)
increased the spectral resolution of the templates by fitting the SEDs
of the galaxies in the reference spectroscopic sample with models of
the stellar population and dust emission (probing more than $10^{11}$
different models).

In Appendix~\ref{models_app}, we describe the stellar and dust
emission modelling procedure, the building of the reference template
set, and the procedure to get photometric redshift, stellar mass, and
SFR estimates for each galaxy in our entire sample. In this Appendix,
we also evaluate the goodness of our photometric redshift, stellar
mass, and SFR estimates. We show that our photometric redshifts for
galaxies at z$<$1.5 are better than $\sigma_\mathrm{z}$/(1+z)$<$0.1
(where $\sigma_\mathrm{z}$ is the absolute value of
$\delta$z$=$z$_\mathrm{spec}$-z$_\mathrm{photo}$) for approximately
87\% of the galaxies in our complete sample, and better than
$\sigma_\mathrm{z}$/(1+z)$<$0.2 for 95\%. At z$>$1.5, we test our
photometric redshifts distributions for different samples of high
redshift galaxies (LBGs, DRGs, and $BzK$ sources, see
Section~\ref{comparison} for more details), obtaining acceptable
results, in good agreement with other spectroscopic and photometric
redshift analysis.

The distributions of photometric redshift uncertainties (as derived
from the comparison with spectroscopic redshifts in
Appendix~\ref{models_app}) for different magnitude and redshift
intervals are used in Section~\ref{stellarmass} to estimate the
uncertainties in the stellar mass functions. In addition, the redshift
intervals in that Section and the following are constructed assuming
that the typical photometric redshift error is
$\sigma_\mathrm{z}$/(1+z)$\sim$0.1 (equation valid for more than 85\%
of our sample). We would like to stress that the results described in
the following Sections are more robust at z$<$1.5, where the
photometric redshifts are well tested, and the photometry is very
accurate, than at z$>$1.5, where the unavailability of spectroscopic
redshifts does not allow a characterization of the photometric
redshifts as thorough as at low z, and photometric errors are
generally larger.

In Appendix~\ref{models_app}, we also discuss the goodness of our
stellar mass estimates. We conclude that the choices of a single
population or a two component population in the stellar emission
models, the use of distinct stellar population libraries, different
IMFs, or different extinction recipes produce changes in the derived
stellar masses of a factor of 2--3, which is also the typical error in
any stellar population synthesis analysis linked to the degeneracies
of the solutions to the problem. Therefore, our stellar mass estimates
are good within a factor of 2--3.

The estimations of the SFRs for each galaxy are also proved to be good
within a factor of 2 in Appendix~\ref{models_app}, an uncertainty
which is consistent with other evaluations of UV- and IR-based SFRs
\citep[e.g., ][]{2002ApJ...579L...1P,
2005ApJ...632..169L,2006ApJ...637..727C}.

Finally, Appendix~\ref{models_app} also discusses the validity of our
estimated parameters for galaxies harboring an AGN. We conclude that
photometric redshifts and stellar masses should not be affected
dramatically for most AGNs (except for very bright Type 1 AGNs), but
IR-based SFRs can be overestimated. For this reason, we exclude AGNs
from the analysis of SFRs performed in Section~\ref{sect_sfr}, but we
keep most of them (only excluding very bright X-ray sources) in our
calculations of the stellar mass functions and densities.

\section{REDSHIFT DISTRIBUTION OF OUR SAMPLE}
\label{photozsect}

\begin{figure}
\begin{center}
\includegraphics[angle=-90,width=9.cm]{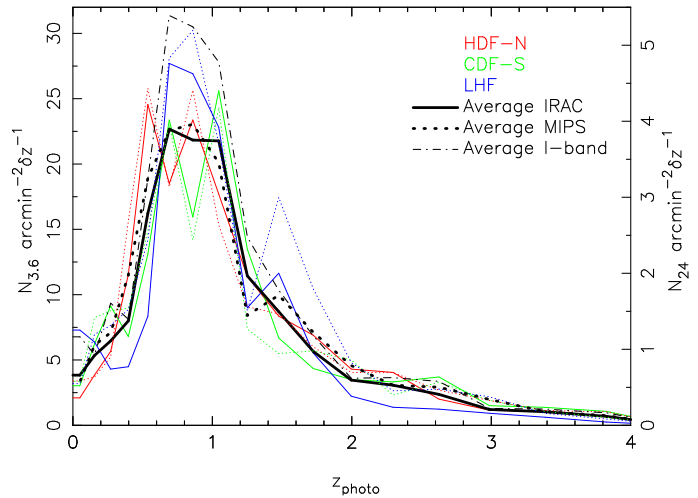}
\figcaption{\label{photoz_distro}Redshift distribution of our IRAC, MIPS, 
and $I$-band selected samples (including all galaxies above the
completeness level). For the IRAC and MIPS samples, the three fields
used in this paper are plotted with different colors, and the average
number densities are plotted in black. Continuous lines refer to the
number densities for the entire IRAC selected sample (scale on the
left vertical axis). The dashed lines refer to the subsample (within
the IRAC sample) detected also at 24\mic\, (scale on the right
vertical axis). The dash-dotted line show the redshift distribution of
the $I$-band selected sample, with the same scale as the IRAC
distribution. }
\end{center}
\end{figure}

Figure~\ref{photoz_distro} shows the photometric redshift distribution
of our IRAC selected sample (average number density and number
densities in each field), the subsample detected also by MIPS at
24\mic, and the $I$-band selected sample. Only sources with fluxes
above our 75\% completeness levels are included. The distributions
have been constructed taking into account the typical photometric
redshift error ($\sigma_\mathrm{z}$/(1+z)$\sim$0.1), i.e,
Figure~\ref{photoz_distro} represents the real redshift distribution
convolved with the photometric redshift probability distribution.

\slugcomment{Please, plot this figure with the width of one column}
\placefigure{photoz_distro}

Figure~\ref{photoz_distro} demonstrates the importance of cosmic
variance effects on deep photometric surveys. Indeed, large scale
structures are clearly visible and located at different redshifts for
our 3 fields, especially at z$\lesssim$1. Number density variations of
up to a factor of 2 can exist at a given redshift from one field to
another. The HDF-N shows two very prominent density peaks at
z$\sim$0.5 and z$\sim$0.9, consistent with the spectroscopic redshift
histogram found in figure 16 of
\citet{2004AJ....127.3121W}. There are also minor prominences 
at z$=$1.5--2.0 and z$=$2.0--2.5, which are also seen in the
spectroscopic follow-up of UV-selected galaxies in
\citet{2006ApJ...653.1004R}. The CDF-S presents very prominent density
peaks at z$\sim$ 0.3, z$\sim$0.7, and z$\sim$1.1 (the latter broadens
up to z$\sim$1.4), which coincides (after convolution with the typical
photometric redshift uncertainty) with the most prominent
spectroscopically confirmed peaks found in figure 7 of
\citet{2006A&A...454..423V}. The LHF shows an enhanced density at 
z$\lesssim$0.3, z$\sim$0.7-1.0 and a very prominent peak at
z$=$1.5--1.8. These peaks are consistent with the high density of
X-ray sources found by \citet{2002astro.ph..2211M} and
\citet{2005A&A...434..801Z} at z$\sim$0.8 and z$\sim$1.6--1.8, and the 
analysis of the shallower IRAC SWIRE data in
\citet{2005AJ....129.1183R}.

Only by combining data for several fields are we able to smooth out
cosmic variance effects. Indeed, the average redshift distributions
for IRAC sources (black continuous line) and MIPS sources (black
dashed line) are much smoother than the analogous curves for the
individual fields. The shape of the redshift distribution for the IRAC
sample is typical of a flux limited sample with a roughly homogeneous
detection probability, i.e, the detection of a source depends only on
its magnitude
\citep[see, e.g.,][]{2000ApJ...536..571B}. The detection probability 
of our IRAC survey peaks at around z$=$0.8--1.0. For z$<$0.6, the
detection of sources is dominated by the surveyed volume, and after
z$\sim$1.0, the detection probability decreases exponentially up to
z$\sim$4. About half of our sample lies at z$\gtrsim$0.9, $\sim$40\%
at z$>$1, and $\sim$20\% at z$>$1.5. The bulk of the galaxies in this
study ($\sim$90\%) lie at z$<$2. This implies that our results about
stellar mass functions and densities are very robust up to z$\sim$2,
just where our photometric redshifts are empirically well
tested. Beyond that point, we still include $\sim$3000 galaxies,
enough to still obtain statistically meaningful results (although
systematic errors such as redshift outliers will also contribute more
to the errors above z$=$2). 

The statistics for the $I$-band selected sample are very similar to
those for the IRAC sample: the average distribution peaks at around
z$=$0.7, and then decays exponentially, enclosing about 50\% of the
sources below z$=$0.9, 80\% at z$<$1.5, and 10\% at
z$>$2.0. Figure~\ref{photoz_distro} shows that most of the galaxies
included in the $I$-band selected sample and missed by IRAC lie at
z$\lesssim$1.5. At higher redshifts, the number densities of the
$I$-band and IRAC samples are almost identical, which is consistent
with the fact that more than 90\% of the IRAC sources were detected in
our deep Subaru $I$-band images (see Appendix~\ref{catalog}). This
means that the $I$-band mass completeness level is very similar to the
IRAC level, except for z$\lesssim$1.5, where the $I$-band should help
to probe (slightly) lower masses than the IRAC selection (see
Figure~\ref{massfunctions}). We would need optical images deeper than
$I$$\sim$25.5 to detect less massive systems at high redshift.

Figure~\ref{photoz_distro} also shows the redshift distribution of the
IRAC sources detected by MIPS at 24\mic\, and having
$F(24)$$=$80\micJy\, (dashed lines). The redshift distribution is
similar to that presented in \citet{2005ApJ...630...82P}, but the
improvement in the photometric redshift estimations reveals a more
pronounced density bump at z$\sim$1.7 and a weak bump at
z$\sim$2.6. The origin of these bumps can be found in the increase of
the detection probability induced by prominent PAH features entering
the MIPS 24\mic\, filter as we move to higher redshifts \citep[see
also][]{2006ApJ...637..727C}. Indeed, a typical PAH spectrum shows an
absence of features around $\lambda$$=$10\mic, which produces the
detection local minimum at z$\sim$1.3 observed in
Figure~\ref{photoz_distro}. At 6$\lesssim$$\lambda$$\lesssim$10\mic,
there are several PAH features (the most prominent at 5.5\mic\, and
7.7\mic) that are responsible for the bumps in the redshift
distribution. Note that the final detected density for MIPS sources is
a convolution of the real redshift distribution of galaxies (affected
by large scale structure), the detection probability (dependent on the
limiting flux of the survey and the spectra of the galaxies), and the
photometric redshift uncertainty distribution. These two effects
(detection probability and redshift uncertainties) result in blurring
out redshift-dependent features so they are at lower contrast to the
overall real distribution.

\section{STELLAR MASS FUNCTIONS AND DENSITIES}
\label{stellarmass}

\subsection{Completeness of the sample}
\label{completeness}

\begin{figure}
\begin{center}
\includegraphics[angle=-90,width=8.cm]{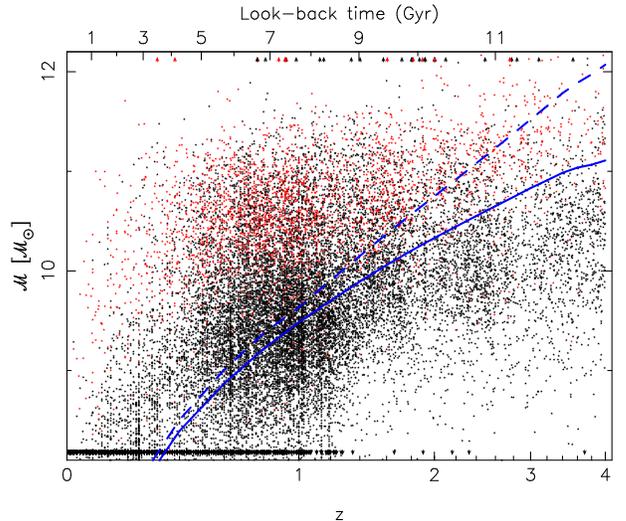}
\figcaption{\label{mass_limits}Distribution of the stellar masses of
all individual galaxies in the IRAC (all symbols) and MIPS (red
symbols) selected samples as a function of redshift (shown with a
logarithmic scale in the quantity 1+z in the bottom horizontal axis
and the corresponding look-back times in the top axis). The continuous
blue line shows the stellar mass value at each redshift above which
our IRAC survey is 75\% complete for passively evolving galaxies. The
dashed blue line shows the completeness for highly extincted
[$E(B-V)$$=$1.1] bursts. Sources whose stellar mass is beyond the
vertical axis scale are plotted with arrows at the source redshift.}
\end{center}
\end{figure}

\slugcomment{Please, plot this figure with the width of one column}
\placefigure{mass_limits}

Figure~\ref{mass_limits} shows the distribution of stellar masses of
individual galaxies in our IRAC survey as a function of redshift. The
blue continuous line shows the stellar mass corresponding to a
passively evolving galaxy formed in a single instantaneous burst of
star formation occurred at z$=$$\infty$ and having a 3.6\mic\, flux
equal to the 75\% completeness level of our IRAC sample. The stellar
mass calculated in this way assumes the maximum mass-to-light ratio
given by the oldest instantaneously formed stellar population possible
at each redshift. Any burst occurring after the primary placed at
z$=$$\infty$ should decrease the observed mass-to-light ratio (unless
it presents a high attenuation, see below), thus giving a smaller
stellar mass. Therefore, the values given by the blue continuous curve
in Figure~\ref{mass_limits} are the minimum stellar masses that a
maximally old galaxy with a flux equal to the 3.6\mic\, 75\%
completeness level should present, and our survey must be complete
(actually, at least 75\% complete) against passively evolving galaxies
with masses above the continuous curve. As noted by
\citet{2006A&A...459..745F}, high mass-to-light ratios can also be
found in galaxies with very extincted bursts. The dashed blue line in
Figure~\ref{mass_limits} shows the completeness level of our survey
for instantaneous star-forming bursts extincted by $E(B-V)$$=$1.1
magnitudes (as used by \citealt{2006A&A...459..745F}, based on the
typical extinction of highly obscured high redshift galaxies) and
following a \citet{2000ApJ...533..682C} extinction law.

Note that if the density of galaxies of a given stellar mass at a
certain redshift is very small, our surveyed volume may not be large
enough to enclose any galaxy of that mass (we would not detect any,
although there might exist galaxies of that mass in the Universe at
that redshift). This is the effect seen in Figure~\ref{mass_limits} at
high stellar masses: at z$<$0.2, our surveyed volume is not enough to
detect galaxies with
$\mathcal{M}$$\gtrsim$$10^{11.0}$~$\mathcal{M}_\sun$, and we can only
detect galaxies with
$\mathcal{M}$$\gtrsim$$10^{12.0}$~$\mathcal{M}_\sun$ at
z$\gtrsim$0.6. It is also interesting to notice that the most massive
galaxies with $\mathcal{M}$$\gtrsim$$10^{12.0}$~$\mathcal{M}_\sun$ are
only found in the regions presenting the highest densities, just where
the redshift distribution for individual fields peak (see
Figure~\ref{photoz_distro}).

The estimations of the stellar mass functions in the following
sections will be carried out for stellar masses above the completeness
level (against passively evolving galaxies) shown in
Figure~\ref{mass_limits} (with a continuous line), i.e., no
completeness correction will be carried out to try to recover the
stellar mass function at smaller masses (below the blue continuous
curve in Figure~\ref{mass_limits}).

\subsection{Stellar mass function estimation procedure}

The entire redshift range 0$<$z$<$4 was divided into 12 intervals, the
size of each bin chosen to have a statistically representative number
of galaxies and taking into account the typical photometric redshift
errors. Our goal was to estimate stellar mass functions at each
redshift bin. Classical methods to obtain luminosity functions or mass
function \citep[see][ for a discussion about
them]{1997AJ....114..898W} rely heavily on the use of a flux band on
which the selection of the studied sample is based. If the band where
the selection is based is far from the band where we want to estimate
the luminosity function (or, in the case of estimating stellar mass
functions, the magnitude is not directly and easily linked to the
stellar mass of each galaxy), significant systematic errors are
introduced \citep[see,
e.g.,][]{2000MNRAS.312..557L,2004MNRAS.351..541I}. In our case, our
selection is carried out in luminosity at 3.6-4.5\mic, but we want to
obtain a stellar mass function, which is linked to that luminosity
(but not directly proportional). To solve this problem, we estimated a
bivariate luminosity-stellar mass function for each redshift bin. The
procedure is identical to that used in \citet{2003ApJ...587L..27P},
and accounts for the fact that the selection of the sample is carried
out in a certain photometric band, while we eventually want the number
density function relative to a different parameter (in our case, the
stellar mass). The bivariate luminosity-stellar mass function, BLMF or
$\Phi(L,\mathcal{M})$, is defined as the number density of galaxies
(in a limited co-moving volume given by our surveyed area and the
redshift interval considered) with a given luminosity in a certain
band and a given stellar mass. This definition is an extension of the
bivariate luminosity function
\citep{2000MNRAS.312..557L}. The estimation of the BLMF was performed
with a stepwise maximum likelihood (SWML) technique \citep[][see also
\citealt{1997AJ....114..898W}]{1988MNRAS.232..431E}, extended to consider 
two independent variables. 

To estimate stellar mass functions, we used the IRAC 3.6\mic\, band as
the luminosity variable in the BLMF, given that this is the filter
where the selection of the sample was carried out. For the $I$-band
selected sample, we used the $I$ filter as the selection band. Once
the BLMF is estimated, if we integrate it through all luminosities, we
can estimate the number density of galaxies with a given stellar mass,
i.e., the stellar mass function, SMF or
$\phi_\mathrm{SM}(\mathcal{M})$. We only estimated the stellar mass
function down to the completeness threshold of the stellar mass
discussed in Section~\ref{completeness}.

In the classical SWML method, the errors in the BLMF are estimated
from the covariance matrix. In our case, the estimation of the BLMF
uncertainties was carried out by combining the SWML technique with a
Montecarlo method to take into account the photometric redshift errors
and outliers, as we did in
\citet{2005ApJ...630...82P} following the procedure described in 
\citet{2003ApJ...586..745C}. We considered the
photometric redshift as a statistical variable whose error comes from
the comparison with spectroscopic redshifts. These errors depend on
the actual redshift of the galaxy, so we considered different
photometric redshift uncertainty distributions for different redshift
intervals. We also considered the dependence of the redshift
uncertainties on the apparent brightness of the source (more accurate
photometry allows better estimations of the photometric redshift),
dividing the redshift dependent distribution of redshift uncertainties
into magnitude bins. For z$>$1.5, where very few spectroscopic
redshifts are available to test our photometric redshifts, we only
considered one single redshift and magnitude interval. The Montecarlo
method uses the redshift uncertainties based on the comparison with
spectroscopy, given that they are more reliable (they directly test
the goodness of the photometric redshifts) than the errors derived
from the probability distribution based on the $\chi^2$ minimization,
and they include the effect of outliers. The Montecarlo extension to
the SWML method consists in calculating the stellar mass function by
randomly varying the redshifts of the whole sample according to the
distribution of uncertainties (which are usually non-Gaussian), and
calculate the stellar mass function again. After 100 iterations, the
average and standard deviations of each point in the stellar mass
function are taken as the final results.

The results for the SMF (data points and uncertainties\footnote{Note
that the data point at $\mathcal{M}$$=$$10^{12.0}$~$\mathcal{M}_\sun$
in each SMF, which accounts for the very few high-mass sources
discussed in Section~\ref{completeness}, presents a very large
uncertainty (as it includes very few sources) and have a negligible
effect on the Schechter fits.}) were fitted with a smooth function
using a
\citet{1976ApJ...203..297S} parametrization, to facilitate comparison 
with similar fits in the literature. Both the IRAC and $I$-band SMF
estimations were used in the fits, down to the IRAC completeness
level. For the five bins at highest redshifts, the faint-end slope of
the SMF was poorly constrained by our data, so we combined our results
with other estimations of the stellar mass functions found in the
literature.  These estimations are based on the analysis of galaxy
samples typically selected at optical wavelengths, which is more
effective at probing the low mass regime of the stellar mass
function. We only used literature data points at masses
$\mathcal{M}$$>$$10^{8}$~$\mathcal{M}_\sun$ at z$<$1.3 and
$\mathcal{M}$$>$$10^{9}$~$\mathcal{M}_\sun$ at z$>$1.3, where the
completeness levels of these optically based samples are supposed to
be high (based on the sharp turnovers of these SMFs). Note that the
completeness of the optically selected samples at masses around
$\mathcal{M}$$\sim$$10^{8-9}$~$\mathcal{M}_\sun$ is difficult to
estimate (due, for example, to the significant effect of the
extinction, and the need of extremely deep data to probe this mass
regime) and is not well understood (it is not discussed in the
reference papers), so the low mass slopes at z$\gtrsim$2 should be
taken with caution. The SMF points, errors, and Schechter fits for
each redshift bin are shown in Figures~\ref{local_massfunction} and
~\ref{massfunctions} (black filled and open stars for the IRAC and
$I$-band selected samples, respectively). These figures also show
other SMF estimations found in the literature (color points, see
captions for references). The plots also depict the SMFs and fits for
the subsample of galaxies detected simultaneously by IRAC and MIPS at
24\mic\, (filled circles). The data points and Schechter fit
parameters are given in the electronic Tables~\ref{tablesmf} and
\ref{schechter}, respectively.

\subsection{The local stellar mass function and density}

\slugcomment{Please, plot this figure with the width of one columns}
\placefigure{local_massfunctions}
\begin{figure}
\begin{center}
\includegraphics[angle=-90,width=8.cm]{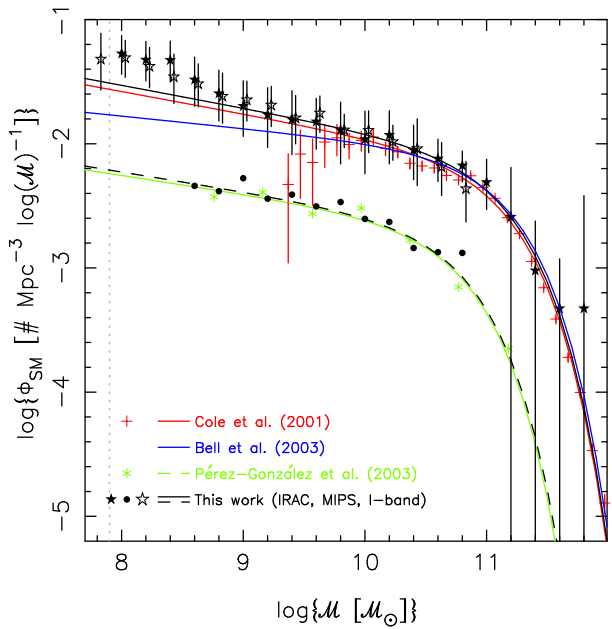}
\figcaption{\label{local_massfunction}Local stellar mass function
estimated with the IRAC selected (filled stars), $I$-band selected
(open stars), and MIPS selected (filled circles) samples at
z$<$0.2. For clarity, the $I$-band data points have been artificially
drifted from the original x-position (the same as the ones for the
IRAC selected sample) and we do not show the uncertainties for the
MIPS data points. The vertical gray dashed line shows the completeness
level of our IRAC survey in the local Universe. The Schechter fit to
the IRAC and $I$-band data (for masses
$\mathcal{M}$$>$$10^{9}$~$\mathcal{M}_\sun$) is shown with a black
continuous line. Our estimation of the local stellar mass function is
compared with the one estimated by \citet[][red crosses and
line]{2001MNRAS.326..255C}, and by \citet[][blue
line]{2003ApJS..149..289B}. The best Schechter fit to the data for the
MIPS sample (i.e., for local star-forming galaxies) is plotted with a
dashed line. This SMF is compared with the one published by
\citet[][green asterisks and line]{2003ApJ...587L..27P} for
H$\alpha$-selected local star-forming galaxies.}
\end{center}
\end{figure}

Figure~\ref{local_massfunction} shows our estimations of the local
stellar mass function (including sources at 0.0$<$z$<$0.2) based on
both the IRAC (filled stars) and $I$-band (open stars) selected
samples. Given that we are surveying a very limited volume in the
local Universe, we do not detect many sources with
$\mathcal{M}$$>$$10^{11.0}$~$\mathcal{M}_\sun$ (this explains the
large errors in this mass regime), but our statistics are much better
at low masses. Our Schechter fit only refers to the data points above
$\mathcal{M}$$>$$10^{9}$~$\mathcal{M}_\sun$, where our data is
directly comparable with previously published mass functions. Our
results are very similar to those published by
\citet{2001MNRAS.326..255C} and \citet{2003ApJS..149..289B} based on
NIR 2MASS data down to the completeness limit of their surveys
($\mathcal{M}$$\sim$$10^{9.5}$~$\mathcal{M}_\sun$). Our deeper data
confirm the faint-end slope estimated by \citet{2001MNRAS.326..255C}
down to even smaller masses,
$\mathcal{M}$$\sim$$10^{9.0}$~$\mathcal{M}_\sun$. We also find a
steepening of the stellar mass function at
$\mathcal{M}$$\lesssim$$10^{9.0}$~$\mathcal{M}_\sun$ (at least for
$\mathcal{M}$$\gtrsim$$10^{7.9}$~$\mathcal{M}_\sun$, our completeness
level at z$\sim$0).

By integrating our local SMF, we obtain a value for the local stellar
mass density of
$\rho_*$$=$$10^{8.75\pm0.12}$~$\mathcal{M}_\sun$\,Mpc$^{-3}$, in
excellent agreement with the values found in
\citet{1999MNRAS.309..923S},
\citet{2001MNRAS.326..255C}, and \citet{2003ApJS..149..289B}
($\rho_*$$=$$10^{8.75, 8.76, 8.74}$~$\mathcal{M}_\sun$\,Mpc$^{-3}$,
respectively). The good agreement of our local stellar mass function
and density with previous estimations found in the literature
demonstrates that aperture effects in our photometric catalogs are not
critical, i.e., they do not affect our results even at low redshifts
where the galaxies present relatively large angular sizes. The
steepening of the stellar mass function at
$\mathcal{M}$$\lesssim$$10^{9.0}$~$\mathcal{M}_\sun$ has no
significant effect on the integrated stellar mass density (justifying
the exclusion of these points in the Schechter fit): the galaxies with
$10^{7.8}$$<$$\mathcal{M}$$<$$10^{9.0}$~$\mathcal{M}_\sun$ contribute
less than 2\% to the total stellar mass density.

Figure~\ref{local_massfunction} also shows the SMF of the sources
detected by MIPS at 24\mic, i.e., the galaxies with active star
formation (filled circles and dashed line fit). The MIPS results (data
points and fit) are in excellent agreement with those published by
\citet{2003ApJ...587L..27P} for a  H$\alpha$-selected sample of
star-forming galaxies in the local Universe. The local stellar mass
density locked in star-forming galaxies is
$\rho^{SF}_*$$=$$10^{7.85\pm0.07}$~$\mathcal{M}_\sun$\,Mpc$^{-3}$,
i.e., 13$\pm$4\% of the global stellar mass density in the local
Universe is found in active star-forming
galaxies. Figure~\ref{local_massfunction} also shows that
approximately 1 of every 4 galaxies in the local Universe with
$\mathcal{M}$$\lesssim$$10^{10.5}$~$\mathcal{M}_\sun$ is forming stars
currently and would be detected in the IR or with a SFR tracer such as
the H$\alpha$ emission. At higher stellar masses, the fraction of
star-forming galaxies decreases by more than a factor of 2 (e.g.,
$\sim$10\% of all galaxies with
$\mathcal{M}$$=$$10^{11.0}$~$\mathcal{M}_\sun$ are forming stars
actively).

\subsection{The evolution of the stellar mass function}
\label{sectsmf}

\begin{figure*}
\begin{center}
\includegraphics[angle=-90,width=18.cm]{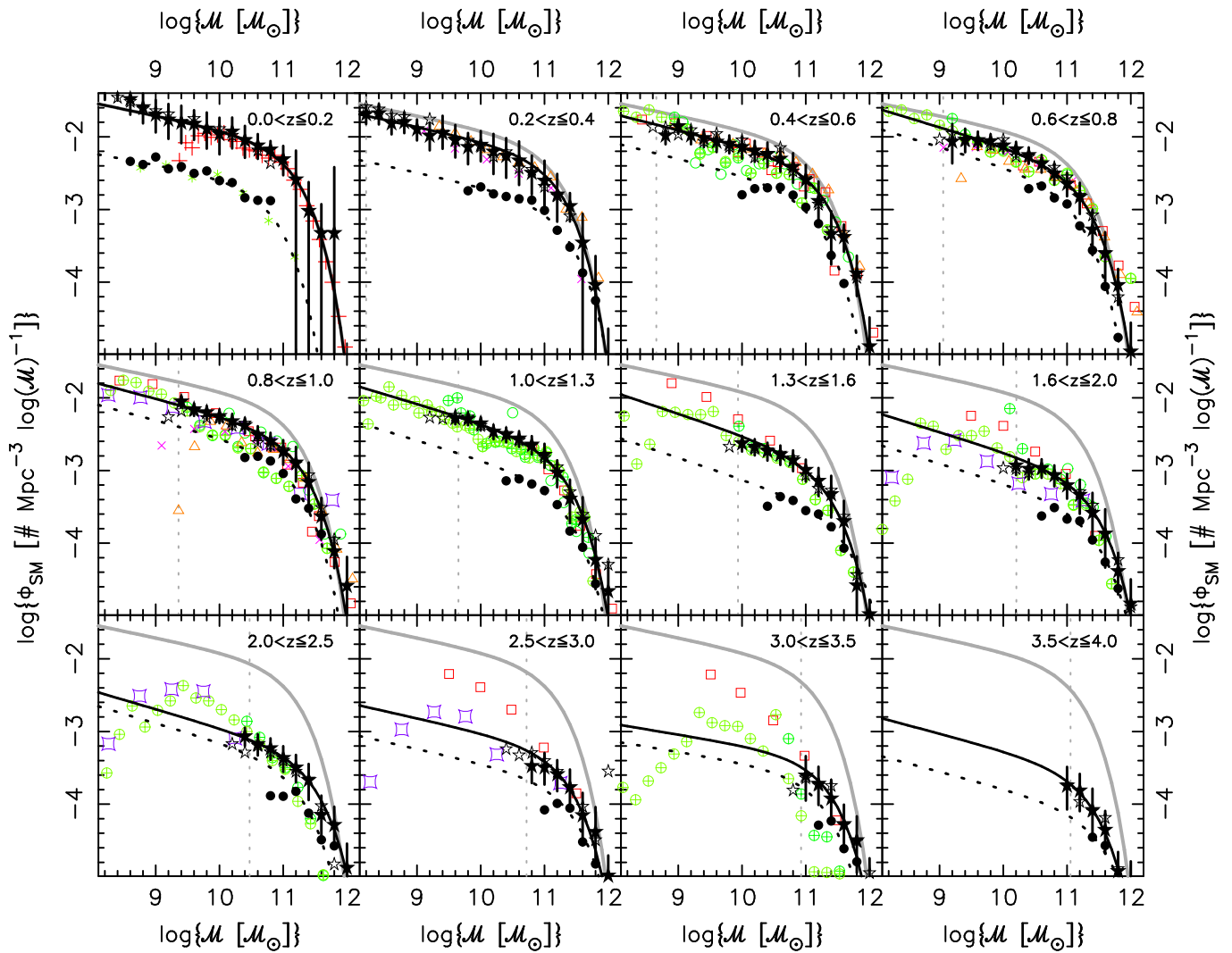}
\figcaption{\label{massfunctions}Stellar mass functions for 12 redshift 
intervals from z$=$0 to z$=$4. Our estimations at each redshift
interval are plotted with black filled stars and errors for the IRAC
selected sample, and with open black stars for the $I$-band selected
sample (errors for this sample are not plotted for clarity). Filled
circles show the SMF for galaxies detected by MIPS at 24\mic. The SMF
data (our estimations and others) are fitted with a
\citet{1976ApJ...203..297S} function (black continuous line for the
global SMF, and dashed line for the SMF for 24\mic\, sources). All
panels show the local SMF from
\citet{2001MNRAS.326..255C} with a gray curve. The vertical dotted 
line shows our 75\% completeness limit for the IRAC selected sample
(continuous curve in Figure~\ref{mass_limits}). Color points show
estimations from other papers: red crosses come from
\citet[][C01]{2001MNRAS.326..255C}; orange open triangles from
\citet[][B06]{2006A&A...453..869B}; magenta crosses from
\citet{2006ApJ...639L...1P}; red squares from
\citet[][D04]{2004ApJ...608..742D} 
and \citet[][D05]{2005ApJ...619L.131D}; green crossed circles from
\citet[][FC03]{2003ApJ...594L...9F}, \citet[][F04]{2004A&A...424...23F}, and 
\citet[][F06]{2006A&A...459..745F}; and purple squares from 
\citet[][C05]{2005ApJ...620..564C}. Green asterisks at 0.0$<$z$<$0.2 
show the stellar mass function of local star-forming galaxies
\citep[][P03]{2003ApJ...587L..27P}.}
\end{center}
\end{figure*}

\slugcomment{Please, plot this figure with the width of two columns}
\placefigure{massfunctions}

Figure~\ref{massfunctions} presents the global stellar mass functions
estimated in the 12 redshift intervals up to z$=$4. We show the
results obtained with the IRAC selected (filled black stars), the
$I$-band selected (open black stars), and the MIPS selected (filled
circles) samples. Other estimations found in the literature are also
plotted at each redshift interval (normalized to the
\citealt{1955ApJ...121..161S} IMF). We have fitted our SMF data points, 
and for z$>$1.5, also the data points from other works below our
completeness level to better constrain the slope at lighter masses, to
a Schechter function. In the case of the 24\mic\, galaxies, we assumed
the same faint-end slope estimated for the global SMF (based on the
IRAC and $I$-band samples).

Our estimation of the stellar mass function is consistent (within
errors and for the same mass ranges) with previous estimations found
in the literature. It is interesting to notice that the faint-end
slope derived by \citet{2005ApJ...619L.131D} at z$>$1.5 is
significantly larger than that found by other authors
\citep{2004A&A...424...23F,2005ApJ...620..564C}. Our results are in good
agreement with Drory's in the stellar mass range probed by both
surveys. The possible explanations for the discrepancy at low masses
are field-to-field variations, an overestimation of the faint-end
slope in \citet{2005ApJ...619L.131D} due to the use of the
$V_\mathrm{max}$ method \citep[see][]{2004MNRAS.351..541I}, or
systematic errors in the determination of the stellar masses. Due to
this discrepancy, we did not use these points in our SMF fits. At
z$>$2, we estimate number densities of massive galaxies which can be
up to 0.8dex higher than those estimated by
\citet{2006A&A...459..745F}. Some of this discrepancy (up to $\sim$20\%) 
may be due to the under-density observed in CDF-S (see
Section~\ref{sectdensity}).

Our results show that the local density of galaxies (shown with a gray
line in all panels) with masses
$\mathcal{M}$$\gtrsim$$10^{12}$~$\mathcal{M}_\sun$ was already reached
by the SMF at z$=$2.5-3.0, i.e., the most massive galaxies were
already in place at that redshift (approximately 11~Gyr ago). The mass
assembly of galaxies shifts to smaller masses as we move to lower
redshifts. By z$\sim$1, the SMF has reached nearly the local density
for galaxies with
$\mathcal{M}$$\gtrsim$$10^{11.8}$~$\mathcal{M}_\sun$. At z$<$1, the
star formation in the Universe occurs mainly in galaxies with
$\mathcal{M}$$\lesssim$$10^{11.5}$~$\mathcal{M}_\sun$. It is also
interesting to notice the significant evolution (approximately a
factor of 0.2dex or 60\%, as shown by our data, and also confirmed by
the results of \citealt{2006ApJ...639L...1P} and
\citealt{2006A&A...453..869B}) of the SMF between z$\sim$0.4 and z$=$0
(i.e., a period of 4~Gyr) for stellar masses
$10^{9}\lesssim\mathcal{M}$$\lesssim$$10^{11}$~$\mathcal{M}_\sun$. We
will comment more on this recent evolution in
Section~\ref{sectdensity}.

Figure~\ref{massfunctions} also shows that the slope of the SMF at low
masses remains approximately constant up to at least z$\sim$2 at a
value $\alpha$$=$-1.2$\pm$0.1 \citep[consistent with the models
in][]{2005ApJ...618...23N}. Only at very low masses
($\mathcal{M}$$\lesssim$$10^{9.0}$~$\mathcal{M}_\sun$ at z$<$1 and
$\mathcal{M}$$\lesssim$$10^{10.0}$~$\mathcal{M}_\sun$ at higher
redshifts), the SMF seems to become steeper (based on our results and
those from other surveys), but this steepening has a minor effect on
the global stellar mass density.

\subsection{The evolution of the cosmic stellar mass density}
\label{sectdensity}

\begin{figure*}
\begin{center}
\includegraphics[angle=-90,width=14.cm]{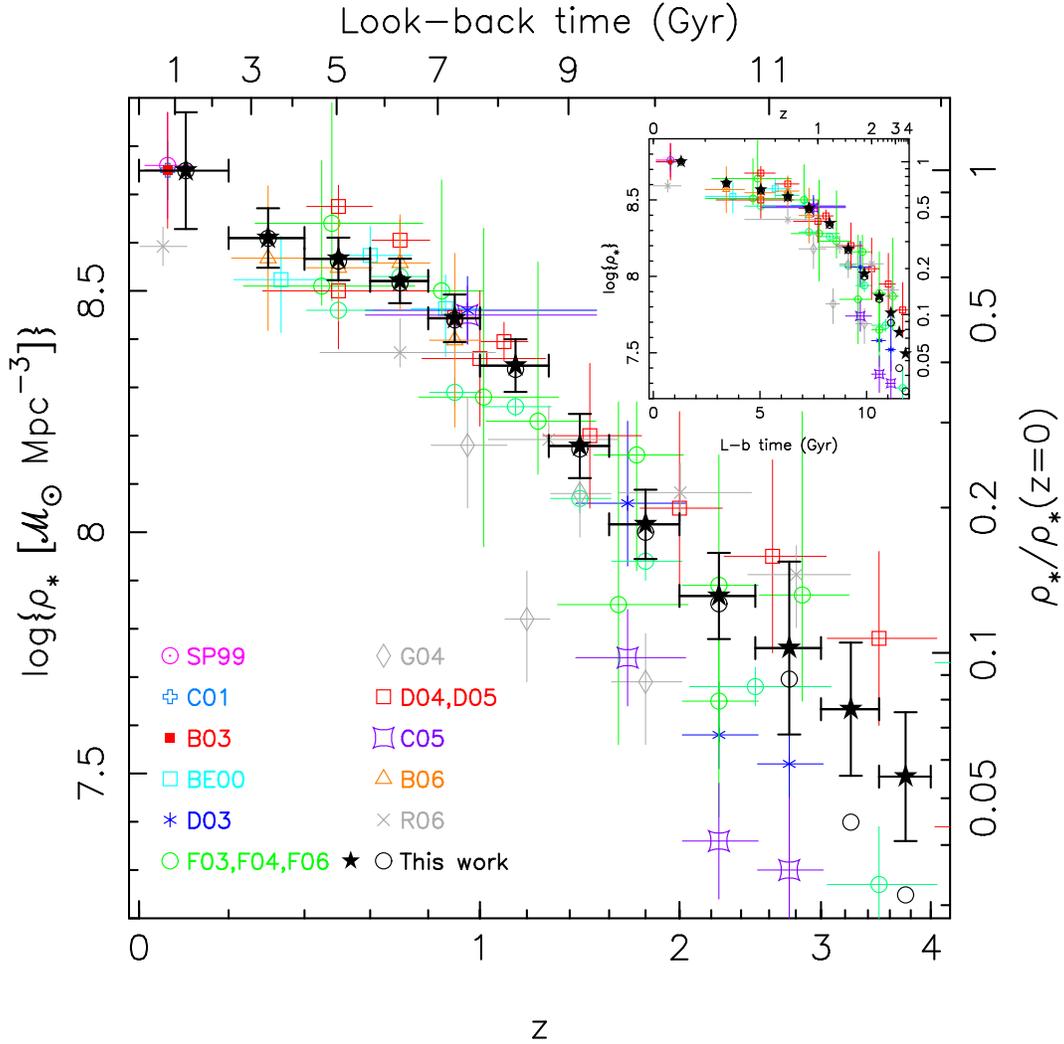}
\figcaption{\label{massdensity}Evolution of the stellar mass density 
of the Universe as function of redshift (shown with a logarithmic
scale in the quantity 1+z in the bottom horizontal axis and the
corresponding look-back times in the top axis). Our estimations are
plotted with black filled stars (based on the integration of the
stellar mass functions with a Schechter parametrization) and open
circles (observed values down to the completeness level). Color points
and error bars show other estimations found in the literature. To the
references mentioned in the caption of Figure~\ref{massfunctions}, we
have also added estimations from
\citet[][SP99]{1999MNRAS.309..923S},
\citet[][BE00]{2000ApJ...536L..77B}, \citet[][B03]{2003ApJS..149..289B}, 
\citet[][D03]{2003ApJ...587...25D},
\citet[][G04]{2004Natur.430..181G}, and \citet[][R06]{2006ApJ...650..624R}. 
The inset shows the same evolution of the stellar mass density of the
Universe, but this time with a linear scale in look-back time in the
horizontal axis.}
\end{center}
\end{figure*}

The SMFs were integrated for all masses above the completeness level
to obtain the observed cosmic co-moving stellar mass density. We also
integrated the Schechter fits to estimate an extrapolated value of the
cosmic stellar mass density at each redshift interval. In
Figure~\ref{massdensity}, we present these results, comparing them
with other estimations of the stellar mass density available in the
literature (see the captions of Figures~\ref{massfunctions} and
\ref{massdensity} for references). Note that the observed density values 
are very similar to the extrapolated ones up to z$\sim$2, i.e., our
survey is detecting most of the galaxies that dominate the global
stellar mass density at z$<$2. We calculate field-to-field variations
of the stellar mass density of the order of 30\%-40\% (depending on
the redshift) with respect to the average density. For example, the
stellar mass density locked in galaxies with
$\mathcal{M}$$>$$10^{11}$~$\mathcal{M}_\sun$ at z$>$2 is 15\%-20\%
lower in CDF-S than the average of the 3 fields, a slightly lower
under-density than that observed by \citet{2006ApJ...638L..59V}
comparing three $\sim$100~arcmin$^2$ fields (they calculate a 40\%
difference of the CDF-S with the average).


Figure~\ref{massdensity} shows that there is a relatively large
increase (by a factor of $\sim$1.4) in the stellar mass density of the
Universe in the last 4~Gyr (from z$\sim$0.4 to z$=$0). This large
difference could be due to an overestimation of the local stellar mass
density \citep[suggested by, for example,][]{2004A&A...424...23F} or
an underestimation of the density at z$\sim$0.3 (for example, if low
mass objects below our detection limit have a non-negligible
contribution to the stellar mass density at this redshift). However,
all the estimations of the local density are very similar (differences
of less than 5\% between our value and those found by
\citealt{1999MNRAS.309..923S}, \citealt{2001MNRAS.326..255C}, and
\citealt{2003ApJS..149..289B}; even higher values are found by
\citealt{1998ApJ...503..518F}, \citealt{2001ApJ...560..566K}, and
\citealt{2003ApJ...587...55G}), and the same occurs for the different
estimations at 0.2$<$z$<$0.4 (\citealt{2000ApJ...536L..77B} and
\citealt{2006A&A...453..869B}). As we discussed in
Section~\ref{sectsmf}, this significant recent evolution of the
stellar mass density is mainly due to a $\sim$60\% increase in the
number density of galaxies with
$10^{9}\lesssim\mathcal{M}$$\lesssim$$10^{11}$~$\mathcal{M}_\sun$. Assuming
an average value of the cosmic SFR density of approximately
0.03~$\mathcal{M}_\sun$yr$^{-1}$ at 0.0$<$z$<$0.4 \citep[][see also
\citealt{2007A&A...472..403T} and Figure~\ref{lillymadau}]{2006ApJ...651..142H},
and a 28\% gas recycle factor (see Section~\ref{role24mic} for
details), we calculate that the stellar mass density of the Universe
has grown in 10$^{8.0\pm0.1}$~$\mathcal{M}_\sun$\,Mpc$^{-3}$ from
z$=$0.4 to z$=$0.0 (in $\sim$4.3~Gyr) by just star formation. This is
55$\pm$10\% of the stellar mass density change at z$<$0.4.
Therefore, the remaining change in stellar mass density
($\sim$10$^{7.9\pm0.1}$~$\mathcal{M}_\sun$\,Mpc$^{-3}$) must have
occurred by either accretion of small satellite galaxies or by major
mergers between gas-depleted galaxies (i.e., mergers accompanied by
very little star formation), as also suggested by \citet{2007A&A...472..403T}. In
addition, given that both in the local Universe and at 0.2$<$z$<$0.4
the SMFs steepen at low stellar masses
($\mathcal{M}$$\lesssim$$10^{9.0}$~$\mathcal{M}_\sun$), the minor
merger possibility (accretion of
$\mathcal{M}$$\lesssim$$10^{9.0}$~$\mathcal{M}_\sun$ galaxies
producing very few or even no new stars at all) seems to be favored in
detriment of the existence of major mergers.

\slugcomment{Please, plot this figure with the width of two columns}
\placefigure{massdensity}


The evolution in the previous 3--4~Gyr (between z$\sim$1.0 and
z$\sim$0.4) was slightly slower. About 25\% of the local stellar mass
density was assembled in that period, adding up a total decrease of
about 50\% in the stellar mass density from z$=$0 to z$=$1.

At z$\sim$1.0 (8~Gyr ago), the evolution of the stellar mass density
of the Universe becomes faster (approximately a factor of 2), just
when the cosmic SFR density reaches a maximum
\citep[see, e.g.,][]{2005ApJ...630...82P} and the galaxies with 
$\mathcal{M}$$\gtrsim$$10^{10.5}$~$\mathcal{M}_\sun$ dominate the
production of stars in the Universe. The rate at which the Universe is
creating stars stays at approximately a constant level or decays very
slowly from z$\sim$1 up to at least z$\sim2$ (10~Gyr ago). Between
z$\sim$1 and z$\sim$2, the density of galaxies with
$\mathcal{M}$$\gtrsim$$10^{10.5}$~$\mathcal{M}_\sun$ decreases
significantly (by a factor of 3--4). This population of galaxies
evolving rapidly at 1$<$z$<$2 (in about 2~Gyr) seem to be dominated by
early-type objects \citep[see, e.g.,][]{2007ApJ...669..184A}.

Beyond z$\sim$2, the errors in the stellar mass density estimates and
the differences between the observed and extrapolated values become
increasingly larger. We find that the rate at which stars are being
formed remains constant or even increases slightly, while the giant
galaxies with $\mathcal{M}$$\sim$$10^{12.0}$~$\mathcal{M}_\sun$ are
finishing the assembly of most of their stellar mass.

These different steps in the assembly of the cosmic stellar mass
density depicted in Figure~\ref{massdensity} are consistent with the
latest results on the evolution of the observed UV luminosity density
of the Universe \citep{2007A&A...472..403T} and the evolution of the SFR density
\citep{2005ApJ...630...82P,2006ApJ...651..142H} up to z$\sim$5. 
The luminosity density presents a maximum at around z$=$1.2 with a
value approximately 6 times larger than the local UV luminosity
density. At z$>$1.2, the luminosity and SFR density evolution is
consistent with a constant. Our results are also consistent with the
hydrodynamical models of \citet{2006ApJ...653..881N}, which predict
that $\sim$60\% of the present stellar mass density was already formed
by z$=$1. However, the discrepancy is significant at z$>$1, where
these models predict a larger stellar mass density than any
observation (i.e., they predict a quicker formation of the most
massive galaxies). The semi-analytic models of
\citet{2000MNRAS.319..168C} match our results better at z$=$3--4,
where they predict a stellar mass density of about 10\% the present
value, but they fail to reproduce the evolution at low redshift.

\subsection{Quantifying ``downsizing''}
\label{downsizing}

\begin{figure}
\begin{center}
\includegraphics[angle=-90,width=8.cm]{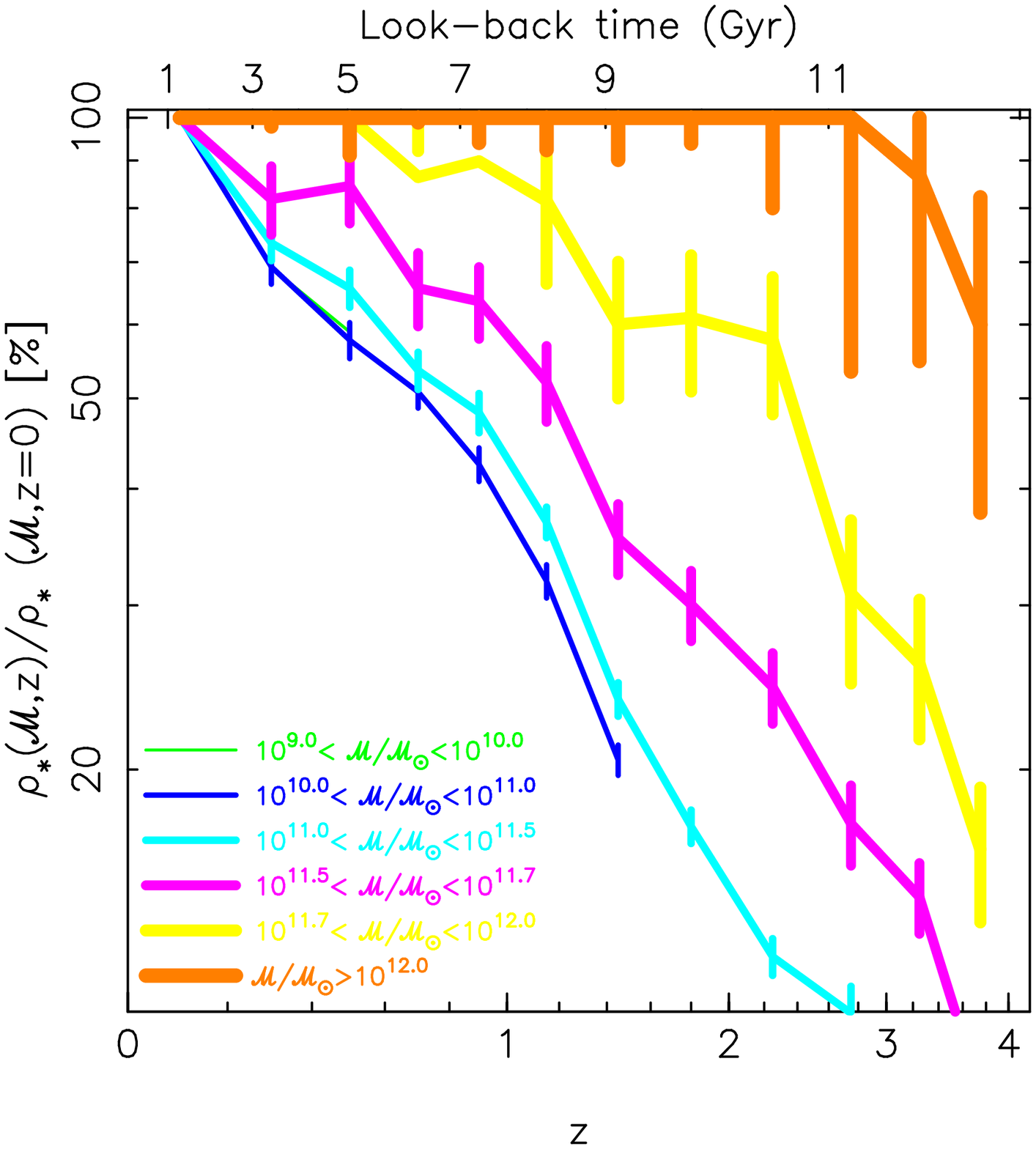}
\figcaption{\label{ageformation}Fraction of the local stellar mass 
density already assembled at a given redshift for several mass
intervals (wider lines referring to more massive systems). Only
results for masses above our 75\% completeness level at each redshift
are shown. }
\end{center}
\end{figure}

\slugcomment{Please, plot this figure with the width of one column}
\placefigure{ageformation}


The previous discussion about the evolution of the cosmic stellar mass
density is clearly consistent with a ``downsizing'' scenario for
galaxy formation. We quantify some properties of this ``downsizing''
theory in Figure~\ref{ageformation}, were we plot the fraction of the
total local stellar mass density already assembled in galaxies of a
given stellar mass at each redshift. This Figure shows that the most
massive systems ($\mathcal{M}$$\gtrsim$$10^{12.0}$~$\mathcal{M}_\sun$,
orange widest continuous line) formed first (they assembled more than
80\% of their total stellar mass before z$=$3) and very rapidly (about
40\% of their mass was assembled in 1~Gyr between z$=$4 and
z$=$3). Systems with masses
$10^{11.7}$$<$$\mathcal{M}$$<$$10^{12.0}$~$\mathcal{M}_\sun$ assembled
their stellar mass more slowly: from z$\sim$4 to z$\sim$2.5 (1.5~Gyr),
they assembled around 50\% of their stars, and then evolved more
slowly to reach the local density at low redshift. Less massive
galaxies assembled their mass at even a slower speed, reaching the
local density at very recent epochs. Again, this plot shows the rapid
recent evolution of the galaxies with masses
$\mathcal{M}$$\sim$$10^{10.5}$~$\mathcal{M}_\sun$, which assembled
$\sim$30\% of their mass in the last 3~Gyr. 


Our results are consistent with the stellar population models assumed
by \citet{2007ApJ...654..858B} for the most luminous (and probably
most massive) red galaxies. According to that paper, red massive
galaxies start forming at an early epoch, at z$=$4, following an
exponential SF law with a short
$\tau$$=$0.6~Gyr. \citet{2007ApJ...669..947J} find that the most
massive early-type galaxies in the local Universe formed at z$>$2.5
and experienced a very rapid chemical enrichment, lasting
1-2~Gyr. Also \citet{2005ApJ...631..145V} find signs of the formation
of massive ($\mathcal{M}$$>$$2\times10^{11}$~$\mathcal{M}_\sun$,
according to these authors) early-type galaxies at z$\gtrsim$2, while
less massive systems present lower formation redshifts
(1$<$z$<$2). The analysis of optical spectra for spheroidal and
bulge-dominated galaxies at 0.2$<$z$<$1.2 by
\citet{2005ApJ...633..174T} also reveals that most of the mass (99\%)
in systems with $\mathcal{M}$$>$$10^{11.5}$~$\mathcal{M}_\sun$ formed
at z$>$2, while most recent bursts (at z$\sim$1.2) can account for
20\%-40\% of the total stellar mass of galaxies with
$\mathcal{M}$$<$$10^{11.0}$~$\mathcal{M}_\sun$. \citet{2004Natur.430..181G}
estimates that 38$\pm$18\% of the stellar mass density in galaxies
with $\mathcal{M}$$>$$10^{10.8}$~$\mathcal{M}_\sun$ were already in
place at z$=$1, consistent with our value of $\sim$45\%. At z$=$1.8,
\citet{2004Natur.430..181G} obtain 16$\pm$6\%, also in agreement with 
our own estimation of $\sim$21\%.

If we consider a high value of the fraction of the stellar mass
density already assembled at a given redshift, above which the star
formation in a galaxy should be relatively low, e.g., 70\%,
Figure~\ref{ageformation} shows that galaxies with
$\mathcal{M}$$\sim$$10^{10.5}$~$\mathcal{M}_\sun$ reached that level
at z$\sim$0.2, systems with
$\mathcal{M}$$\sim$$10^{11.25}$~$\mathcal{M}_\sun$ at z$\sim$0.4, and
galaxies with $\mathcal{M}$$\sim$$10^{11.75}$~$\mathcal{M}_\sun$
around z$\sim$0.7. These numbers are roughly consistent with the
(1+z)$^{3.5}$ evolution estimated by
\citet{2006ApJ...651..120B} for the quenching stellar mass 
($\mathcal{M}_\mathrm{Q}$), a mass limit above which the star
formation appears to be suppressed. 

According to Figure~\ref{ageformation}, 50\% of the stars in galaxies
with $\mathcal{M}$$>$$10^{11.0}$~$\mathcal{M}_\sun$ were already in
place at z$\sim$0.9. This compares well with the prediction from the
models in \citet{2006MNRAS.366..499D} which establish that half of the
stars in objects of this mass are assembled into single objects at
z$\sim$0.8. However, these models also predict that most of these
stars were already formed at z$\sim$2.5, but were placed in several
objects that would coalesce into a single object later in a
hierarchical way. Our results favor a dual scenario where the most
massive systems with
$\mathcal{M}$$\gtrsim$$10^{12.0}$~$\mathcal{M}_\sun$ formed most of
their stars at z$>$2.5 and assembled very rapidly in a way closer to a
monolithic collapse than to a hierarchical coalescence. At the same
time, less massive systems
($10^{11.0}$$\gtrsim$$\mathcal{M}$$\gtrsim$$10^{12.0}$~$\mathcal{M}_\sun$)
could have formed their stars later and/or assembled half of their
mass from several progenitors (where stars were already formed at
z$\sim$2.5) in the time interval between z$\sim$2.5 and z$\sim$1
(about 4~Gyr), and most of their mass (80\%) not before z$\sim$0.5. To
confirm this scenario, it would be necessary to probe the stellar mass
function at low masses (for objects that would act as building blocks
for the galaxies with
$\mathcal{M}$$\gtrsim$$10^{11.0}$~$\mathcal{M}_\sun$), but the scatter
of the currently available SMF estimations at low masses in this
redshift range is too large to obtain robust results (maybe due to
cosmic variance effects). Indeed, our estimations of the cosmic
stellar mass density at z$>$3 are affected by the large uncertainties
at masses below $10^{11}$~$\mathcal{M}_\sun$ (this explains the large
difference between the observed and extrapolated values of the density
at z$>$3). The dual galaxy formation scenario (quasi-monolithic and
rapid collapse of the most massive galaxies which cease to form stars
at a certain epoch, and hierarchical collapse for less massive
systems) has been reproduced by other models where AGNs are supposed
to quench the star formation in very massive halos \citep[see,
e.g.,][]{2006MNRAS.365...11C,2006MNRAS.370..645B}.

\section{THE NATURE OF THE IRAC SAMPLE: COMPARISON WITH OTHER SURVEYS}
\label{comparison}

In this Section, we will discuss the main properties of the sources in
our IRAC sample, comparing them with the populations of galaxies
detected with different selection techniques by other surveys. The
results discussed in this Section are summarized in
Tables~\ref{type_fractions} and \ref{type_densities}.

Based on the observed photometric data points and the SED fit for each
galaxy in our sample, we estimated synthetic observed magnitudes in 9
bands ($FUV$, $NUV$, $U_n$, $B$, $\mathcal{G}$, $\mathcal{R}$, $z$,
$J$, and $K_s$) in order to test which of our galaxies would qualify
as Lyman Break Galaxies (LBGs) at z$=$1.5--2.5
\citep[LBG-BM and LBG-BX galaxies in][]{2004ApJ...604..534S}, at
z$\sim$3 \citep[``classical'' LBGs, ][]{2003ApJ...592..728S}, and at
z$\sim$1 \citep[GALEX LBGs, ][]{2006A&A...450...69B}, and which of our
galaxies would qualify as Distant Red Galaxies
\citep[DRGs;][]{2003ApJ...587L..79F,2003ApJ...587L..83V} or $BzK$ sources 
\citep{2004ApJ...617..746D}. Our analysis is similar to that used by 
\citet{2007AJ....134.1103Q} and \citet{2007A&A...465..393G}.

We identified LBGs following \citet{2003ApJ...592..728S} and
\citet{2004ApJ...604..534S}, which establish the locus of LBGs in a
$Un-\mathcal{G}$ vs. $\mathcal{G}-\mathcal{R}$ color-color diagram,
and adopt the magnitude cut ($R$$<$25.5) for their survey. We
identified GALEX LBGs with an analog procedure, but this time using a
color criteria based on GALEX UV photometric bands
\citep[see][]{2006A&A...450...69B}. Following
\citet{2003ApJ...587L..79F}, we defined DRGs as the galaxies
presenting a color $J-K_s$$>$1.37 [corresponding to
$(J-K_s)_\mathrm{Vega}$$>$2.3]. Finally, we identified star-forming
$BzK$ galaxies ($BzK$-SF) and passively evolving $BzK$ galaxies
($BzK$-PE) using Equations (2) and (3) in
\citet{2004ApJ...617..746D}. In our IRAC selected sample, we
identified 6,656 sources as LBGs with $R$$<$25.5 (summing up all
types), 763 sources as DRGs with $K$$<$22.9, and 2,426 as $BzK$
sources with $K$$<$22.9 and z$>$1.4 (summing up the two types).

The average surface density of LBGs (including all sub-types) with
$R$$<$25.5 detected by our IRAC survey is 10.0~LBGs~arcmin$^{-2}$. We
detect 0.7~LBGs~arcmin$^{-2}$ with the GALEX bands and $NUV$$<$25.0, a
higher density than the one given by
\citet[][0.3~$\mathrm{arcmin}^{-2}$]{2006A&A...450...69B}, but closer
to the density given in
\citet[][1.0~$\mathrm{arcmin}^{-2}$]{2007astro.ph..1322B}. We find
4.6~LBG-BMs~$\mathrm{arcmin}^{-2}$ (5.3~LBG-BMs~$\mathrm{arcmin}^{-2}$ without any optical
magnitude cut), 3.1~LBG-BXs~$\mathrm{arcmin}^{-2}$ (3.6~LBG-BXs~$\mathrm{arcmin}^{-2}$
without any optical magnitude cut), and 1.6~``classical''
LBGs~$\mathrm{arcmin}^{-2}$ (2.0~``classical'' LBGs~$\mathrm{arcmin}^{-2}$ at any
$R$-magnitude), very similar values to those found in
\citet[][5.3~LBG-BMs~$\mathrm{arcmin}^{-2}$ and 3.6~LBG-BXs~$\mathrm{arcmin}^{-2}$]
{2004ApJ...604..534S} and
\citet[][1.7~``classical'' LBGs~$\mathrm{arcmin}^{-2}$]{2003ApJ...592..728S}.
\citet{2004ApJ...607..226A} and \citet{2007A&A...465..393G} also 
find very similar surface densities for the 3 types of LBGs at
z$>$1. The median magnitudes for the LBG sub-sample are $R$$=$24.6 and
$K$$=$23.1, a very faint NIR magnitude only reachable by the deepest
ground-based or IRAC surveys. The average photometric redshifts for
the LBGs in our sample are consistent with the literature, as
discussed in Appendix~\ref{models_app}, jointly with the results for
the other populations of high redshift galaxies.

The average surface density of DRGs in our IRAC survey is
1.8~DRGs~$\mathrm{arcmin}^{-2}$ (1.1~DRGs~$\mathrm{arcmin}^{-2}$ for
sources with $K$$<$22.9), a value in between the densities quoted by
\citet[][3.0~DRGs~$\mathrm{arcmin}^{-2}$]{2003ApJ...587L..79F}, 
\citet[][1.0--1.6~DRGs~$\mathrm{arcmin}^{-2}$]{2004ApJ...616...40F}, and 
\citet[][0.8~DRGs~$\mathrm{arcmin}^{-2}$]{2006ApJ...640...92P}. The median 
magnitudes for the DRG sub-sample are $R$$=$25.7 (a very faint optical
magnitude beyond the reach of most UV/optical surveys) and
$K$$=$22.6. 

The average surface density of $BzK$ galaxies at z$>$1.4 down to
$K$$=$21.9 is 1.3~$BzK$s~$\mathrm{arcmin}^{-2}$, divided into
0.2~$BzK$-PE~$\mathrm{arcmin}^{-2}$ and 1.1~$BzK$-SF~$\mathrm{arcmin}^{-2}$. This is
consistent with the densities given in
\citet[][0.22~$BzK$-PE~$\mathrm{arcmin}^{-2}$ and
0.91~$BzK$-SF~$\mathrm{arcmin}^{-2}$]{2004ApJ...617..746D} for the same
brightness limit. At fainter magnitudes, $K$$<$22.9, we identify
0.4~$BzK$-PE~$\mathrm{arcmin}^{-2}$ and 3.3~$BzK$-SF~$\mathrm{arcmin}^{-2}$, close to
the values found by \citet[][0.24~$BzK$-PE~$\mathrm{arcmin}^{-2}$ and
3.1~$BzK$-SF~$\mathrm{arcmin}^{-2}$] {2006ApJ...644..792R} and
\citet[][0.65~$BzK$-PE~$\mathrm{arcmin}^{-2}$ and
3.2~$BzK$-SF~$\mathrm{arcmin}^{-2}$]{2007A&A...465..393G} for the same
magnitude cut. At even fainter $K$-band magnitudes, the source density
of galaxies identified as $BzK$ continues rising (especially the SF
sub-type) as redshift interlopers become more numerous (up to
40\%). 

These figures (densities and average redshifts) demonstrate that our
IRAC survey constitute an almost complete census of the previously
detected galaxies at 1.5$\lesssim$z$\lesssim$4, including most of the
LBGs, DRGs, and $BzK$ sources, the most important populations of
galaxies selected at z$>$1. Still, some of the IRAC sources are not
recovered by any of these selection criteria (even when no magnitude
cut is performed for LBGs, DRGs, or $BzK$ sources). The numbers of
these galaxies recovered only by the deep IRAC observations are given
in Table~\ref{type_fractions}.

The LBG population accounts for a negligible fraction (less than 10\%)
of the entire IRAC sample at 0.4$<$z$<$1.0. At z$<$0.4, $\sim$30\% of
IRAC galaxies are classified as LBGs (40\% at 0.0$<$z$<$0.2 and 20\%
at 0.2$<$z$<$0.4), most of them within the LBG-BX sub-type, which has
a significant fraction of z$<$1 interlopers at bright apparent
magnitudes \citep[see][]{2004ApJ...604..534S}. LBGs selected with
GALEX bands are also a minor fraction (around 5\%) of the total number
of IRAC galaxies at 0.8$<$z$<$1.3. However, at z$>$1, other LBG
sub-type start to be very numerous and even dominate the IRAC galaxy
counts: $\sim$35\% of all the sources in our IRAC survey at
1.0$<$z$<$1.3 are LBGs (80\% of them LBG-BMs), 50\%--60\% at
1.3$<$z$<$3.0 (with similar contributions from the different
sub-types), 65\% at 3.0$<$z$<$3.5 (all of them ``classical'' LBGs),
and 50\% at 3.5$<$z$<$4.0 (all of them ``classical'' LBGs). These
fractions are slightly higher for LBGs not limited by any $R$-band
magnitude.

The median stellar masses of LBGs range from
$10^{9.6}$~$\mathcal{M}_\sun$ to $10^{10.2}$~$\mathcal{M}_\sun$ at
1$<$z$<$2.5. These values are 0.1--0.2dex lower than the median
stellar masses for the global population of IRAC sources at each
redshift interval. For this reason, although their numbers are
relatively large (even dominate the number counts), LBGs have a less
important contribution to the global stellar mass density. Indeed, at
1.0$<$z$<$1.6, they harbor less than 25\% of the total stellar mass
density\footnote{This percentage has been calculated by adding the
total stellar masses of LBGs in that redshift interval, and dividing
it by the total stellar masses of all galaxies. This must be analogous
to dividing the stellar mass densities of both galaxy populations for
a fixed volume (that enclosed at the given redshift interval).}. At
1.6$<$z$<$4.0, they account for 35\%--45\% of the total stellar mass
density (roughly consistent with the estimations from the models in
\citealt{2005ApJ...618...23N}). These percentages increase by
5\%--15\% if we consider all LBGs without any $R$-band cut, then
making our estimations consistent with those in
\citet[][where they do not apply any magnitude
cut]{2007A&A...465..393G}.

Around 10\% of LBGs are detected by MIPS at 24\mic\, above the 75\%
completeness flux level (20\% with any flux), especially at
1.6$<$z$<$2.5, where MIPS is more efficient detecting sources due to
the pass of the 7.7\mic\, PAH feature through the filter. At z$>$2.5,
the fraction of MIPS detections is about 10\%, consistent with
\citet{2005ApJ...634..137H}. MIPS detections are more common for the
highest mass galaxies: the median stellar mass for 24\mic\, detected
LBGs ($10^{10.9}$~$\mathcal{M}_\sun$ at z$\sim$2.5 and
$10^{11.1}$~$\mathcal{M}_\sun$ at z$\sim$3.0 for sources with
$F(24)$$=$80\micJy) is $\sim$0.8dex larger than the median for all
LBGs. LBGs with MIPS detections account for 10\%--20\% of the total
stellar mass at z$>$1.5.


In contrast with the previous figures for LBGs, DRGs are less numerous
but more massive. DRGs only account for 15\% of the sources at
2.0$<$z$<$2.5, and $\sim$30\% at 2.5$<$z$<$4.0. However, their median
stellar masses are larger than those of LBGs at each redshift, and
even larger than the median for the entire population of IRAC
sources. For 1$<$z$<$2, their median masses are 0.3-0.5dex larger than
those for the entire IRAC population, and at z$>$2 they remain
$\sim$0.6dex larger (with a median of
$10^{11.0}$~$\mathcal{M}_\sun$). This translates to DRGs accounting
for 70\% of the total stellar mass density at z$>$2.5, $\sim$35\% at
2.0$<$z$<$2.5, and less than 10\% below z$=$2. These figures are very
similar to those found by \citet{2006ApJ...650..624R}, who find that
DRGs contribute 30\% and 64\% to the stellar mass density at z$\sim$2
and z$=$2.8, respectively. They are also consistent with the results
obtained by \citet{2007A&A...465..393G}, \citet{2007ApJ...656...42M}
and \citet{2006ApJ...638L..59V}. Note that most of the stellar mass
density of the Universe at z$>$2.5 would not be detected by optical
surveys reaching depths brighter than $R$$\sim$25.5. Consistently with
\citet{2006ApJ...640...92P} and \citet{2006ApJ...636L..17W}, we find 
that within the DRG population, about 40\% are detected by MIPS at
24\mic\, (up to 50\% at 2.0$<$z$<$3.5), and these objects have median
stellar masses 0.1-0.3dex larger than the median for all DRGs. DRGs
with MIPS detections account for more than 40\% of the total stellar
mass at z$>$2.5 (20\% at 2.0$<$z$<$2.5, and less than 5\% at z$<$2.0).

The $BzK$ criterium is very effective in detecting massive galaxies at
z$>$1.5, even more than the $J-K$ selection of DRGs. Up to 75\%--95\%
of the IRAC sources at 1.3$<$z$<$2.5 are recovered by the $BzK$
selection, 80\% at 2.5$<$z$<$3.0, 55\% at 3.0$<$z$<$3.5, and 30\%
beyond z$=$3.5. If we only consider $BzK$ galaxies with $K$$<$22.9,
these percentages decrease by a factor of $\sim$2. Most of the $BzK$
galaxies are classified as star-forming (typically 90\%). Median
masses for $BzK$ galaxies range from 10$^{10.2}$ at z$=$1.5 to
10$^{10.6}$ at z$=$3.0 and 10$^{10.9}$ at z$=$4.0, 0.1--0.4 dex larger
than median stellar masses for the whole IRAC sample. This translates
into $BzK$ galaxies tracing a large fraction of the stellar mass
density at z$>$1.5: more than 55\% and up to 97\% at z$>$1.3. Again,
if we only consider $BzK$ galaxies with $K$$<$22.9, these percentages
decrease by 15\%-20\%. These fractions are comparable to the 94\%
contribution of $BzK$ sources to the total stellar mass density at
z$\sim$1.8 found by \citet{2007A&A...465..393G}. Typically, 30\% or
more $BzK$ galaxies are detected by MIPS at 24\mic, with a
predilection for the passively evolving sub-type at z$>$2 ($\sim$60\%
and $\sim$30\% of $BzK$-PE and $BzK$-SF galaxies are detected by
MIPS). This means that passively evolving $BzK$ galaxies may still
harbor significant star formation or obscured AGNs.

Very few galaxies are identified as LBGs and DRGs simultaneously in
our IRAC survey: just 5\% of all galaxies at 2.0$<$z$<$3.0, $\sim$8\%
at z$>$3, and less than 1\% elsewhere. However, this does not mean
that the 2 selection criteria are completely orthogonal. Indeed, about
20\% of the DRGs at z$=$2--3 and 30\% of the DRGs at z$>$3 qualify as
LBGs, and 15\% of LBGs at z$>$2.5 are DRGs. Most of the LBGs that also
qualify as DRGs lie in the ``classical'' sub-type (more than 95\% of
them), which makes our results also consistent with the fractions
found in \citet{2007A&A...465..393G}, who only discussed BM-BX
objects. Note that if we only consider the DRGs brighter than
K$=$21.9, the fraction of LBGs that are also DRGs drops below the 5\%
level, in good agreement with the 10\% upper limit prediction from the
hydrodynamic models of \citet{2005ApJ...627..608N}.

The $BzK$ and DRG selection criteria present a large overlap. Around
20\%--30\% of all IRAC galaxies at z$>$2 are recovered by both
selection techniques, especially by the $BzK$-SF criterium. Indeed,
more than 95\% of all DRGs at 1.5$<$z$<$3.0 are $BzK$ galaxies, most
of them ($\sim$90\%) within the star-forming $BzK$ sub-type (in
agreement with \citealt{2007A&A...465..393G}). DRGs are only a minor
contributor to the $BzK$ population at z$<$3, where less than 35\% of
$BzK$ sources are DRGs, but this percentage rises to 70\% at
z$>$3.5. It is also interesting to mention that 50\% of $BzK$-PE
galaxies at 2.0$<$z$<$2.5 and all the $BzK$-PE galaxies at z$>$2.5 are
DRGs.

None of the LBGs lie in the $BzK$-PE type (as also noted by
\citealt{2007A&A...465..393G}), but the $BzK$-SF type also has a large
overlap with the LBG population: more than 95\% of BM-BX galaxies up
to z$\sim$2 are recovered in the $BzK$ diagram
\citep[consistent with ][]{2005ApJ...633..748R}, and less than 40\% 
in the case of ``classical'' LBGs at z$>$3.


\section{LINKING STELLAR MASSES AND STAR FORMATION RATES UP TO z$=$4}
\label{sfrs}

\subsection{The evolution of the cosmic star formation rate density}
\label{role24mic}

\begin{figure}
\begin{center}
\includegraphics[angle=-90,width=8.cm]{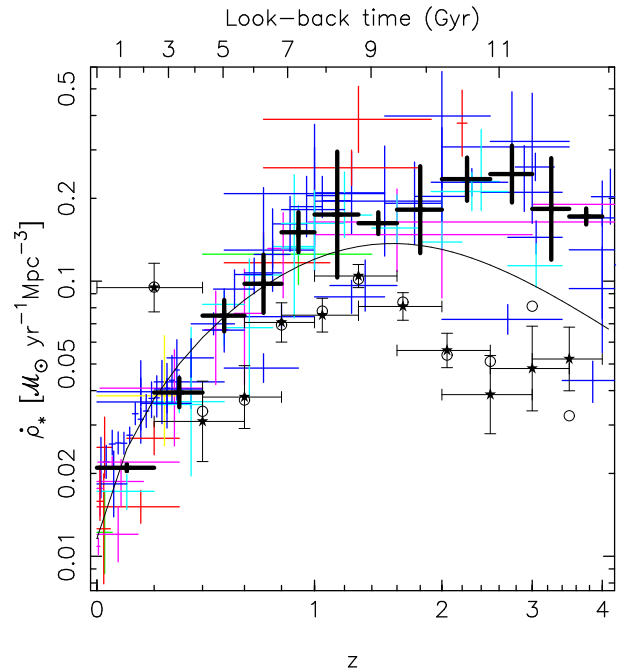}
\figcaption{\label{lillymadau}Evolution of the co-moving SFR density of 
the Universe (Lilly-Madau plot,
\citealt{1996ApJ...460L...1L,1996MNRAS.283.1388M}). Filled stars and thin 
error bars show the SFR density estimations based on the time
derivative of the stellar mass density evolution shown in
Figure~\ref{massdensity}. Open circles show the derivative for the
observed values of the stellar mass density. The colored points (shown
with error bars) are extracted from different sources in the
literature (using different SFR tracers), compiled and normalized to
the same cosmology by \citet{2004ApJ...615..209H} and
\citet{2006ApJ...651..142H}. To this compilation, we have also added the 
SFR density estimations at z$\sim$2 and z$\sim$3 found in
\citet{2007arXiv0706.4091R}. Red symbols are estimations based on
Hydrogen emission-lines, and green points on $[OII]\lambda3737$ SFR
estimations. UV-based data points are plotted in blue. Cyan
estimations are based on mid-infrared data. Magenta points are based
on sub-mm and radio observations. The yellow point is based on X-ray
data. Thick black error bars show weighted averages and standard
deviations of the literature data points for the 12 redshift intervals
considered for the stellar mass functions in this paper. The black
line shows the evolution of the cosmic SFR density as parametrized in
\citet{2001MNRAS.326..255C}.}
\end{center}
\end{figure}

\slugcomment{Please, plot this figure with the width of two columns}
\placefigure{lillymadau}


The time derivative of the stellar mass density function plotted in
Figure~\ref{massdensity} can be used to obtain the evolution of the
SFR density of the Universe, i.e., the well-known Lilly-Madau plot
\citep{1996ApJ...460L...1L,1996MNRAS.283.1388M}. For each pair of 
stellar mass density points in Figure~\ref{massdensity}, we have
estimated the SFR density (averaged through the time interval enclosed
by the points) necessary to produce the stellar mass density
difference between the corresponding redshifts. This SFR density must
be corrected upwards by some amount to account for the mass loss due
to stellar winds and supernova ejecta. For a Salpeter IMF, the
correction is 28\%
\citep{2001MNRAS.326..255C,2003ApJ...587...25D,2006ApJ...651..142H}. 
These SFR density estimations are plotted in Figure~\ref{lillymadau}
with filled black stars (open circles show the average SFR densities
derived from the observed stellar mass density values), and compared
with other cosmic SFR density estimations (based on direct SFR
measurements) found in the literature. Surprisingly, our estimations
of the cosmic SFR density are systematically smaller than the
previously published results (on average, a factor of $\sim$1.7 at
z$<$2 and a factor of 4.5 at higher redshifts, compared to individual
estimations). This discrepancy has also been remarked by
\citet{2006ApJ...650..624R}, \citet{2006ApJ...651..142H}, and
\citet{2006A&A...453..869B}, who integrate the time evolution of the 
cosmic SFR density to obtain the evolution of the stellar mass
density. 

\citet{2006ApJ...650..624R} find a factor of 1.6--2.5 offset
at z$<$2 between the measured stellar mass densities and the values
derived from the integration of the SFR density evolution published by
\citet{2001MNRAS.326..255C}. At z$>$2, they find very good agreement. 
However, the fit of the SFR density evolution published by
\citet{2001MNRAS.326..255C} gives a factor of 2--3 lower SFR densities 
than the latest estimations at z$\gtrsim$2 (see Figure~\ref{lillymadau}).

\citet{2006ApJ...651..142H} argue that the difference can be
related to a limitation in our understanding of the IMF, given that
the direct SFR estimations are sensitive to the high mass end of the
IMF, while the stellar mass estimations are sensitive to the low mass
end. Indeed, \citet{2006A&A...453..869B} find good agreement between
the SFR and the stellar mass density evolution at z$<$1 by choosing a
\citet{2003ApJ...586L.133C} IMF, which gives stellar masses similar to a 
\citet{1993MNRAS.262..545K} IMF ($\sim$1.7 times lower than our masses, 
based on a Salpeter IMF), but with lower SFRs (by a factor of 3). In
our case, an offset of 3/1.7$\sim$1.8 would also make consistent the
results of the evolution of the SFR density and stellar mass density
up to z$\sim$2.  However, at z$\gtrsim$2, the same
\citet{2003ApJ...586L.133C} IMF  fails to match the SFR and stellar mass 
densities: using the same Chabrier IMF, a good fit to the stellar mass
density evolution gives SFR densities lower than the latest
observations by up to a factor of 2--3. A top-heavy IMF (compared to a
Chabrier IMF) at high redshifts (i.e., an evolution of the IMF) could
make the SFR and stellar mass density evolutions at z$\gtrsim$2
match. A top-heavy IMF has also been proposed by the galaxy evolution
models in \citet{2005MNRAS.356.1191B} and
\citet{2007arXiv0704.1562L} to explain the number counts of galaxies
in the IR and sub-mm spectral ranges \citep[see also, among others,
][]{2004MNRAS.354..367E,2005MNRAS.363L..31N,2005MNRAS.364.1337S,
2006MNRAS.365..712L,2007MNRAS.374L..29K}. Theoretical calculations of
formation of stars also predict a top-heavy IMF for starbursts
\citep[e.g.,][]{1997MNRAS.288..145P,1998MNRAS.301..569L,2007arXiv0708.1164K}. 
Finally, some observational evidence also suggests a top-heavy IMF for
certain stellar populations
\citep{2001A&A...378....1F,2007ApJ...655...30V,2007ApJ...669.1024M}.

The discrepancy in Figure~\ref{lillymadau} could also be solved if the
SFRs estimated (with different star formation tracers, mainly the UV,
IR, and sub-mm emission) for the massive galaxies at z$\gtrsim$2 were
overestimated due, for example, to the presence of strong obscured AGN
in most of these sources \citep[see][and references
therein]{2007ApJ...670..173D}, which would imply that a significant
fraction of their UV or IR emission arise from the AGN, i.e., it is
not linked to star formation.

Finally, the reader should also note the very high SFR density derived
for 0.0$<$z$<$0.4 from the stellar mass density derivative, directly
related to the significant increase in the stellar mass density of the
Universe in this time period, as discussed in previous
Sections. Direct SFR density measurements at z$<$0.4 are a factor of
2--4 smaller, which indicates that the evolution of the stellar mass
function at low redshift is not only governed by star formation (but
also by mergers) and/or we may be underestimating the stellar mass
density at z$\sim$0.4 (if there is a numerous population of low mass
galaxies below our detection limit, which can merge together in the
last 4~Gyr to increase the density of galaxies with mass
$\mathcal{M}$$=$$10^{9-11}$~$\mathcal{M}_\sun$) or overestimating the
local value. In Section~\ref{sectdensity}, we estimated a change of
the stellar mass density of
10$^{7.9\pm0.1}$~$\mathcal{M}_\sun$\,Mpc$^{-3}$ from z$=$0.4 to
z$=$0.0 due to dry mergers. Since these mergers are affecting galaxies
with $\mathcal{M}$$=$$10^{9-11}$~$\mathcal{M}_\sun$ and the average
number density of these systems is
10$^{-2.1}$~Mpc$^{-3}$\,$\log(\mathcal{M})$ at 0.2$<$z$<$0.4, we
calculate a mass accretion rate of
$\Delta\mathcal{M}/\mathcal{M}$=0.12$\pm$0.03~Gyr$^{-1}$ from z$=$0.4
to z$=$0.0. This value is in good agreement with the accretion rate of
z$=$0.1 galaxies in the red sequence, estimated by
\citet{2005AJ....130.2647V} in 0.09$\pm$0.04~Gyr$^{-1}$ (our value is
just 30\% higher, but consistent within errors), who also calculate
that the median mass ratio of the mergers in nearby early-type
galaxies is 1:4.

\subsection{Specific star formation rates}
\label{sect_sfr}

\begin{figure}
\begin{center}
\includegraphics[angle=-90,width=9.cm]{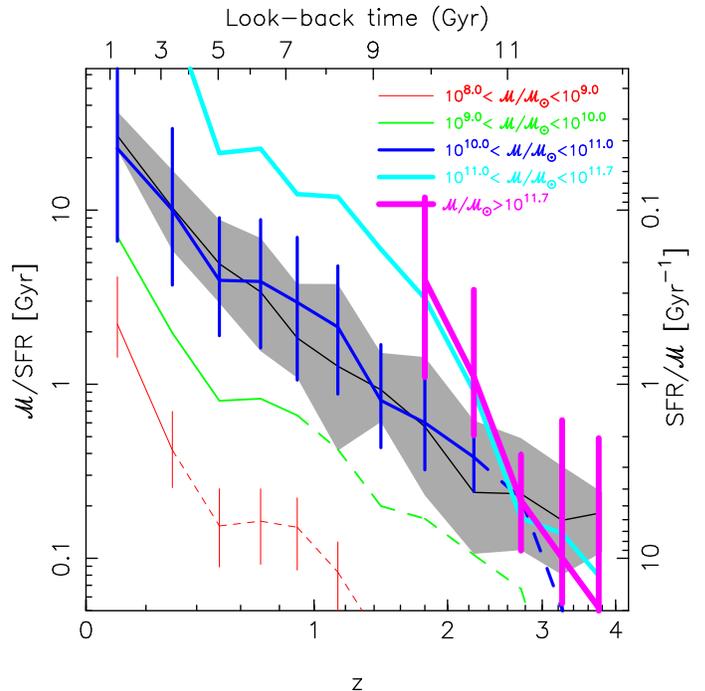}
\figcaption{\label{sfr/m}Evolution of the cosmic specific SFR.  The
black continuous line shows the evolution obtained from the the
stellar mass and SFR density estimates plotted in
Figures~\ref{massdensity} and \ref{lillymadau}. The gray shaded area
depicts the typical uncertainties in the calculation of the cosmic
specific SFR. Color lines join the median values of the distribution
of specific SFRs of our IRAC sample for several mass intervals
(excluding all X-ray detected sources), while vertical error bars show
the quartiles of that distribution. For clarity, we have only depicted
the quartiles for non-consecutive mass intervals. We only show the
median and quartiles for redshift bins where we detect more than 10
galaxies within a given stellar mass interval. Dashed lines mark the
redshift ranges where our sample is less than 75\% complete for the
given mass interval.}
\end{center}
\end{figure}

\slugcomment{Please, plot this figure with the width of two columns}
\placefigure{sfr/m}

Figure~\ref{sfr/m} shows the evolution of the specific SFR (SFR per
stellar mass unit) of the Universe (gray shaded area), calculated by
dividing the average values of the cosmic SFR density (given in
Figure~\ref{lillymadau}) by our stellar mass density estimates (given
in Figure~\ref{massdensity}). There is a continuous increase of the
specific SFR of the Universe as we move to higher redshifts. If we
consider the evolution of the specific SFR for different stellar mass
intervals (color lines in Figure~\ref{sfr/m}), we clearly see that the
most massive galaxies ($\mathcal{M}$$>$$10^{11.7}$~$\mathcal{M}_\sun$)
presented very large specific SFRs at high redshift. These galaxies
exhibit values of the SFR that are so large that they could double
their stellar mass in just 0.1~Gyr (see the scale on the left axis) at
z$=$3--4\footnote{The individual SFRs and median specific SFRs for the
entire sample derived from the extrapolated rest-frame 24\mic\,
luminosities and the calibration given by \citet[][see also
\citealt{2007ApJ...666..870C}]{2006ApJ...650..835A} are a factor of
2--3 smaller at z$\gtrsim$2 than those derived directly from
extrapolated estimations of the total IR luminosity.
\citet{2007ApJ...668...45P} also find (based on 70\mic\, observations) 
that SFRs of z$\sim$2 galaxies obtained from total IR luminosities
(extrapolated from fits of dust emission models to single 24\mic\,
detections) are overestimated by factors of a few. Using the SFR
estimations based on rest-frame 24\mic\, luminosities, the median
specific SFRs of the most massive galaxies are very similar to the
Universe average (the shaded area in Figure~\ref{sfr/m}) at z$=$3--4,
and our global results presented in Section~\ref{sect_sfr} are not
affected.}. As we move to lower redshifts, their specific SFRs
decrease considerably, by a factor of 10 from z$\sim$4 to z$\sim$2.5
(in less than 1.5~Gyr), and by a factor of $\sim$100 from z$\sim$4 to
z$\sim$1.5 (in 3~Gyr), in agreement with the results from
\citet{2006ApJ...640...92P}. For lower stellar masses, the evolution
is less pronounced. For example, galaxies with
$10^{10.0}$$<$$\mathcal{M}$$<$$10^{11.0}$~$\mathcal{M}_\sun$, whose
evolution is very similar to the cosmic average, present a decrease in
the specific SFR of a factor of 10 from z$\sim$2.5 to z$\sim$0.5 (in
6~Gyr). The evolution at z$<$1 is very similar for all the stellar
mass intervals, as already noted by \citet{2007ApJ...661L..41Z}, with
a change of about a factor of 10 in this period of $\sim$8~Gyr. In
this period, there is a significant luminosity (and maybe density)
evolution of luminous IR galaxies
\citep{2001ApJ...556..562C,2005ApJ...630...82P}. This evolution is
also detected in our stellar mass analysis, given that the fraction of
the total stellar mass density locked in galaxies emitting strongly in
the thermal IR (and being detected by MIPS) increase from about 10\%
at z$=$0 to $\sim$50\% at z$=$0.7--1.0, where LIRGs dominate the
cosmic SFR density. The fraction of the total stellar mass density
locked in MIPS sources remains approximately constant from z$\sim$1 up
to z$=$4, where LIRGs and even ULIRGs have a significant contribution
to the total SFR density, as the comparison with other SFR tracers
show \citep{2005ApJ...630...82P}.

The evolution of the cosmic specific SFR is also consistent with the
``downsizing'' picture described in Section~\ref{downsizing}, where
the most massive galaxies formed most of their mass at z$>$3 in very
intense and rapid episodes of star formation, presenting high specific
SFRs which would double the stellar mass of these systems in time
scales shorter than 1~Gyr. Less massive systems assembled more slowly,
presenting specific SFRs which would double their mass in time scales
comparable to the look-back time of the Universe at each redshift.





\section{CONCLUSIONS}
\label{conclusions}

We characterize the mass assembly of galaxies in the last 12~Gyr
($\sim$90\% of the Hubble time) by analyzing the stellar mass
functions and densities estimated from a sample of $\sim$28,000
sources selected in the rest-frame near-infrared. The sample has been
built from \spitzer/IRAC (the selection being made at 3.6\mic\, and
4.5\mic) observations of 3 fields: the Hubble Deep Field North, the
Chandra Deep Field South, and the Lockman Hole. The total surveyed
area is 664~arcmin$^2$. This IRAC sample is 75\% complete down to
1.6\micJy\, ($[3.6]$=23.4), which translates to an approximate stellar
mass completeness level (for passively evolving galaxies formed at
z$=$$\infty$) of at least 10$^{9}$~$\mathcal{M}_\sun$ up to z$=$0.5,
10$^{10}$~$\mathcal{M}_\sun$ up to z$=$1.0, and
10$^{11}$~$\mathcal{M}_\sun$ up to z$=$4.0. In order to analyze the
effects on our results of low mass galaxies faint at rest-frame
near-infrared wavelengths, we complement the IRAC survey with an
optically ($I$$<$25.5) selected sample of a similar size.

We estimate photometric redshifts, stellar masses, and star formation
rates for all galaxies using a set of templates built by modelling the
stellar population and dust emission of galaxies with known
spectroscopic redshifts. The quality of our photometric redshifts is
very good for more than 85\% of the sample, and good for nearly 95\%,
according to the comparison with spectroscopic redshift at z$<$1.5. At
z$>$1.5, where spectroscopic redshifts are scarce, we test our
photometric redshifts by comparing the average values and standard
deviations of the redshift distributions of several galaxy
populations: Lyman Break Galaxies (LBGs), Distant Red Galaxies (DRGs),
and $BzK$ galaxies. We find very good agreement with spectroscopic
surveys of these sources, and other photometric redshift analysis. We
also analyze the goodness of the stellar mass and star formation rate
estimates, finding that they are accurate within a factor of 2--3 (see
Appendix~\ref{models_app}).

Our estimation of the local stellar mass function is in good agreement
with previous estimations based on 2MASS data
\citep{2001MNRAS.326..255C,2003ApJS..149..289B}. We find a slope of 
$\alpha$$\sim$--1.2 at low stellar masses (similar values are also
found for stellar mass functions at all redshifts up to z$=$4), and a
pronounced steepening of the stellar mass function at
$\mathcal{M}$$<$10$^{9}$~$\mathcal{M}_\sun$. Approximately 1 out of 4
local galaxies are actively forming stars. Around 10--15\% of the
global stellar mass density in the local Universe is found in active
star-forming galaxies (in agreement with
\citealt{2003ApJ...587L..27P}), and this percentage rises to $\sim$50\% 
at z$\sim$1, remaining approximately constant beyond that redshift.

Our results indicate that the most massive systems
($\mathcal{M}$$>$$10^{12.0}$~$\mathcal{M}_\sun$) assembled the bulk of
their stellar mass in a very rapid collapse (half of their stellar
mass in less than 1~Gyr) at early epochs (z$>$3 or 11~Gyr ago), close
to what can be regarded as a monolithic collapse. The formation was
characterized by large specific SFRs with doubling times of about
0.1~Gyr. Galaxies with
$10^{11.5}$$<$$\mathcal{M}$$<$$10^{12.0}$~$\mathcal{M}_\sun$ formed
more slowly, assembling half of their stellar mass before z$\sim$1.5
(more than 9~Gyr ago) and more than 90\% of their stellar mass beyond
z$\sim$0.6. Less massive systems (with
$10^{9.0}$$<$$\mathcal{M}$$<$$10^{11.0}$~$\mathcal{M}_\sun$) formed at
even a slower speed: half of their stellar mass was assembled beyond
z$\sim$1 (more than 7~Gyr ago), and they experienced a significant
increase in their stellar mass (20\%--40\%) recently (at z$<$0.4 or in
the last 4~Gyr), probably by dry accretion of small satellite galaxies
(with an accretion rate 
$\Delta\mathcal{M}/\mathcal{M}$=0.12$\pm$0.03~Gyr$^{-1}$). The
specific SFRs of these galaxies evolved (closely to the cosmic
average) from 10~Gyr$^{-1}$ at z$\sim$4 to less than 0.1~Gyr$^{-1}$ in
the local Universe.

We find that approximately half of the local stellar mass density was
already formed at z$\sim$1 (8~Gyr ago), which translates to an average
assembling rate of 0.048~$\mathcal{M}_\sun$yr$^{-1}$Mpc$^{-3}$ (taking
into account a 28\% recycle factor, i.e., the fraction of the total
stellar mass re-injected in the interstellar medium in the form of
stellar winds and supernova ejecta). At least another 40\% of the
local stellar mass density assembled from z$=$1 to z$=$4 (in 4~Gyr) at
an average rate of 0.074~$\mathcal{M}_\sun$yr$^{-1}$Mpc$^{-3}$. We
find that the cosmic SFR densities estimated by differentiating the
evolution of the cosmic stellar mass density do not match the
observations based on direct SFR tracers, as also noticed by
\citet{2006ApJ...650..624R},
\citet{2006ApJ...651..142H}, and \citet{2006A&A...453..869B}. The
mismatch up to z$\sim$2 (a factor of $\sim$1.7) could be explained by
changing the IMF to a \citet{2003ApJ...586L.133C} IMF (instead of a
\citealt{1955ApJ...121..161S} IMF, the default used in this paper). At
z$\gtrsim$2, the discrepancy is larger (a factor of 4--5), and can
only be solved if the IMF is top-heavy (i.e., a different IMF at high
redshift) and/or if the SFRs of the most massive galaxies at z$=$3--4
(calculated with different star formation tracers, mainly the UV, IR,
and sub-mm emission) are overestimated due, for example, to the
presence of strong obscured AGN in most of these sources (which would
imply that a significant fraction of their UV or IR emission arises
from the AGN, i.e., it is not linked to star formation).


We confirm that galaxy formation follows a ``downsizing'' scenario
\citep{1996AJ....112..839C}. Our results are broadly consistent with
previous observational works that confirm this formation theory
\citep{2004Natur.428..625H,2005ApJ...619L.135J,2004Natur.430..181G,
2005ApJ...630...82P,2006A&A...459..745F}, and with models of galaxy
formation
\citep[e.g.,][]{2004ApJ...610...45N,2006MNRAS.365...11C,2006MNRAS.370..645B}.
At low redshift (z$<$1), there is also an ``upsizing'' effect, when
intermediate mass galaxies
($\mathcal{M}$$=$$10^{9-11}$~$\mathcal{M}_\sun$) increase their
density by accretion or coalescence of previously formed smaller
galaxies (producing little star formation).

We have also analyzed the nature of the galaxies in our sample,
comparing them with the populations of sources detected with different
selection techniques by other surveys. Based on the measured number
densities and redshifts, we conclude that our survey constitutes an
almost complete census of the different populations of previously
known galaxies at high redshifts, including most of the LBGs at
1$<$z$<$3.5, most of the DRGs at z$\gtrsim$2, and most of the $BzK$
galaxies at z$>$1.4. LBGs dominate the number counts of IRAC galaxies
at high redshift, being about a factor of 2--3 more numerous than DRGs
and $BzK$ galaxies, but most of the stellar mass density (more than
50\% and up to 97\%) at z$>$2.5 resides in the latter, while LBGs
account for less than 50\% of the total stellar mass density.

\acknowledgments
We acknowledge support from the Spanish Programa Nacional de
Astronom\'{\i}a y Astrof\'{\i}sica under grant AYA 2006--02358.
Support was also provided by NASA through Contract no.  1255094 issued
by JPL/Caltech.  This work is based in part on observations made with
the {\it Spitzer} Space Telescope, which is operated by the Jet
Propulsion Laboratory, Caltech under NASA contract 1407. GALEX is a
NASA Small Explorer launched in 2003 April. We gratefully acknowledge
NASA's support for construction, operation, and scientific analysis of
the GALEX mission. This research has made use of the NASA/IPAC
Extragalactic Database (NED) which is operated by the Jet Propulsion
Laboratory, California Institute of Technology, under contract with
the National Aeronautics and Space Administration. Based in part on
data collected at Subaru Telescope and obtained from the SMOKA, which
is operated by the Astronomy Data Center, National Astronomical
Observatory of Japan. This work is based in part on data obtained as
part of the UKIRT Infrared Deep Sky Survey.  Funding for the Sloan
Digital Sky Survey (SDSS) and SDSS-II has been provided by the Alfred
P. Sloan Foundation, the Participating Institutions, the National
Science Foundation, the U.S. Department of Energy, the National
Aeronautics and Space Administration, the Japanese Monbukagakusho, and
the Max Planck Society, and the Higher Education Funding Council for
England. The SDSS Web site is http://www.sdss.org/. This publication
makes use of data products from the Two Micron All Sky Survey, which
is a joint project of the University of Massachusetts and the Infrared
Processing and Analysis Center/California Institute of Technology,
funded by the National Aeronautics and Space Administration and the
National Science Foundation. PGP-G acknowledges support from the
Ram\'on y Cajal Fellowship Program financed by the Spanish Government,
and AGdP from the MAGPOP EU Marie Curie Research Training Network. We
are grateful to Andrew Hopkins for providing useful data for this
paper.
\bibliographystyle{apj}
\bibliography{referencias}

\clearpage

\placetable{tablesmf}
\setcounter{table}{0}
\begin{deluxetable*}{lccccc}
\tabletypesize{\tiny}
\tablewidth{300pt}
\tablecaption{\label{tablesmf}Stellar mass functions for the global and star-forming population of galaxies.}
\tablehead{
\colhead{Redshift range} & \colhead{$\log(\mathcal{M})$$^a$} & \colhead{$\log(\phi_\mathrm{IRAC})$$^b$}  & \colhead{$\log(\phi_\mathrm{I-band})$$^b$} & \colhead{$\log(\phi_\mathrm{MIPS})$$^b$}
}
\startdata
$0.0<z<0.2$     &  8.0 & $-1.275^{+0.147}_{-0.165}$ &  $-1.308^{+0.140}_{-0.157}$ &         \nodata              \\
                &  8.2 & $-1.325^{+0.150}_{-0.169}$ &  $-1.377^{+0.152}_{-0.173}$ &         \nodata              \\
                &  8.4 & $-1.328^{+0.155}_{-0.175}$ &  $-1.462^{+0.184}_{-0.215}$ &         \nodata              \\
                &  8.6 & $-1.485^{+0.183}_{-0.213}$ &  $-1.518^{+0.162}_{-0.184}$ &  $-2.339^{+0.053}_{-0.061}$  \\
                &  8.8 & $-1.596^{+0.189}_{-0.221}$ &  $-1.618^{+0.186}_{-0.218}$ &  $-2.382^{+0.049}_{-0.055}$  \\
                &  9.0 & $-1.698^{+0.205}_{-0.244}$ &  $-1.647^{+0.151}_{-0.170}$ &  $-2.278^{+0.045}_{-0.050}$  \\
                &  9.2 & $-1.767^{+0.219}_{-0.263}$ &  $-1.690^{+0.152}_{-0.172}$ &  $-2.442^{+0.062}_{-0.072}$  \\
                &  9.4 & $-1.806^{+0.232}_{-0.283}$ &  $-1.792^{+0.180}_{-0.209}$ &  $-2.410^{+0.068}_{-0.080}$  \\
                &  9.6 & $-1.823^{+0.189}_{-0.221}$ &  $-1.754^{+0.139}_{-0.155}$ &  $-2.504^{+0.050}_{-0.044}$  \\
                &  9.8 & $-1.890^{+0.225}_{-0.273}$ &  $-1.890^{+0.127}_{-0.141}$ &  $-2.469^{+0.054}_{-0.062}$  \\
                & 10.0 & $-1.965^{+0.228}_{-0.277}$ &  $-1.893^{+0.157}_{-0.179}$ &  $-2.605^{+0.055}_{-0.063}$  \\
                & 10.2 & $-1.931^{+0.196}_{-0.231}$ &  $-1.984^{+0.134}_{-0.149}$ &  $-2.630^{+0.070}_{-0.083}$  \\
                & 10.4 & $-2.052^{+0.201}_{-0.238}$ &  $-2.041^{+0.242}_{-0.297}$ &  $-2.839^{+0.062}_{-0.073}$  \\
                & 10.6 & $-2.122^{+0.159}_{-0.181}$ &  $-2.186^{+0.196}_{-0.231}$ &  $-2.873^{+0.072}_{-0.087}$  \\
                & 10.8 & $-2.177^{+0.119}_{-0.131}$ &  $-2.363^{+0.221}_{-0.267}$ &  $-2.879^{+0.039}_{-0.043}$  \\
                & 11.0 & $-2.311^{+0.187}_{-0.218}$ &         \nodata             &         \nodata              \\
                & 11.2 & $-2.591^{+0.400}_{-9.999}$ &         \nodata             &         \nodata              \\
                & 11.4 & $-3.022^{+0.400}_{-9.999}$ &         \nodata             &         \nodata              \\
                & 11.6 & $-3.327^{+0.400}_{-9.999}$ &         \nodata             &         \nodata              \\
                & 11.8 & $-3.328^{+0.910}_{-9.999}$ &         \nodata             &         \nodata              \\
$0.2<z<0.4$     &  8.2 & $-1.690^{+0.123}_{-0.136}$ &  $-1.591^{+0.160}_{-0.183}$ &         \nodata              \\
                &  8.4 & $-1.681^{+0.135}_{-0.150}$ &  $-1.623^{+0.144}_{-0.162}$ &         \nodata              \\
                &  8.6 & $-1.817^{+0.136}_{-0.151}$ &  $-1.712^{+0.143}_{-0.160}$ &         \nodata              \\
                &  8.8 & $-1.800^{+0.135}_{-0.151}$ &  $-1.729^{+0.144}_{-0.162}$ &         \nodata              \\
                &  9.0 & $-1.897^{+0.137}_{-0.153}$ &  $-0.971^{+0.026}_{-0.026}$ &         \nodata              \\
                &  9.2 & $-1.987^{+0.173}_{-0.200}$ &  $-1.760^{+0.130}_{-0.144}$ &         \nodata              \\
                &  9.4 & $-1.953^{+0.242}_{-0.212}$ &  $-1.942^{+0.144}_{-0.163}$ &         \nodata              \\
                &  9.6 & $-2.042^{+0.205}_{-0.243}$ &  $-2.055^{+0.144}_{-0.162}$ &         \nodata              \\
                &  9.8 & $-2.145^{+0.198}_{-0.234}$ &  $-2.053^{+0.121}_{-0.133}$ &  $-2.742^{+0.076}_{-0.092}$  \\
                & 10.0 & $-2.113^{+0.193}_{-0.226}$ &  $-2.136^{+0.130}_{-0.144}$ &  $-2.690^{+0.078}_{-0.095}$  \\
                & 10.2 & $-2.207^{+0.247}_{-0.306}$ &  $-2.274^{+0.183}_{-0.213}$ &  $-2.786^{+0.091}_{-0.116}$  \\
                & 10.4 & $-2.263^{+0.245}_{-0.302}$ &  $-2.212^{+0.142}_{-0.159}$ &  $-2.822^{+0.091}_{-0.116}$  \\
                & 10.6 & $-2.329^{+0.267}_{-0.336}$ &  $-2.434^{+0.211}_{-0.252}$ &  $-2.852^{+0.104}_{-0.137}$  \\
                & 10.8 & $-2.499^{+0.240}_{-0.294}$ &  $-2.453^{+0.178}_{-0.206}$ &  $-2.872^{+0.087}_{-0.109}$  \\
                & 11.0 & $-2.605^{+0.276}_{-0.350}$ &  $-2.689^{+0.187}_{-0.218}$ &  $-3.012^{+0.080}_{-0.099}$  \\
                & 11.2 & $-2.798^{+0.288}_{-0.370}$ &  $-2.817^{+0.210}_{-0.250}$ &  $-3.286^{+0.126}_{-0.179}$  \\
                & 11.4 & $-2.955^{+0.181}_{-0.211}$ &  $-3.092^{+0.254}_{-0.469}$ &  $-3.517^{+0.085}_{-0.106}$  \\
                & 11.6 & $-3.458^{+0.400}_{-9.999}$ &         \nodata             &  $-3.872^{+3.872}_{-3.872}$  \\
                & 11.8 & $-4.040^{+0.400}_{-5.950}$ &         \nodata             &  $-4.252^{+4.252}_{-4.252}$  \\
                & 12.0 & $-5.037^{+0.400}_{-4.953}$ &         \nodata             &         \nodata              \\
\enddata
\tablecomments{$^a$ In units of $\mathcal{M}_\odot$. $^b$ In units of
$\mathrm{Mpc}^{-3}\,(\log\mathcal{M})^{-1}$.}
\end{deluxetable*}
\setcounter{table}{0}
\begin{deluxetable*}{lccccc}
\tabletypesize{\tiny}
\tablewidth{300pt}
\tablecaption{Stellar mass functions for the global and star-forming population of galaxies.}
\tablehead{
\colhead{Redshift range} & \colhead{$\log(\mathcal{M})$$^a$} & \colhead{$\log(\phi_\mathrm{IRAC})$$^b$}  & \colhead{$\log(\phi_\mathrm{I-band})$$^b$} & \colhead{$\log(\phi_\mathrm{MIPS})$$^b$}
}
\startdata
$0.4<z<0.6$     &  8.6 &         \nodata            &  $-1.868^{+0.121}_{-0.133}$ &         \nodata              \\
                &  8.8 & $-1.984^{+0.109}_{-0.118}$ &  $-1.876^{+0.113}_{-0.124}$ &         \nodata              \\
                &  9.0 & $-1.858^{+0.095}_{-0.102}$ &  $-1.969^{+0.135}_{-0.150}$ &         \nodata              \\
                &  9.2 & $-1.981^{+0.113}_{-0.123}$ &  $-1.907^{+0.121}_{-0.133}$ &         \nodata              \\
                &  9.4 & $-2.019^{+0.114}_{-0.125}$ &  $-2.067^{+0.143}_{-0.161}$ &         \nodata              \\
                &  9.6 & $-2.070^{+0.128}_{-0.142}$ &  $-1.961^{+0.113}_{-0.124}$ &         \nodata              \\
                &  9.8 & $-2.099^{+0.118}_{-0.130}$ &  $-2.139^{+0.137}_{-0.153}$ &         \nodata              \\
                & 10.0 & $-2.145^{+0.138}_{-0.155}$ &  $-2.048^{+0.102}_{-0.111}$ &  $-2.799^{+0.082}_{-0.101}$  \\
                & 10.2 & $-2.193^{+0.136}_{-0.152}$ &  $-2.282^{+0.180}_{-0.209}$ &  $-2.715^{+0.067}_{-0.079}$  \\
                & 10.4 & $-2.238^{+0.157}_{-0.179}$ &  $-2.242^{+0.137}_{-0.153}$ &  $-2.702^{+0.075}_{-0.091}$  \\
                & 10.6 & $-2.305^{+0.155}_{-0.176}$ &  $-2.155^{+0.102}_{-0.111}$ &  $-2.695^{+0.068}_{-0.081}$  \\
                & 10.8 & $-2.414^{+0.187}_{-0.219}$ &  $-2.476^{+0.177}_{-0.205}$ &  $-2.802^{+0.073}_{-0.088}$  \\
                & 11.0 & $-2.596^{+0.208}_{-0.248}$ &  $-2.571^{+0.205}_{-0.243}$ &  $-2.969^{+0.097}_{-0.125}$  \\
                & 11.2 & $-2.886^{+0.271}_{-0.342}$ &  $-2.932^{+0.236}_{-0.289}$ &  $-3.194^{+0.092}_{-0.118}$  \\
                & 11.4 & $-3.339^{+0.316}_{-0.418}$ &  $-3.270^{+0.348}_{-0.476}$ &  $-3.632^{+0.087}_{-0.108}$  \\
                & 11.6 & $-3.382^{+0.175}_{-0.202}$ &  $-3.321^{+0.287}_{-0.369}$ &  $-4.018^{+4.018}_{-4.018}$  \\
                & 11.8 & $-3.883^{+0.248}_{-0.307}$ &  $-3.936^{+0.400}_{-9.999}$ &         \nodata              \\
                & 12.0 & $-4.885^{+0.400}_{-5.105}$ &         \nodata             &         \nodata              \\
$0.6<z<0.8$     &  9.0 &         \nodata            &  $-2.037^{+0.114}_{-0.125}$ &         \nodata              \\
                &  9.2 & $-2.067^{+0.244}_{-0.300}$ &  $-1.959^{+0.075}_{-0.080}$ &         \nodata              \\
                &  9.4 & $-2.041^{+0.153}_{-0.173}$ &  $-2.059^{+0.106}_{-0.115}$ &         \nodata              \\
                &  9.6 & $-2.050^{+0.120}_{-0.132}$ &  $-2.066^{+0.123}_{-0.136}$ &         \nodata              \\
                &  9.8 & $-2.064^{+0.098}_{-0.106}$ &  $-2.123^{+0.125}_{-0.139}$ &         \nodata              \\
                & 10.0 & $-2.124^{+0.093}_{-0.100}$ &  $-2.044^{+0.076}_{-0.080}$ &         \nodata              \\
                & 10.2 & $-2.174^{+0.105}_{-0.114}$ &  $-2.269^{+0.134}_{-0.149}$ &         \nodata              \\
                & 10.4 & $-2.254^{+0.107}_{-0.117}$ &  $-2.278^{+0.124}_{-0.138}$ &  $-2.713^{+0.060}_{-0.070}$  \\
                & 10.6 & $-2.367^{+0.116}_{-0.128}$ &  $-2.376^{+0.141}_{-0.158}$ &  $-2.678^{+0.053}_{-0.060}$  \\
                & 10.8 & $-2.509^{+0.134}_{-0.150}$ &  $-2.424^{+0.145}_{-0.463}$ &  $-2.840^{+0.065}_{-0.077}$  \\
                & 11.0 & $-2.650^{+0.150}_{-0.170}$ &  $-2.559^{+0.189}_{-0.221}$ &  $-2.924^{+0.068}_{-0.081}$  \\
                & 11.2 & $-2.830^{+0.198}_{-0.233}$ &  $-2.806^{+0.213}_{-0.254}$ &  $-3.227^{+0.105}_{-0.139}$  \\
                & 11.4 & $-3.280^{+0.274}_{-0.348}$ &  $-3.080^{+0.158}_{-0.180}$ &  $-3.561^{+0.117}_{-0.160}$  \\
                & 11.6 & $-3.607^{+0.296}_{-0.383}$ &  $-3.597^{+0.400}_{-9.999}$ &  $-4.058^{+4.058}_{-4.058}$  \\
                & 11.8 & $-4.042^{+0.225}_{-0.273}$ &  $-4.205^{+0.124}_{-0.137}$ &  $-4.761^{+0.063}_{-0.073}$  \\
$0.8<z<1.0$     &  9.2 &         \nodata            &  $-2.265^{+0.411}_{-0.178}$ &         \nodata              \\
                &  9.4 & $-2.053^{+0.078}_{-0.083}$ &  $-2.182^{+0.113}_{-0.124}$ &         \nodata              \\
                &  9.6 & $-2.162^{+0.083}_{-0.088}$ &  $-2.181^{+0.102}_{-0.111}$ &         \nodata              \\
                &  9.8 & $-2.203^{+0.089}_{-0.095}$ &  $-2.278^{+0.123}_{-0.135}$ &         \nodata              \\
                & 10.0 & $-2.254^{+0.090}_{-0.096}$ &  $-2.278^{+0.115}_{-0.127}$ &         \nodata              \\
                & 10.2 & $-2.326^{+0.103}_{-0.112}$ &  $-2.385^{+0.143}_{-0.161}$ &         \nodata              \\
                & 10.4 & $-2.379^{+0.104}_{-0.113}$ &  $-2.353^{+0.129}_{-0.144}$ &  $-2.821^{+0.061}_{-0.071}$  \\
                & 10.6 & $-2.509^{+0.118}_{-0.130}$ &  $-2.621^{+0.176}_{-0.204}$ &  $-2.805^{+0.051}_{-0.058}$  \\
                & 10.8 & $-2.599^{+0.118}_{-0.130}$ &  $-2.654^{+0.176}_{-0.203}$ &  $-2.871^{+0.061}_{-0.070}$  \\
                & 11.0 & $-2.742^{+0.159}_{-0.181}$ &  $-2.720^{+0.179}_{-0.207}$ &  $-3.041^{+0.078}_{-0.096}$  \\
                & 11.2 & $-2.893^{+0.216}_{-0.259}$ &  $-2.912^{+0.184}_{-0.214}$ &  $-3.391^{+0.106}_{-0.141}$  \\
                & 11.4 & $-3.160^{+0.270}_{-0.341}$ &  $-3.065^{+0.144}_{-0.162}$ &  $-3.520^{+0.109}_{-0.146}$  \\
                & 11.6 & $-3.634^{+0.257}_{-0.321}$ &  $-3.513^{+0.160}_{-0.182}$ &  $-3.876^{+0.124}_{-0.174}$  \\
                & 11.8 & $-4.113^{+0.195}_{-0.230}$ &  $-3.951^{+0.306}_{-0.401}$ &         \nodata              \\
\enddata
\tablecomments{$^a$ In units of $\mathcal{M}_\odot$. $^b$ In units of
$\mathrm{Mpc}^{-3}\,(\log\mathcal{M})^{-1}$.}
\end{deluxetable*}
\setcounter{table}{0}
\begin{deluxetable*}{lccccc}
\tabletypesize{\tiny}
\tablewidth{300pt}
\tablecaption{Stellar mass functions for the global and star-forming population of galaxies.}
\tablehead{
\colhead{Redshift range} & \colhead{$\log(\mathcal{M})$$^a$} & \colhead{$\log(\phi_\mathrm{IRAC})$$^b$}  & \colhead{$\log(\phi_\mathrm{I-band})$$^b$} & \colhead{$\log(\phi_\mathrm{MIPS})$$^b$}
}
\startdata
$1.0<z<1.3$     &  9.2 &         \nodata            &  $-2.271^{+0.167}_{-0.192}$ &         \nodata              \\
                &  9.4 &         \nodata            &  $-2.286^{+0.173}_{-0.199}$ &         \nodata              \\
                &  9.6 & $-2.285^{+0.076}_{-0.081}$ &  $-2.245^{+0.126}_{-0.139}$ &         \nodata              \\
                &  9.8 & $-2.294^{+0.080}_{-0.086}$ &  $-2.268^{+0.132}_{-0.147}$ &         \nodata              \\
                & 10.0 & $-2.355^{+0.089}_{-0.095}$ &  $-2.387^{+0.138}_{-0.154}$ &         \nodata              \\
                & 10.2 & $-2.468^{+0.096}_{-0.103}$ &  $-2.467^{+0.157}_{-0.178}$ &         \nodata              \\
                & 10.4 & $-2.508^{+0.098}_{-0.105}$ &  $-2.494^{+0.151}_{-0.170}$ &  $-3.143^{+0.060}_{-0.070}$  \\
                & 10.6 & $-2.563^{+0.104}_{-0.113}$ &  $-2.591^{+0.172}_{-0.199}$ &  $-3.117^{+0.075}_{-0.090}$  \\
                & 10.8 & $-2.647^{+0.121}_{-0.133}$ &  $-2.661^{+0.180}_{-0.209}$ &  $-3.186^{+0.070}_{-0.083}$  \\
                & 11.0 & $-2.788^{+0.155}_{-0.175}$ &  $-2.827^{+0.207}_{-0.246}$ &  $-3.275^{+0.072}_{-0.087}$  \\
                & 11.2 & $-3.032^{+0.173}_{-0.199}$ &  $-2.956^{+0.209}_{-0.249}$ &  $-3.471^{+0.085}_{-0.106}$  \\
                & 11.4 & $-3.392^{+0.265}_{-0.333}$ &  $-3.298^{+0.162}_{-0.185}$ &  $-3.834^{+0.130}_{-0.186}$  \\
                & 11.6 & $-3.678^{+0.312}_{-0.410}$ &  $-3.608^{+0.267}_{-0.336}$ &  $-4.057^{+0.132}_{-0.190}$  \\
                & 11.8 & $-4.232^{+0.306}_{-0.400}$ &  $-3.899^{+0.400}_{-9.999}$ &  $-4.560^{+0.133}_{-0.193}$  \\
                & 12.0 & $-4.663^{+0.400}_{-5.327}$ &  $-4.295^{+0.400}_{-5.753}$ &         \nodata              \\
$1.3<z<1.6$     &  9.8 &         \nodata            &  $-2.674^{+0.183}_{-0.212}$ &         \nodata              \\
                & 10.0 & $-2.628^{+0.103}_{-0.112}$ &  $-2.633^{+0.154}_{-0.175}$ &         \nodata              \\
                & 10.2 & $-2.677^{+0.106}_{-0.115}$ &  $-2.702^{+0.305}_{-0.398}$ &         \nodata              \\
                & 10.4 & $-2.736^{+0.117}_{-0.128}$ &  $-2.719^{+0.161}_{-0.184}$ &  $-3.491^{+0.071}_{-0.085}$  \\
                & 10.6 & $-2.783^{+0.127}_{-0.141}$ &  $-2.766^{+0.148}_{-0.168}$ &  $-3.361^{+0.072}_{-0.086}$  \\
                & 10.8 & $-2.860^{+0.136}_{-0.152}$ &  $-2.828^{+0.149}_{-0.169}$ &  $-3.414^{+0.067}_{-0.079}$  \\
                & 11.0 & $-2.994^{+0.143}_{-0.161}$ &  $-3.066^{+0.242}_{-0.298}$ &  $-3.553^{+0.099}_{-0.129}$  \\
                & 11.2 & $-3.145^{+0.180}_{-0.208}$ &  $-3.250^{+0.239}_{-0.293}$ &  $-3.600^{+0.079}_{-0.096}$  \\
                & 11.4 & $-3.329^{+0.256}_{-0.319}$ &  $-3.277^{+0.227}_{-0.275}$ &  $-3.773^{+0.108}_{-0.144}$  \\
                & 11.6 & $-3.700^{+0.281}_{-0.359}$ &  $-3.636^{+0.202}_{-0.239}$ &  $-4.068^{+0.112}_{-0.150}$  \\
                & 11.8 & $-4.582^{+0.400}_{-5.408}$ &  $-4.439^{+0.095}_{-0.103}$ &  $-5.156^{+5.156}_{-5.156}$  \\
                & 12.0 & $-4.982^{+0.203}_{-0.241}$ &  $-5.191^{+0.400}_{-4.790}$ &         \nodata              \\
$1.6<z<2.0$     & 10.0 &         \nodata            &  $-2.968^{+0.199}_{-0.235}$ &         \nodata              \\
                & 10.2 & $-2.935^{+0.117}_{-0.128}$ &  $-2.979^{+0.266}_{-0.335}$ &         \nodata              \\
                & 10.4 & $-2.979^{+0.133}_{-0.148}$ &  $-3.034^{+0.251}_{-0.311}$ &         \nodata              \\
                & 10.6 & $-2.967^{+0.121}_{-0.134}$ &  $-2.980^{+0.236}_{-0.288}$ &  $-3.626^{+0.073}_{-0.088}$  \\
                & 10.8 & $-3.066^{+0.136}_{-0.152}$ &  $-3.073^{+0.280}_{-0.357}$ &  $-3.510^{+0.057}_{-0.065}$  \\
                & 11.0 & $-3.207^{+0.159}_{-0.182}$ &  $-3.210^{+0.227}_{-0.276}$ &  $-3.670^{+0.084}_{-0.105}$  \\
                & 11.2 & $-3.340^{+0.215}_{-0.258}$ &  $-3.298^{+0.304}_{-0.398}$ &  $-3.695^{+0.071}_{-0.084}$  \\
                & 11.4 & $-3.585^{+0.277}_{-0.353}$ &  $-3.481^{+0.210}_{-0.251}$ &  $-3.956^{+0.099}_{-0.128}$  \\
                & 11.6 & $-3.873^{+0.341}_{-0.462}$ &  $-3.868^{+0.189}_{-0.222}$ &  $-4.259^{+0.098}_{-0.126}$  \\
                & 11.8 & $-4.388^{+0.255}_{-0.318}$ &  $-4.231^{+0.466}_{-0.490}$ &  $-4.624^{+0.079}_{-0.097}$  \\
                & 12.0 & $-4.834^{+0.337}_{-0.456}$ &  $-4.881^{+0.504}_{-0.242}$ &         \nodata              \\
$2.0<z<2.5$     & 10.2 &         \nodata            &  $-3.173^{+0.209}_{-0.249}$ &         \nodata              \\
                & 10.4 & $-3.068^{+0.119}_{-0.132}$ &  $-3.289^{+0.219}_{-0.263}$ &         \nodata              \\
                & 10.6 & $-3.176^{+0.139}_{-0.156}$ &  $-3.197^{+0.335}_{-0.452}$ &         \nodata              \\
                & 10.8 & $-3.234^{+0.139}_{-0.155}$ &  $-3.287^{+0.208}_{-0.247}$ &  $-3.884^{+0.069}_{-0.082}$  \\
                & 11.0 & $-3.367^{+0.165}_{-0.188}$ &  $-3.404^{+0.211}_{-0.251}$ &  $-3.891^{+0.089}_{-0.112}$  \\
                & 11.2 & $-3.499^{+0.176}_{-0.204}$ &  $-3.551^{+0.185}_{-0.215}$ &  $-3.824^{+0.070}_{-0.084}$  \\
                & 11.4 & $-3.672^{+0.226}_{-0.273}$ &  $-3.659^{+0.283}_{-0.362}$ &  $-4.123^{+0.102}_{-0.134}$  \\
                & 11.6 & $-4.151^{+0.268}_{-0.339}$ &  $-4.025^{+0.214}_{-0.257}$ &  $-4.490^{+0.107}_{-0.142}$  \\
                & 11.8 & $-4.289^{+0.254}_{-0.316}$ &  $-4.827^{+0.274}_{-0.347}$ &  $-4.574^{+0.094}_{-0.120}$  \\
                & 12.0 & $-4.878^{+0.336}_{-0.454}$ &         \nodata             &         \nodata              \\
\enddata
\tablecomments{$^a$ In units of $\mathcal{M}_\odot$. $^b$ In units of
$\mathrm{Mpc}^{-3}\,(\log\mathcal{M})^{-1}$.}
\end{deluxetable*}
\setcounter{table}{0}
\begin{deluxetable*}{lccccc}
\tabletypesize{\tiny}
\tablewidth{300pt}
\tablecaption{Stellar mass functions for the global and star-forming population of galaxies.}
\tablehead{
\colhead{Redshift range} & \colhead{$\log(\mathcal{M})$$^a$} & \colhead{$\log(\phi_\mathrm{IRAC})$$^b$}  & \colhead{$\log(\phi_\mathrm{I-band})$$^b$} & \colhead{$\log(\phi_\mathrm{MIPS})$$^b$}
}
\startdata
$2.5<z<3.0$     & 10.4 &         \nodata            &  $-3.243^{+0.208}_{-0.248}$ &         \nodata              \\
                & 10.6 &         \nodata            &  $-3.327^{+0.197}_{-0.232}$ &         \nodata              \\
                & 10.8 & $-3.477^{+0.198}_{-0.233}$ &  $-3.296^{+0.389}_{-0.558}$ &         \nodata              \\
                & 11.0 & $-3.495^{+0.200}_{-0.236}$ &  $-3.405^{+0.194}_{-0.228}$ &  $-4.079^{+0.087}_{-0.109}$  \\
                & 11.2 & $-3.591^{+0.197}_{-0.232}$ &  $-3.553^{+0.325}_{-0.434}$ &  $-3.989^{+0.088}_{-0.110}$  \\
                & 11.4 & $-3.770^{+0.247}_{-0.306}$ &  $-3.767^{+0.200}_{-0.236}$ &  $-4.055^{+0.082}_{-0.101}$  \\
                & 11.6 & $-4.156^{+0.313}_{-0.413}$ &  $-4.032^{+0.198}_{-0.234}$ &  $-4.519^{+0.100}_{-0.130}$  \\
                & 11.8 & $-4.385^{+0.344}_{-0.468}$ &  $-4.495^{+0.248}_{-0.306}$ &  $-4.823^{+0.102}_{-0.134}$  \\
                & 12.0 & $-4.986^{+0.279}_{-0.356}$ &  $-3.550^{+0.179}_{-0.208}$ &         \nodata              \\
$3.0<z<3.5$     & 10.8 &         \nodata            &  $-3.811^{+1.336}_{-0.991}$ &         \nodata              \\
                & 11.0 & $-3.610^{+0.270}_{-0.341}$ &  $-3.642^{+0.467}_{-0.203}$ &         \nodata              \\
                & 11.2 & $-3.721^{+0.237}_{-0.290}$ &  $-3.741^{+1.389}_{-0.298}$ &  $-4.288^{+0.080}_{-0.098}$  \\
                & 11.4 & $-3.925^{+0.262}_{-0.328}$ &  $-3.748^{+0.110}_{-0.649}$ &  $-4.231^{+0.097}_{-0.125}$  \\
                & 11.6 & $-4.277^{+0.271}_{-0.342}$ &  $-4.290^{+0.699}_{-1.009}$ &  $-4.612^{+0.128}_{-0.182}$  \\
                & 11.8 & $-4.498^{+0.331}_{-0.445}$ &  $-4.516^{+0.634}_{-2.205}$ &  $-4.793^{+0.112}_{-0.152}$  \\
                & 12.0 & $-5.142^{+0.400}_{-4.848}$ &  $-4.947^{+1.621}_{-0.317}$ &         \nodata              \\
$3.5<z<4.0$     & 11.0 & $-3.748^{+0.260}_{-0.325}$ &  $-3.737^{+0.564}_{-1.017}$ &         \nodata              \\
                & 11.2 & $-3.816^{+0.232}_{-0.282}$ &  $-3.967^{+0.191}_{-0.224}$ &         \nodata              \\
                & 11.4 & $-4.084^{+0.298}_{-0.387}$ &  $-4.096^{+0.104}_{-0.113}$ &  $-4.455^{+0.071}_{-0.086}$  \\
                & 11.6 & $-4.355^{+0.263}_{-0.330}$ &  $-4.188^{+0.160}_{-0.182}$ &  $-4.568^{+0.100}_{-0.130}$  \\
                & 11.8 & $-4.923^{+0.265}_{-0.333}$ &  $-4.891^{+0.174}_{-0.201}$ &  $-5.029^{+0.068}_{-0.080}$  \\
                & 12.0 & $-5.101^{+0.315}_{-0.416}$ &         \nodata             &         \nodata              \\
\enddata
\tablecomments{$^a$ In units of $\mathcal{M}_\odot$. $^b$ In units of
$\mathrm{Mpc}^{-3}\,(\log\mathcal{M})^{-1}$.}
\end{deluxetable*}

\begin{turnpage}
\placetable{schechter}
\begin{deluxetable*}{lccccccccccc}
\tabletypesize{\scriptsize}
\tablewidth{600pt}
\tablecaption{\label{schechter}Results of the Schechter (1976) fits (including stellar mass densities) to the global and star-forming stellar mass functions.}
\tablehead{ & \multicolumn{5}{c}{GLOBAL} & & \multicolumn{5}{c}{STAR-FORMING}\\
\cline{2-6} \cline{8-12} \\
\colhead{Redshift range} & \colhead{$\alpha$} & \colhead{$\log(\mathcal{M}^*)$$^a$}  & \colhead{$\log(\phi^*)$$^b$} & \colhead{$\log(\rho_*)$$^c$} & \colhead{$\log(\rho_*^{\mathrm{obs}})$$^c$} & & \colhead{$\alpha_\mathrm{SF}$} & \colhead{$\log(\mathcal{M}^*_\mathrm{SF})$$^a$}  & \colhead{$\log(\phi^*_\mathrm{SF})$$^b$} & \colhead{$\log(\rho^\mathrm{SF}_*)$$^c$} & \colhead{$\log(\rho_*^{\mathrm{obs}})$$^c$} 
}
\startdata
$0.0<z\leq0.2$ & -1.18$\pm$0.12 & 11.16$\pm$0.25 & -2.47$\pm$0.22 & 8.75$\pm$0.12 & 8.75 &  & -1.16$\pm$0.12 & 10.84$\pm$0.17 & -3.04$\pm$0.16 & 7.85$\pm$0.07 & 7.85 \\
$0.2<z\leq0.4$ & -1.19$\pm$0.08 & 11.20$\pm$0.10 & -2.65$\pm$0.15 & 8.61$\pm$0.06 & 8.61 &  & -1.19$\pm$0.08 & 11.33$\pm$0.09 & -3.30$\pm$0.06 & 8.09$\pm$0.05 & 8.07 \\
$0.4<z\leq0.6$ & -1.22$\pm$0.07 & 11.26$\pm$0.11 & -2.76$\pm$0.13 & 8.57$\pm$0.04 & 8.56 &  & -1.22$\pm$0.07 & 11.18$\pm$0.06 & -3.14$\pm$0.05 & 8.11$\pm$0.04 & 8.07 \\
$0.6<z\leq0.8$ & -1.26$\pm$0.08 & 11.25$\pm$0.08 & -2.82$\pm$0.12 & 8.52$\pm$0.05 & 8.52 &  & -1.26$\pm$0.08 & 11.16$\pm$0.11 & -3.07$\pm$0.09 & 8.18$\pm$0.04 & 8.04 \\
$0.8<z\leq1.0$ & -1.23$\pm$0.09 & 11.27$\pm$0.09 & -2.91$\pm$0.14 & 8.44$\pm$0.05 & 8.44 &  & -1.23$\pm$0.09 & 11.20$\pm$0.09 & -3.19$\pm$0.08 & 8.10$\pm$0.04 & 7.95 \\
$1.0<z\leq1.3$ & -1.26$\pm$0.04 & 11.31$\pm$0.11 & -3.06$\pm$0.11 & 8.35$\pm$0.05 & 8.34 &  & -1.26$\pm$0.04 & 11.35$\pm$0.05 & -3.57$\pm$0.06 & 7.87$\pm$0.03 & 7.75 \\
$1.3<z\leq1.6$ & -1.29$\pm$0.08 & 11.34$\pm$0.10 & -3.27$\pm$0.18 & 8.18$\pm$0.07 & 8.17 &  & -1.29$\pm$0.08 & 11.62$\pm$0.16 & -3.96$\pm$0.09 & 7.77$\pm$0.08 & 7.61 \\
$1.6<z\leq2.0$ & -1.27$\pm$0.11 & 11.40$\pm$0.18 & -3.49$\pm$0.22 & 8.02$\pm$0.07 & 8.00 &  & -1.27$\pm$0.11 & 11.45$\pm$0.05 & -3.93$\pm$0.08 & 7.62$\pm$0.04 & 7.49 \\
$2.0<z\leq2.5$ & -1.26$\pm$0.08 & 11.46$\pm$0.15 & -3.69$\pm$0.22 & 7.87$\pm$0.09 & 7.85 &  & -1.26$\pm$0.08 & 11.26$\pm$0.10 & -3.83$\pm$0.16 & 7.52$\pm$0.07 & 7.29 \\
$2.5<z\leq3.0$ & -1.20$\pm$0.27 & 11.34$\pm$0.39 & -3.64$\pm$0.43 & 7.76$\pm$0.18 & 7.70 &  & -1.20$\pm$0.27 & 11.42$\pm$0.07 & -4.08$\pm$0.11 & 7.40$\pm$0.06 & 7.21 \\
$3.0<z\leq3.5$ & -1.14$\pm$0.21 & 11.33$\pm$0.31 & -3.74$\pm$0.43 & 7.63$\pm$0.14 & 7.40 &  & -1.14$\pm$0.21 & 11.26$\pm$0.26 & -3.97$\pm$0.36 & 7.33$\pm$0.17 & 7.01 \\
$3.5<z\leq4.0$ & -1.23$\pm$0.05 & 11.36$\pm$0.17 & -3.94$\pm$0.25 & 7.49$\pm$0.13 & 7.25 &  & -1.23$\pm$0.05 & 11.53$\pm$0.10 & -4.51$\pm$0.15 & 7.10$\pm$0.07 & 6.86 \\
\enddata
\tablecomments{$^a$ In units of $\mathcal{M}_\odot$. $^b$ In units of
$\mathrm{Mpc}^{-3}\,(\log\mathcal{M})^{-1}$. $^c$ In units of
$\mathcal{M}_\odot\mathrm{Mpc}^{-3}$.}
\end{deluxetable*}

\placetable{type_fractions}
\setcounter{table}{2}
\begin{deluxetable*}{llccccccccrrrr}
\tabletypesize{\scriptsize}
\tablewidth{570pt}
\tablecaption{\label{type_fractions}The IRAC sample: Comparison with other surveys.}
\tablehead{& \multicolumn{8}{c}{Number of sources/MIPS detections} & & \multicolumn{4}{c}{Number of sources}\\
\cline{2-9} \cline{11-14}\\
\colhead{} & \colhead{IRAC$^a$} & \multicolumn{4}{c}{LBG$^b$} &\colhead{DRG$^a$} &\multicolumn{2}{c}{$BzK$$^a$} 
         & & \colhead{LBG$^b$} & \colhead{DRG$^a$} &\colhead{$BzK$$^a$} & \multicolumn{1}{c}{JUST} \\
  \cline{3-6} \cline{8-9}\\
\colhead{Redshift} & \colhead{} & \colhead{GALEX} & \colhead{BM}  
         & \colhead{BX}   &\colhead{''classic''} &\colhead{} &\colhead{PE} 
         & \colhead{SF} & &\colhead{DRG$^a$} &\colhead{$BzK$$^a$} &\colhead{LBG$^b$} & \multicolumn{1}{c}{IRAC$^{a,c}$}}
\startdata
$(0.0,0.2]$ & 1473/242   & 0/0     & 81/11     & 510/94    & 2/0     & 9/4     & 0/0    & 468/62     & & 3   & 9   & 243  & 655    \\
$(0.2,0.4]$ & 1745/375   & 0/0     & 121/10    & 183/30    & 5/0     & 8/1     & 0/0    & 318/54     & & 1   & 8   & 134  & 1252   \\
$(0.4,0.6]$ & 2473/660   & 0/0     & 26/1      & 6/1       & 9/1     & 37/4    & 1/0    & 303/39     & & 1   & 32  & 29   & 2152   \\
$(0.6,0.8]$ & 3953/979   & 0/0     & 55/2      & 6/0       & 1/0     & 38/9    & 2/1    & 403/55     & & 1   & 27  & 39   & 3514   \\
$(0.8,1.0]$ & 4226/1065  & 300/117 & 273/43    & 12/2      & 0/0     & 78/28   & 2/0    & 508/109    & & 7   & 46  & 192  & 3314   \\
$(1.0,1.3]$ & 4140/1012  & 189/66  & 1100/189  & 117/25    & 1/1     & 118/48  & 21/5   & 1060/130   & & 13  & 87  & 540  & 2233   \\
$(1.3,1.6]$ & 2040/619   & 0/0     & 817/213   & 141/48    & 5/2     & 61/26   & 105/32 & 1458/439   & & 2   & 58  & 804  & 315    \\
$(1.6,2.0]$ & 1640/532   & 0/0     & 414/105   & 412/129   & 24/8    & 55/18   & 104/33 & 1417/468   & & 8   & 55  & 821  &  90    \\
$(2.0,2.5]$ & 1404/439   & 0/0     & 144/39    & 551/154   & 95/29   & 231/109 & 49/17  & 1274/406   & & 43  & 230 & 762  & 52     \\
$(2.5,3.0]$ & 882/264    & 0/0     & 4/0       & 135/31    & 278/82  & 253/127 & 23/15  & 677/197    & & 59  & 234 & 322  & 68     \\
$(3.0,3.5]$ & 558/162    & 0/0     & 0/0       & 1/1       & 365/85  & 171/93  & 15/10  & 294/104    & & 49  & 144 & 194  & 52     \\
$(3.5,4.0]$ & 529/95     & 0/0     & 0/0       & 4/0       & 276/36  & 165/59  & 33/20  & 115/23     & & 40  & 102 & 65   & 117    \\ 
\enddata
\tablecomments{$^a$ Any magnitude. 
$^b$ Magnitude limited to $R$$<$25.5. $^c$ IRAC sources not recovered
by any other selection criteria (i.e., they are not LBGs, DRGs, or
$BzK$ galaxies).}
\end{deluxetable*}

\placetable{type_densities}
\begin{deluxetable*}{lrrrrrrrrrrrrrrrrrr}
\tabletypesize{\scriptsize}
\tablewidth{560pt}
\tablecaption{\label{type_densities}The IRAC sample: stellar mass statistics and contribution to the stellar mass density.}
\tablehead{& \multicolumn{18}{c}{Stellar masses$^a$  and percentage of total stellar mass density} \\
\colhead{Redshift} & \multicolumn{1}{c}{ALL} &  \multicolumn{5}{c}{LBG} & & \multicolumn{5}{c}{DRG} & & \multicolumn{5}{c}{$BzK$}\\
\cline{3-7} \cline{9-13} \cline{15-19}\\
 & \multicolumn{1}{c}{IRAC} & \multicolumn{2}{c}{Any magnitude} & & \multicolumn{2}{c}{$R$$<$25.5} & & \multicolumn{2}{c}{Any magnitude} & & \multicolumn{2}{c}{$K$$<$22.9} & & \multicolumn{2}{c}{Any magnitude} & & \multicolumn{2}{c}{$K$$<$22.9}\\
\cline{3-4} \cline{6-7} \cline{9-10} \cline{12-13} \cline{15-16} \cline{18-19}
}
\startdata
$(0.0,0.2]$ &  7.4$^{8.3}_{6.5}$   &  7.2$^{8.1}_{6.2}$   & 12\%  & &  7.4$^{8.2}_{6.5}$   & 12\%  & &  7.7$^{8.5}_{7.6}$    &  0\% &  &  8.3$^{8.5}_{7.8}$   &  0\% & &  6.6$^{7.3}_{5.8}$   &  1\% & &  7.3$^{8.0}_{6.6}$   &  1\% \\
$(0.2,0.4]$ &  8.8$^{9.5}_{8.2}$   &  8.1$^{8.8}_{7.5}$   &  6\%  & &  8.2$^{8.9}_{7.6}$   &  6\%  & &  8.5$^{9.2}_{8.3}$    &  0\% &  &  8.9$^{9.4}_{8.4}$   &  0\% & &  7.9$^{8.6}_{7.5}$   &  4\% & &  8.7$^{9.1}_{8.0}$   &  3\% \\
$(0.4,0.6]$ &  9.6$^{10.1}_{9.1}$  &  9.0$^{9.3}_{8.7}$   &  0\%  & &  9.1$^{9.3}_{8.8}$   &  0\%  & &  8.6$^{9.7}_{8.4}$    &  1\% &  &  9.5$^{10.0}_{8.5}$  &  1\% & &  9.0$^{9.3}_{8.4}$   &  3\% & &  9.4$^{9.7}_{8.6}$   &  2\% \\
$(0.6,0.8]$ &  9.7$^{10.2}_{9.2}$  &  9.1$^{9.5}_{8.3}$   &  5\%  & &  9.2$^{9.5}_{8.8}$   &  5\%  & &  9.7$^{10.2}_{8.8}$   &  1\% &  & 10.1$^{10.2}_{9.7}$  &  0\% & &  9.1$^{9.5}_{8.4}$   &  6\% & &  9.7$^{10.0}_{9.3}$  &  6\% \\
$(0.8,1.0]$ &  9.8$^{10.3}_{9.3}$  &  9.7$^{10.1}_{9.2}$  & 18\%  & &  9.7$^{10.1}_{9.3}$  & 18\%  & &  9.9$^{10.4}_{9.2}$   &  1\% &  & 10.2$^{10.5}_{9.8}$  &  1\% & &  9.4$^{9.9}_{8.7}$   &  8\% & & 10.0$^{10.3}_{9.5}$  &  7\% \\ 
$(1.0,1.3]$ &  9.8$^{10.3}_{9.4}$  &  9.5$^{9.9}_{9.1}$   & 22\%  & &  9.6$^{10.0}_{9.2}$  & 21\%  & & 10.1$^{10.6}_{9.5}$   &  8\% &  & 10.4$^{10.7}_{9.9}$  &  8\% & &  9.6$^{10.1}_{9.1}$  & 19\% & & 10.2$^{10.5}_{9.8}$  & 17\% \\
$(1.3,1.6]$ & 10.2$^{10.7}_{9.8}$  &  9.9$^{10.2}_{9.6}$  & 24\%  & & 10.0$^{10.3}_{9.6}$  & 23\%  & & 10.7$^{11.3}_{10.4}$  &  7\% &  & 11.0$^{11.3}_{10.7}$ &  7\% & & 10.2$^{10.6}_{9.8}$  & 72\% & & 10.6$^{10.9}_{10.3}$ & 67\% \\
$(1.6,2.0]$ & 10.3$^{10.7}_{10.0}$ & 10.1$^{10.5}_{9.8}$  & 44\%  & & 10.2$^{10.5}_{9.8}$  & 42\%  & & 10.7$^{11.0}_{10.4}$  &  7\% &  & 10.9$^{11.1}_{10.8}$ &  6\% & & 10.3$^{10.7}_{9.9}$  & 93\% & & 10.7$^{11.0}_{10.4}$ & 87\% \\
$(2.0,2.5]$ & 10.4$^{10.9}_{10.0}$ & 10.2$^{10.6}_{10.0}$ & 52\%  & & 10.2$^{10.6}_{10.0}$ & 45\%  & & 11.0$^{11.2}_{10.7}$  & 34\% &  & 11.1$^{11.3}_{10.9}$ & 31\% & & 10.4$^{10.9}_{10.0}$ & 97\% & & 10.9$^{11.2}_{10.6}$ & 83\% \\
$(2.5,3.0]$ & 10.4$^{10.9}_{10.0}$ & 10.3$^{10.8}_{9.9}$  & 45\%  & & 10.3$^{10.7}_{9.9}$  & 34\%  & & 11.0$^{11.3}_{10.8}$  & 63\% &  & 11.2$^{11.4}_{11.0}$ & 53\% & & 10.5$^{11.0}_{10.1}$ & 87\% & & 11.0$^{11.3}_{10.8}$ & 69\% \\ 
$(3.0,3.5]$ & 10.5$^{10.9}_{10.0}$ & 10.4$^{10.9}_{9.8}$  & 62\%  & & 10.3$^{10.8}_{9.8}$  & 44\%  & & 11.1$^{11.4}_{10.8}$  & 73\% &  & 11.4$^{11.6}_{11.2}$ & 62\% & & 10.7$^{11.2}_{10.3}$ & 81\% & & 11.3$^{11.6}_{11.1}$ & 64\% \\
$(3.5,4.0]$ & 10.5$^{10.9}_{10.0}$ & 10.4$^{10.8}_{10.0}$ & 45\%  & & 10.4$^{10.7}_{9.9}$  & 34\%  & & 11.0$^{11.3}_{10.7}$  & 69\% &  & 11.3$^{11.5}_{11.1}$ & 50\% & & 10.9$^{11.3}_{10.6}$ & 56\% & & 11.4$^{11.6}_{11.1}$ & 39\% \\ 
\enddata
\tablecomments{$^a$ Logarithms of the median and quartiles of the distribution of stellar 
masses in units of $[\mathcal{M}_\sun]$. }
\end{deluxetable*}

\end{turnpage}

\clearpage

\begin{appendix}
\section{THE MERGED PHOTOMETRIC CATALOG}
\label{catalog}

This Appendix describes how we selected and measured multi-wavelength
photometry for the galaxies included in the IRAC and $I$-band selected
samples. First, we characterize the reduction, detection and
photometry procedures in the \spitzer\, images. Then, we outline how
we merged this photometry with the fluxes estimated in ground-based
optical and NIR images. Special details are given for the spectroscopy
compiled for our sources. We also discuss the methods used to remove
stars from our catalogs. Finally, we discuss the presence of Active
Galactic Nuclei (AGN) in our samples.

\subsection{IRAC and MIPS detection and photometry}
\label{irac_mips}

We compiled all the IRAC and MIPS data available in the HDF-N, the
CDF-S, and the LHF, including the GTO data in the 3 fields, the GOODS
data in HDF-N and CDF-S, and the data around the GOODS footprint in
the CDF-S taken as a \spitzer\, Legacy Survey (PI: van Dokkum). All
the reduced data (Basic Calibrated Data products delivered by the {\it
Spitzer Science Center}) were mosaicked together using the procedure
developed by \citet{2004ApJS..154...44H}. This procedure includes
pointing refinement, distortion correction, drizzling to a scale half
of the original (approximately $0.6\, \mathrm{arcsec/pixel}$), and
correction of detector artifacts (more noticeably, mux-bleeding).

Detection of sources in the IRAC images was performed with {\sc
sextractor} \citep{1996A&AS..117..393B}. Given that the FWHM of the
IRAC Point Spread Function (PSF) is 1.8-2.0$\arcsec$
\citep{2004ApJS..154...39F}, and that the PSFs are very sharp and
stable, almost all sources are point-like in the 4 channels, and
objects can be resolved for separations of the order of
$\sim$1$\arcsec$. The crowdedness of the our very deep images in the
two bluer IRAC bands is very high, mostly at 3.6\mic\, and especially
near bright stars, making the deblending of sources hard for automatic
procedures such as that used by {\sc sextractor}. To alleviate this
problem, we detected sources at 3.6\mic\, and also (separately) at
4.5\mic, where the depth is slightly lower and crowdedness is less
severe. The two catalogs built in the two bluer IRAC bands (at 3.6\mic\,
and 4.5\mic) were merged by removing sources whose separation was
smaller than 1$\arcsec$ (roughly, 1.5 pixels in the mosaicked
images). After the selection, we measured aperture photometry in the 4
IRAC images (fixing the positions and forcing the detection in all
bands) following the same technique used by
\citet{2004ApJS..154...44H}. Fluxes were measured in small apertures
of $4\arcsec$ diameter with {\sc sextractor} (obtaining almost
identical results with other software, such as DAOPHOT, which was used
by \citealt{2004ApJS..154...44H}). The final integrated magnitude was
obtained after applying an aperture correction based on empirical IRAC
PSFs. The aperture corrections for this $4\arcsec$ diameter aperture
are [0.32$\pm$0.03,0.36$\pm$0.02, 0.53$\pm$0.02,0.65$\pm$0.02]~mag for
channels [3.6,4.5,5.8,8.0]\mic, respectively, where the uncertainties
include the effects of typical World Coordinate System (WCS) random
alignment errors (always less than 1~pixel). For sources whose Kron
aperture (in optical/NIR bands) was larger than 6$\arcsec$ (a number
chosen by studying our simulations described below), we measured the
photometry with a large enough aperture to enclose the entire object
and checked the results with the {\sc mag\_iso} output given by {\sc
sextractor}. We would like to stress that all the sources in the IRAC
sample have measured fluxes at both 3.6\mic\, and 4.5\mic.

The characterization of the IRAC catalogs (i.e., the analysis of the
effects of confusion on the deblending of sources and the photometry)
was carried out by simulations consisting in adding artificial sources
to the mosaicked images. A given number of sources
(7~sources$/\mathrm{arcmin}^2$, which is the number of sources
corresponding to a Poisson uncertainty in the observed number
densities) of a given magnitude were added to the images, and then the
full detection and photometry procedure was applied. Bulge-dominated
galaxies of different sizes (from 1$\arcsec$ to 10$\arcsec$) were also
added in the simulation to check the photometry of nearby
(z$\lesssim$0.5) extended objects. By measuring the angular sizes of
galaxies in the optical/NIR images, we determined that ``extended
sources'' (defined as sources whose semi-major axis is larger than
6$\arcsec$, see below) are just a minor fraction of the total number
of IRAC sources at z$<$0.5 (less than 3\%), and completely negligible
(less than 0.5\%) at z$>$0.5. By checking the fraction of input
sources recovered by this procedure (in the same position within
1$\arcsec$ or 1.5 pixels), we estimated the completeness levels at
which our catalogs are reliable (in terms of deblending of sources and
photometry) and the accuracy of our photometry. As mentioned in
Section~\ref{data}, our IRAC catalogs are 75\% complete down to
[3.6]$\sim$23.3~mag.

Our simulations also show that for sources whose semi-major axis is
larger than 6$\arcsec$, aperture photometry in a 4$\arcsec$ diameter
aperture corrected to an integrated flux based on empirical PSFs,
underestimated (on average) the total flux of these sources (estimated
from the {\sc mag\_iso} output given by {\sc sextractor}) in more than
10\%, i.e., 1--2 times the typical measurement uncertainty (see
below). Thus, we considered sources larger than 6$\arcsec$ as extended
sources in IRAC (as also recommended in the \spitzer/IRAC cookbook),
and for them we estimated integrated fluxes using large apertures
enclosing the entire objects and the extended source aperture
corrections given in the \spitzer/IRAC cookbook.

The errors of the IRAC photometry were estimated from the sky
uncertainty (estimated with {\sc sextractor} with a box filtering
method), detector readout noise, Poisson noise in the measured fluxes
(using the detector gain and total exposure time per pixel), and the
uncertainty in the aperture corrections (which include the effect of
WCS errors). A 2\% absolute calibration uncertainty was also
considered
\citep{2005PASP..117..978R}.  The final uncertainties were checked
with our simulations. For each input magnitude interval, we analyzed
the output magnitudes obtained with our photometric procedure. For
[3.6]$=$20~mag, the typical uncertainty is 0.05~mag, and for
[3.6]$=$24~mag, the typical uncertainty is 0.3~mag. For
[4.5]$=$20~mag, the typical uncertainty is 0.05~mag, and for
[4.5]$=$24~mag, the typical uncertainty is 0.4~mag. For
[5.8]$=$19~mag, the typical uncertainty is 0.07~mag, and for
[5.8]$=$23~mag, the typical uncertainty is 0.4~mag.  For
[8.0]$=$19~mag, the typical uncertainty is 0.08~mag, and for
[8.0]$=$22~mag, the typical uncertainty is 0.4~mag.

All the MIPS 24\mic\, data for each field (including GTO and GOODS
data) were reduced and mosaicked together using the MIPS Data Analysis
Tool \citep{2005PASP..117..503G}.  We detected sources and measured
integrated fluxes using PSF fitting (with the {\sc DAOPHOT}
IRAF\footnote{IRAF is distributed by the National Optical Astronomy
Observatory, which is operated by the Association of Universities for
Research in Astronomy (AURA), Inc., under cooperative agreement with
the National Science Foundation} package) and aperture
corrections. Sources were detected in three passes to recover the
faintest sources, many of which are hidden by brighter
ones. Photometry was extracted for all the detected sources
simultaneously. For sources of noticeable extent (more than
25$\arcsec$), a large enough aperture was set accordingly. For the
rest, a circular aperture of diameter 15$\arcsec$ (6 pixels) was
utilized. For this aperture, a 17\% correction in flux must be used to
correct to the total flux (based on the theoretical PSF of MIPS). The
sky estimation was carried out in two steps, first removing the
large-scale variation (due to zodiacal light) and then measuring the
background around each source. Based on simulations similar to those
carried out with the IRAC data, we estimate that our catalogs are 75\%
completeness at $F(24)$$=$80\micJy. Uncertainties based on these
simulations are less than 5\% for sources with $F(24)$$>$400\micJy,
and 10\% for sources with $F(24)$$\sim$80\micJy.

\subsection{Optical and NIR photometry}
\label{optical_nir}

The \spitzer\, data were complemented with other publicly available
and proprietary photometric and spectroscopic data in the 3
fields. For the HDF-N and the CDF-S, the dataset is described in
detail in \citet{2005ApJ...630...82P}. For this paper, we added in the
HDF-N the $JK$ data described in Villar et al. (2007, in preparation;
with limiting magnitudes\footnote{Defined as the third quartile of the
magnitude distribution of our sample.}  $J$$=$22.4 and $K$$=$21.4),
GALEX data extracted from the GALEX archive (with limiting magnitudes
$NUV$$=$24.9 and $FUV$$=$25.3), the spectroscopic redshifts published
by \citet{2006ApJ...653.1004R}, and the GOODS IRAC and MIPS data. In
the CDF-S, we added an image of size $37\arcmin\times30\arcmin$ taken
in the $NB816$ filter with the Suprime-Cam instrument on Subaru (with
a limiting magnitude of $NB816$$=$24.8), the spectroscopic redshifts
published by \citet{2006A&A...454..423V}, and the GOODS and \spitzer\,
Legacy Survey (PI: van Dokkum) IRAC and MIPS data. For the LHF field
(not used in \citealt{2005ApJ...630...82P}), we summarize the main
characteristics of the dataset, including the wavelengths, limiting
magnitudes, and references for each filter in Table~\ref{lhdata}. The
Subaru observations in the CDF-S and the LHF were obtained from the
SMOKA Subaru Archive, and reduced using their pipeline {\sc sdfred}
v1.2. The photometric and astrometric calibration was carried out by
comparison with the Sloan Digital Sky Survey (SDSS DR4,
\citealt{2006ApJS..162...38A}) catalogs. The ING data were provided
(fully reduced and calibrated) from the CASU INT Wide Field Camera
Survey
\citep{2002A&A...382..522B,2003AN....324..178Y}. The $H$-band TIFKAM data 
\citep{2004ApJS..154..170L}  were reduced following typical NIR
procedures, and the photometric and astrometric calibration obtained
through comparison with Two Micron All Sky Survey (2MASS)
catalogs. The UKIDSS data were provided (fully reduced and calibrated)
by the UKIRT Infrared Deep Sky Survey (UKIDSS DR2,
\citealt{2007MNRAS.379.1599L}). All the images in our complete dataset
were calibrated photometrically (using direct observations of SDSS and
2MASS catalogs) and astrometrically (using SDSS and 2MASS
catalogs). Typical absolute photometric uncertainties were 0.03~mag,
and the WCS absolute uncertainty was always less than 0.5$\arcsec$.

\placetable{lhdata}
\begin{deluxetable}{lrll}
\tabletypesize{\scriptsize}
\tablewidth{250pt}
\tablecaption{\label{lhdata}Characteristics of the data compiled for the Lockman Hole.}
\tablehead{\colhead{Band}  & \colhead{$\lambda_{\mathrm{eff}}$} & \colhead{$m_{\mathrm{lim}}$}  & \colhead{Source}\\
\colhead{(1)} & \colhead{(2)} & \colhead{(3)} & \colhead{(4)}}
\startdata
IRAC-3.6      &  3.561  & 23.0 & {\it Spitzer} GTO \\
IRAC-4.5      &  4.510  & 23.0 & {\it Spitzer} GTO \\
IRAC-5.8      &  5.689  & 22.3 & {\it Spitzer} GTO \\
IRAC-8.0      &  7.958  & 22.0 & {\it Spitzer} GTO \\
MIPS-24       & 23.844  & 20.0 & {\it Spitzer} GTO \\
$B$           &  0.442  & 26.0 & Subaru Deep imaging$^a$\\
$R$           &  0.652  & 25.4 & Subaru Deep imaging$^a$\\
$I$           &  0.795  & 25.0 & Subaru Deep imaging$^a$\\
$z$           &  0.907  & 24.5 & Subaru Deep imaging$^a$\\
$U$           &  0.361  & 23.1 & ING Archive$^b$\\
$g$           &  0.486  & 24.0 & ING Archive$^b$\\
$i$           &  0.767  & 22.3 & ING Archive$^b$\\
$J$           &  1.251  & 22.5 & UKIDSS$^c$\\
$H$           &  1.649  & 20.4 & TIFKAM$^d$\\
$K$           &  2.208  & 22.9 & UKIDSS$^c$\\
\enddata
\tablecomments{(1)  Name of the observing band. (2) Effective wavelength
(in $\mu$m) of the filter$+$detector.  (3) Limiting AB magnitudes defined 
as the third quartile of the magnitude distribution of our sample. (4) 
Source from where the data were obtained: $^a$ publicly available ultra-deep
optical data from the SMOKA Subaru Archive, taken with the Suprime-Cam 
instrument on the Subaru Telescope; $^b$ data obtained from the 
Archive of the Isaac Newton Group of Telescopes, and taken with the
Wide Field Camera on the 2.5m Isaac Newton Telescope; $^c$ data provided by 
the UKIRT Infrared Deep Sky Survey (UKIDSS), data release 2 (DR2, 
\citealt{2007MNRAS.379.1599L}); $^d$ data taken with 
the TIFKAM instrument on the 2.1 m Telescope at Kitt Peak National
Observatory.  }
\end{deluxetable}

\subsection{Merged photometric catalog}
\label{rainbow}

Aperture matched photometry in all bands was carried out using the
procedure described in \citet{2005ApJ...630...82P}. The coordinates of
the IRAC detected sources are cross-correlated with each one of the
UV, optical and NIR catalogs using a search radius of 2.5$\arcsec$
(roughly two pixels in the original IRAC images) and starting by the
deepest images. Once the source was identified in one of these image
(for most cases, the first one), we took the
\citet{1980ApJS...43..305K} elliptical aperture best enclosing the
entire source from this reference image, and translated it to all the
other bands. The aperture was large enough to enclose the PSF (at
least twice the FWHM of the PSF) in all UV/optical/NIR images (the
seeing was always less than $1.5\arcsec$). By randomly varying the
center of this aperture in each image, we checked that small WCS
errors did not affect the integrated apertures significantly (the
variations were always well within the photometric uncertainties). For
IRAC and MIPS, where the PSFs are comparatively large, we assumed the
integrated magnitude measured in small apertures (applying aperture
corrections), as discussed previously. For GALEX data, given that the
FWHM of the PSF is $6\arcsec-7\arcsec$ (depending on the band,
position in the detector, and brightness of the source) we took the
{\sc mag\_best} magnitude given by {\sc sextractor}. For HST images,
we picked the integrated flux of the closest source measured with {\sc
sextractor}, not carrying out any aperture matching. For this reason,
HST fluxes were not used in the photometric redshift and stellar mass
determination.

Uncertainties of each measured flux were obtained from the sky
pixel-to-pixel variations, detector readout noise, Poisson noise in
the measured fluxes (taking into account the detector gain and total
exposure time per pixel, which were combined to give rms images of the
fields), the errors introduced by the uncertainties in the WCS, and
the uncertainty in the absolute photometric calibration (typically
0.03~mag). Reductions involving drizzling (e.g., in ACS or IRAC
images), non-integer pixel shifts (e.g., NIR images), and also
detector artifacts or unresolved faint sources, produce that
uncertainties derived uniquely from pixel-to-pixel variations of
adjacent sky pixels underestimate the real noise, since these effects
correlate the signal of nearby pixels
\citep[see, e.g.,][]{2000AJ....120.2747C,2003AJ....125.1107L,
2006ApJS..162....1G,2007AJ....134.1103Q}. To account for this, we
estimated the background level and noise in 3 different ways. First,
we measured the average signal per pixel and noise in a circular
corona 5$\arcsec$ wide surrounding the Kron photometric aperture for
each source, scaling the noise with a N$^{1/2}$ factor, where N is the
number of pixels of the source photometric aperture. To get rid of the
effect of correlated noise in this estimation of the uncertainty
introduced by the pixel-to-pixel variance, we also estimated the
background level and noise using 20 artificial apertures of the same
size as the one used for the source. These artificial apertures were
built by randomly selecting (in general, non-adjacent) "sky pixels" in
a 1$\arcmin$$\times$1$\arcmin$ box around the source. Those "sky
pixels" excluded the pixels whose signal was 5$\sigma$ above the rms
value estimated with the first method. The average signal and standard
deviation of the integrated fluxes within these artificial apertures
provided another (less biased) estimation of the background level and
noise. Finally, we also used 20 apertures of the same size, shape, and
orientation as the source photometric aperture in the
1$\arcmin$$\times$1$\arcmin$ box forcing that more than 90\% of their
pixels were "sky pixels" (as defined before) and applying the method
described in \citet{2003AJ....125.1107L}. The final background level
was set to the average of the three estimations, and the background
noise was set to the largest estimation provided by any of the three
previously described methods. In practice, the largest estimation of
the noise was, in most cases, provided by the second method: on
average, the noise was 10\%--20\% higher than what was obtained with
the first method (which proves the importance of correlated noise),
and less than 5\% higher than the third method.

The validity of the method used to obtain merged photometry from the
UV to the MIR bands was tested by comparing the measured colors with
those obtained from images convolved to the same resolution. For this
test, we matched the PSFs of an optical image (the $I$-band) to that
of the IRAC 3.6\mic\, channel (which is worse) using the IRAF {\sc
psf} task (which produces a convolution kernel to match the optical
PSF to the IRAC PSF). We then measured photometry in a $4\arcsec$
diameter aperture in both bands and obtain $I$-[3.6] colors for all
the detected sources. Note that since both images have the same PSF,
any aperture size could in principle be used to obtain colors, but
very small apertures would be more affected by WCS and PSF matching
errors. In the case of resolved nearby sources, very small apertures
could also bias the results if the colors are not uniform across the
galaxy. The colors derived with this method were very similar to those
obtained with our photometric procedure. The absolute mean difference
between both methods was $<$$|$$\Delta$$\{$$I$-[3.6]$\}$$|$$>$=0.02~mag (the
average difference was $<$$\Delta$$\{$$I$-[3.6]$\}$$>$=0.005~mag), and
the scatter 0.11~mag, comparable to the color uncertainties (the
average is 0.15~mag). The average difference is independent of the
integrated magnitude and size. For sources with $I$$<$22, we find
$<$$\Delta$$\{$$I$-[3.6]$\}$$>$=0.004$\pm$0.12~mag, for sources with
22$<$$I$$<$24 we measure
$<$$\Delta$$\{$$I$-[3.6]$\}$$>$=0.01$\pm$0.11~mag, and for sources
with $I$$>$24 we obtain
$<$$\Delta$$\{$$I$-[3.6]$\}$$>$=-0.02$\pm$0.12~mag. For sources
smaller than 6$\arcsec$ we find
$<$$\Delta$$\{$$I$-[3.6]$\}$$>$=-0.01$\pm$0.12~mag, and for larger
galaxies we find $<$$\Delta$$\{$$I$-[3.6]$\}$$>$=0.006$\pm$0.11~mag.

For some of the IRAC sources (10\%-15\% of the entire IRAC sample in
each field), there were several UV/optical/NIR counterparts in
ground-based images for one single IRAC source within the 2.5$\arcsec$
search radius. For these sources, we remeasured the IRAC fluxes by
fixing the positions of the blended objects and deconvolving the
images using the IRAC PSFs. Although the IRAC PSFs have FWHMs of
approximately 2$\arcsec$, the determination of the central position of
each IRAC source can be determined more accurately (the actual value
depending on the brightness of the source) and sources are resolved
for separations of the order of $\sim$1$\arcsec$. This means that if
the source positions are known, we can identify and deblend IRAC
sources separated $\sim$1$\arcsec$ from each other. We adopted a
similar deconvolution method to that used in
\citet{2006A&A...449..951G}. The ground-based optical/NIR reference
image was used to measure the positions of the different blended
sources. Then, the reference image and the IRAC images were realigned
locally (in a $1\arcmin\times1\arcmin$ square region around the
source) to minimize the WCS related errors in the photometry, which
were expected to be large in the very small apertures used in the
deconvolution method. The IRAC photometry in this case was measured by
convolving the IRAC PSF with the reference image PSF and scaling the
flux of each object to match the IRAC fluxes in an aperture of
$0.9\arcsec$ ($\sim$1.5 pixels in the IRAC images). For this aperture
size, the aperture correction of the IRAC bands are
[1.01$\pm$0.07,1.02$\pm$0.08, 1.2$\pm$0.10,1.44$\pm$0.14]~mag for
channels [3.6,4.5,5.8,8.0]\mic, respectively (including WCS
errors). For the separations between the sources which we are trying
to deconvolve (separations larger than $1\arcsec$ and smaller than
$2.5\arcsec$), the flux contamination from the surrounding sources to
a given one was, in most cases, lower than a 10\% of the flux in the
photometric aperture. The artificial source simulations validated this
procedure.

Most of the IRAC selected sources are detected in our deepest Subaru
images in the HDF-N: approximately 90\% are detected in $B$, $R$,
and/or $I$. In these bands, 75\% of our sources are brighter than
$B=25.5$, $R=24.9$, and $I=24.5$. In the CDF-S, 90\% of the sources
are detected in the $NB816$ filter, and 75\% of them are brighter than
$NB816=24.8$. In the same field, about 70\% of sources are detected in
$B$ (75\% of them are brighter than $B=25.3$), 60\% in $R$ (75\% of
them brighter than $R=24.8$), and 40\% in $I$ (75\% of sources
brighter than $I=23.7$). MIPS at 24\mic\, is able to detect about 25\%
of the IRAC sources (75\% of them above $F(24)$$=$40\micJy).

More than 90\% of the $I$-band selected sources (see
Section~\ref{data}) were also detected in deep $BVRz$ imaging. In
these bands, 75\% of our $I$-band sources are brighter than
$B$$=$26.0, $V$$=$25.8, $R$$=$25.5, and $z$$=$25.1. About 50\%-55\% of
the $I$-band sample is detected by IRAC (at 3.6 and 4.5\mic; at 5.8
and 8.0\mic, the fraction drops to 40\%-45\%). MIPS at 24\mic\, is
able to detect about 7\% of the $I$-band sources above
$F(24)$$=$80\micJy.

A negligible fraction of the entire IRAC sample (less than 3\%, not
large enough to change our results significantly) was detected in less
than 5 filters (our limit to calculate a reliable photometric
redshift), all of these galaxies presenting fluxes below the 75\%
completeness level. For the $I$-band selected sample, only 2\% of the
sources are detected in less than 5 filters.

\subsection{The spectroscopic sample}
\label{specz}

Both the HDF-N and the CDF-S include a large compilation of
spectroscopic redshifts obtained by several surveys. Unfortunately,
there is no public spectroscopic survey in the LHF, so this field
could not be used for building templates to estimate photometric
redshifts (see Section~\ref{photozs}). In the HDF-N, we used 1,699
spectroscopic redshifts ($\sim$20\% of the entire sample in that
field) found in \citet{2004AJ....127.3121W},
\citet{2004AJ....127.3137C}, and \citet{2006ApJ...653.1004R}. 
Only a fraction of those redshifts (1,340 sources) are flagged as high
reliability (larger than 80\%). In the CDF-S, we compiled 1,410
spectroscopic redshifts (about 15\% of the sample in that field), 891
of them flagged as reliable with a probability larger than 80\%, from
several sources: \citet{2004A&A...428.1043L},
\citet{2004ApJS..155..271S}, \citet{2005A&A...434...53V}, 
and \citet{2006A&A...454..423V}. More than half of the highly-reliable
spectroscopic redshifts are below z$=$1.0 (55\% in the CDF-S, and 80\%
in the HDF-N), and most of them are below z$=$1.5 (95\% in the CDF-S,
and 97\% in the HDF-N). These spectroscopic redshifts were
complemented with photometric redshifts estimated as explained in
Section~\ref{photozs}.

\subsection{Star-galaxy separation}
\label{stargalaxy}

In order to separate galaxies from stars in the merged photometric
catalogs, we used eight criteria, one based on the {\sc stellarity}
parameter given by {\sc sextractor} \citep{1996A&AS..117..393B}, and
the other seven criteria based on color-color and color-magnitude diagrams
using optical and NIR fluxes. All objects detected in more than one
optical or NIR band, and presenting an average value of the {\sc
stellarity} parameter larger than 0.95 were identified as stars. An
object was also considered a star if it satisfied any of these color
equations (when fluxes were available), extracted from
\citet{2004ApJS..154...48E}, \citet{2005AJ....129.1183R}, and 
\citet{2004ApJ...617..746D}:  a) $[3.6]-[8.0]>-2$ and 
$[3.6]-[8.0]<-1$ and $[8.0]<20.$, or $[3.6]-[4.5]>-1$ and
$[3.6]-[4.5]<-0.5$ and $[4.5]<19.5$; b) $[5.8]-[8.0]>-1$,
$[5.8]-[4.5]<-0.2$ and $[8.0]<20.$; c) $I-[8.0]<-1$ or $I-[3.6]<1$ and
$[3.6]<18.$ or $I-[8.0]<-1$ and $[3.6]-[8.0]<-1$; d)
$B-I>2\times(I-[3.6])+0.070$; e) $J-K+0.956<0.5$; f)
$[3.6]_\mathrm{3\arcsec}-0.460-[3.6]_\mathrm{auto}>-0.25$ and
$[3.6]<15.$ and
$[3.6]_\mathrm{3\arcsec}-0.460-[3.6]_\mathrm{auto}<0.2$, or
$[3.6]_\mathrm{3\arcsec}-[3.6]_\mathrm{auto}<-0.25$, where
$[band]_\mathrm{3\arcsec}$ is the magnitude in a $\mathrm{3\arcsec}$
diameter aperture, and $[band]_\mathrm{auto}$ is the {\sc mag\_auto}
magnitude given by {\sc sextractor} (an estimation of the integrated
magnitude); and g) $z-K$$<$-0.5+0.29$\times$$(B-z)$. The star-galaxy
separation for the IRAC sample was checked against the galactic number
counts published by \citet[][see also the stellar number counts
predicted by \citealt{1998ApJ...508...74A} and
\citealt{1992ApJS...83..111W}] {2004ApJS..154...39F}, finding very
good agreement with our results (absolute differences of less than
0.1dex at all fluxes down to the limits of our survey). Note that
these authors also show that the stars dominate the number counts at
the bright end, but they are a minor contributor at faint magnitudes
(less than 4\% of the sources at $[3.6]$$>$20 are stars), the range
where our extra-galactic analysis is concentrated. We have also
checked that our star detection is able to recover more than 95\% of
the stars in our IRAC sample that have been spectroscopically
confirmed: we identify 222 stars out of 232 spectroscopically
confirmed stars in the HDF-N, and 78 out of 82 sources in the
CDF-S. All the objects considered stars by the spectroscopy and missed
by our algorithm are extended (had effective radii larger than 3
pixels and FWHM larger than 4 pixels) in the ACS images. In the HDF-N,
our star detection algorithm identifies 6 spectroscopically confirmed
galaxies as stars, all of them being point-like in the ACS images. In
the CDF-S, 14 sources with a spectroscopic redshift are identified as
stars by our algorithm, all of them except two being point-like in the
ACS images.

\subsection{AGN identification}
\label{agns}

We used X-ray data (covering our entire surveyed regions in the 3
fields) to select candidates to harbor an AGN within our sample. In
the HDF-N, we used the catalog for the 2 Ms Chandra Deep Field-North
Survey published by \citet{2003AJ....126..539A}, finding an X-ray
counterpart\footnote{Our galaxies were cross-correlated with the X-ray
catalogs using a 2$\arcsec$ search radius, as done by
\citet{2004ApJS..154..160R} to match sources at large off-axis angles
in the Chandra images.} for 5\% of our IRAC sample in that field (3\%
of the $I$-band sample). In the CDF-S, we used the catalogs published
by \citet[][see also \citealt{2006A&A...451..457T}]{2002ApJS..139..369G} 
for the 1 Ms Chandra Deep Field-South Survey, identifying 3\% of our
IRAC sources as X-ray emitters (2\% of the $I$-band sample). In the
LHF, we identified AGN candidates using the XMM catalogs published by
\citet[][see also \citealt{2000A&A...354...35L}, and \citealt{2002A&A...393..425M}]
{2001A&A...365L..45H}, finding an X-ray counterpart for 0.4\% of the
IRAC sources in that field (0.2\% of the $I$-band sample). In the
total IRAC sample, 3\% of galaxies at any redshift were identified as
X-ray emitters (2\% of the entire $I$-band sample), with slightly
larger values (4\%--6\% of all IRAC sources) found for sources at
z$>$1.5. Observations in X-rays are known to miss very obscured AGNs
\citep[e.g., ][]{2004ApJS..154..160R,2005ApJ...634..169D}. 
Other selection procedures have been used to identify
obscured AGNs, such as the presence of a power-law spectrum in the
IRAC bands
\citep{2004ApJS..154..155A,2007ApJ...660..167D}. These power-law
galaxies (some also detected in X-ray or radio data) are also a very
small fraction of our IRAC selected sample, less than 1\% of the total
number of sources. We refer the reader to Section~\ref{agn_pars} for a
discussion about the characterization of the X-ray sources and the
effect of AGN contamination in our results.

\section{ESTIMATION OF PHOTOMETRIC REDSHIFTS, STELLAR MASSES, AND STAR FORMATION RATES}
\label{models_app}

\subsection{Stellar population synthesis models}
\label{subs1}

For the stellar population synthesis models of the SEDs of the
spectroscopic sample, we carried out two sets of fits: 1) one set
assuming that the star formation history of each galaxy can be
described by a declining exponential law with time scale $\tau$, age
$t$ (i.e., $SFR(t)\propto e^{-t/\tau}$), metallicity $Z$, and
attenuated by an amount described by the quantity $A(V)$ (1-POP
models, hereafter, see also
\citealt{2002AJ....123.1864G}); and 2) another set (2-POP models,
hereafter) assuming one recent instantaneous burst of star formation
of age $t_\mathrm{you}$, metallicity $Z_\mathrm{you}$ and extinction
$A(V)_\mathrm{you}$, overimposed on an evolved stellar population
characterized by $\tau_\mathrm{old}$, $t_\mathrm{old}$,
$Z_\mathrm{old}$, and $A(V)_\mathrm{old}$. The attenuation at any
wavelength was calculated from the free parameter $A(V)$ using the
\citet[][CF00 hereafter]{2000ApJ...539..718C} recipe. 
In this work, the attenuation of the gas and stellar emissions is
divided into three components, based on a simple scenario: the light
arising from the newly-formed stars, embedded in a birth cloud, is
attenuated by the material in the HII region, by a surrounding shell
of molecular and/or non-ionized atomic gas and dust, and finally by
the inter-stellar medium. The extinction law is approximated by a
power-law function of the form $A_\lambda\propto\lambda^{n}$ (the
authors suggest $n=-0.7$). There is also a dependence of the birth
cloud extinction on the age of the stars: for stars younger than
10~Myr (the typical lifetime of molecular clouds) the extinction is
$\mu$ times larger than for older stars, where $\mu$$\sim$0.3 (with
significant scatter). We also ran a set of models assuming that the
attenuation law was similar to the one found for local starbursts by
\citet[][CALZ00 recipe, hereafter]{2000ApJ...533..682C}. 
The stellar emission in our models was taken from the PEGASE code
\citep{1997A&A...326..950F}, assuming a  \citet{1955ApJ...121..161S} 
initial mass function (IMF) with
0.1$<$$\mathcal{M}$$<$100~$\mathcal{M}_\sun$ and a single power-law
slope through the entire mass range. We also added the emission from
the Hydrogen gas heated by the stars (emission lines and nebular
continuum) using the emission and recombination coefficients given by
\citet{1980PASP...92..596F} for an electron temperature
$T_e=10^4\,K$, the relations given by
\citet{1971MNRAS.153..471B}, and the theoretical line-ratios expected for
a low density gas ($n_{\mathrm{e}}=10^{2}\,\mathrm cm^{-3}$) with
$T_{\mathrm{e}}=10^{4}\,\mathrm K$ in the recombination Case B
\citep{1989agna.book.....O}. 

The 1-POP models required 4 parameters to fit. Our fitting routine
probed the solution space in the following ranges for the parameters
$[\tau,t,Z,A(V)]$: $i)$ we assumed $\tau$ values from an almost
instantaneous burst ($\tau=1$~Myr) to an almost constant SFR
($\tau=100$~Gyr) using a logarithmic interval of 0.1dex (in yr) for a
total of 51 steps; $ii)$ ages were probed from $t=1$~Myr to
$t=13.5$~Gyr in logarithmic intervals for a total of 60 steps,
constraining the solution for each object so the computed age was not
larger than the age of the Universe at the redshift of the galaxy;
$iii)$ we used the 7 discrete values of the metallicity available in
the PEGASE code $[0.005,0.0.02,0.2,0.4,1.0,2.5,5.0]\times Z_\sun$; and
$iv)$ extinction values ranged from $A(V)=0$~mag to $A(V)=5$~mag in
intervals of 0.10~mag (51 steps).

For the 2-POP models, each one of the 2 stellar populations requires
in principle 4 parameters to fit, but we forced the recent burst to be
instantaneous, so the young stellar population only requires 3 free
parameters to fit. Added to those 7 parameters, one more parameter is
necessary, the burst strength $b$, to describe the fraction of the
total stellar mass of the galaxy that the recent burst has
created. Our fitting routine probed the solution space in the
following ranges for the parameters
$[\tau_\mathrm{old},t_\mathrm{old},Z_\mathrm{old},A(V)_\mathrm{old},
\tau_\mathrm{you},t_\mathrm{you},Z_\mathrm{you},A(V)_\mathrm{you},b]$:
$i)$ $\tau_\mathrm{old}=1$~Myr to $\tau_\mathrm{old}=100$~Gyr using a
logarithmic interval of 0.1dex; $ii)$ $t_\mathrm{old}=1$~Gyr to
$t_\mathrm{old}=13.5$~Gyr in logarithmic intervals (constrained by the
age of the Universe at the redshift of each galaxy); $iii)$
$Z_\mathrm{old}$$=$ $[0.005,0.0.02,0.2,0.4,1.0,2.5,5.0]\times Z_\sun$;
$iv)$ $A(V)_\mathrm{old}=0$~mag to $A(V)_\mathrm{old}=5$~mag in
intervals of 0.1~mag; $v)$ we assumed an instantaneous burst for the
recent star formation event (i.e., $\tau_\mathrm{you}=1$~Myr, so
actually this is not a free parameter); $vi)$ $t_\mathrm{you}=1$~Myr
to $t_\mathrm{you}=1$~Gyr in logarithmic intervals (constrained by the
age of the Universe at the redshift of each galaxy); $vii)$
$Z_\mathrm{you}$$=$ $[0.005,0.0.02,0.2,0.4,1.0,2.5,5.0]\times Z_\sun$;
$viii)$ $A(V)_\mathrm{old}=0$~mag to $A(V)_\mathrm{old}=8$~mag in
intervals of 0.1~mag; and $ix)$ the burst strength could take values
from 0.5\% to 15\% in steps of 0.5\%.

\subsection{Stellar population synthesis fitting procedure}
\label{subs2}

The stellar population synthesis models were compared with the
observed photometric data of the galaxies in the spectroscopic sample
using a maximum likelihood estimator similar to the one defined in
Equation 6 by \citet{2003MNRAS.338..508P}, which takes into account
the uncertainties in each data point. All data points for rest-frame
wavelengths bluer than 4\mic\, (where stars should dominate the
integrated emission of the galaxy in most cases) were included in the
fit.

Given the large number of possible solutions ($1\times10^6$ in the
1-POP case and $3\times10^{11}$ for the 2-POP models), the amount of
photometric data to fit (up to 48 bands in the case of the sources in
the CDF-S, 16 in the HDF-N, and 14 in the LHF), and the number of
galaxies in our samples (more than 50,000 adding IRAC and $I$-band
selected galaxies), the time requirements to probe the complete
solution space for each galaxy (each one at a certain redshift) were
prohibitively high. Therefore, we had to use a minimization procedure
to search for the best solution without evaluating the minimization
function at all points in the grid of solutions. The minimization
procedure was a two step algorithm. First, we used a genetic algorithm
\citep{1995ApJS..101..309C}. This procedure started 
with 200 ``individuals'' (i.e., 200 points in the solution space),
whose ``genome'' was formed by the 4 or 8 free parameters in our
minimization problem. The 200 individuals were ``coupled'' randomly
(obtaining 100 couples). Each one of these couples (formed by
``parents'') produced 2 ``descendants''. Each descendant was built by
combining randomly the parameters of the parents. The ``genome'' of
the descendant had to be a better solution for the minimization
problem than the ``genome'' of its parents. If, after building 10
descendants for a given couple, none or only one of them were better
solutions to the minimization problem, the two best individuals (the
best solutions to the minimization problem) were kept for the next
generation, and the rest discarded. After every 10 combinations of
parameters, we allowed a random mutation in one of them. After all the
couples had produced 2 descendants, we eliminated the parents or
descendants that produced the worst results for the minimization
problem until 200 individuals survived, and then started again the
procedure for another generation with the best 200 individuals. The
total number of generations was set to 100. For the final generation,
we took the 4 best individuals (the best 4 solutions of the
minimization problem) and produced small grids of solutions around
them (with a width equal to one tenth of the full size of the solution
space for each free parameter). We evaluated all the solutions in
these grids, and found the best solution and confidence intervals. Our
minimization procedure was tested for a subsample of 1,000 galaxies in
the 1-POP case by comparing the best solution found by the algorithm
with that obtained by evaluating all the grid points in the entire
solution space. For this test sample, the minimization algorithm
recovered the best solution for $\sim$50\% of the galaxies. For the
rest of sources, the difference between the best value and the value
recovered by the minimization algorithm was always smaller or equal to
one tenth of the size of the grid for each free parameter. We will
come back to the discussion of the goodness of the minimization
algorithm in Section~\ref{evaluation}, when we discuss the quality of
the derived photometric redshifts, stellar masses, and SFRs.

\subsection{Dust emission models}
\label{subs3}

Once the stellar spectrum was modeled, we subtracted the predicted
fluxes from the photometric data points at rest-frame wavelengths
redder than 4\mic\, (if present) to obtain the emission arising from
the dust. This ``IR excess'' was then fitted with one of the dust
emission models of \citet{2001ApJ...556..562C}. We selected the model
best reproducing the colors of the dust emission, if several
photometric points were available (for relatively low redshift; see
the second SED fitting example in Figure~\ref{fig_template}), or the
model giving the closest value to the observed monochromatic
luminosity if only one IR photometric point was available (see the
third example in Figure~\ref{fig_template}). To check the
uncertainties in the derived IR-based SFRs, we also used the models of
\citet{2002ApJ...576..159D} and Rieke et al. (2007, in preparation)
in the fitting of the ``IR excess''. The latter are empirical spectral
templates constructed largely as described in
\citet{2007ApJ...660..167D}, Appendix A.  For ULIRGs, we used spectra
from \citet{2007ApJ...656..148A} for the star-forming galaxies IRAS
1211, 1434, 1525, 2249, in addition to the data on Arp 220 and IRAS
17208 described by \citet{2007ApJ...660..167D}.  For LIRGs, we used
IRS spectra from a MIPS GTO program led by A. Alonso-Herrero. Where
the LIRGs were expected to be extended at the scale of the IRS slit
width, they were mapped. The mapping data were reduced using "{\sc
cubism}" written by J. D. Smith as part of the SINGS legacy
program\footnote{A description of the "{\sc cubism}" software can be
found at
http://ssc.spitzer.caltech.edu/archanaly/contributed/cubism/}. The
mapped spectra were collapsed into a single one to represent the
integrated galaxy properties. For all the spectra, the templates were
extended to shorter wavelengths as described in
\citet{2007ApJ...660..167D}; by constraining spectral segments with
large beam photometry from 2MASS and IRAC, we are able to assemble
reliable templates. Toward long wavelengths, we collected photometry
from IRAS, ISO, Spitzer, and sub-mm facilities from NED.  These data
were fitted with a single blackbody with wavelength dependent
emissivity as $\lambda^\beta$.

\subsection{Photometric redshifts, stellar masses, and SFRs}
\label{photozs}

Our final reference template set is composed of 2,074 galaxies (1,310
galaxies from the HDF-N and 764 from the CDF-S), for which we obtained
1,666 different 1-POP$+$dust models (each one of them with a unique
combination of the free parameters), and 1,958 2-POP$+$dust models. As
mentioned earlier, these galaxies were selected from the spectroscopic
sample, all of them having a spectroscopic redshift measured with a
reliability probability larger than 80\%. In addition, all the
reference sources should have more than 10 different photometric data
points in their SEDs covering the UV, optical, and NIR/MIR spectral
ranges. Three examples of these templates are shown in
Figure~\ref{fig_template}, and discussed in
Section~\ref{evaluation}. The entire template set is available upon
request to the authors.


The photometric redshift estimation for each galaxy in our survey was
carried out with our own code in a similar way to that described in
\citet{2005ApJ...630...82P}. Briefly, the observed data (fluxes and
uncertainties) were compared with the redshifted models (with steps of
$\Delta\mathrm{z}$$=$0.01) using a $\chi^2$ minimization algorithm (as
the one used by \citealt{2000A&A...363..476B}). The method compared
the photometry with the convolutions of the different filters with the
redshifted templates, and determined the best template (the one giving
the lowest $\chi^2$ value) for each redshift.  The technique also
included a preliminary independent detection of the 1.6\mic\, bump
feature (if present), which helped to constrain the final solution and
get rid of outliers. The template giving the best solution at each
redshift also had to provide an age of the stellar population younger
than the age of the Universe at that redshift. The photometric
redshift probability distribution was built with the best values of
the $\chi^2$ estimator (corresponding to the model best reproducing
the observed SED) at each redshift, and the most probable redshift and
uncertainty were estimated from that probability distribution (as a
mean weighted with the probabilities, see
\citealt{2000A&A...363..476B}).

From the best model and most probable photometric redshift, we could
also obtain simultaneously an estimation of the stellar mass, as the
model established the monochromatic luminosity per unit of stellar
mass at all wavelengths. By scaling this model to the observed
monochromatic luminosities (multiplying by a factor), we obtained the
stellar mass of each galaxy. The final stellar mass and associated
uncertainty for each galaxy were obtained as the average and standard
deviation of the stellar masses obtained for each observed photometric
band. The uncertainty includes both the effect of the photometric
errors and the uncertainty in the determination of the redshift. These
errors are estimated for each galaxy by considering the photometric
redshift uncertainty and outliers derived from
Figure~\ref{fig_specz-photoz} for a galaxy in the same redshift and
magnitude intervals, and studying how variations in the redshift
affect the mass-to-light ratios in each band and the final stellar
mass estimate. The average stellar mass uncertainty is 0.2dex, typical
of any stellar population study
\citep[the typical accuracy of stellar masses obtained with stellar
population synthesis models is a factor of 2--3; see,
e.g.,][]{2003MNRAS.338..525P,2003MNRAS.341...33K,
2006ApJ...640...92P,2006A&A...459..745F}.

Star formation rates were estimated from the total IR luminosity
[$L(8-1000)$] calculated by integrating the dust emission models for
each galaxy between 8\mic\, and 1000\mic. Galaxies not detected by
MIPS at 24\mic\, were assumed to have an upper limit flux of
$F(24)=60$\micJy. The final SFR estimation also includes the
contribution from unobscured star formation detected directly in the
UV. According to \citet{2005ApJ...625...23B}, we can estimate the
total SFR for each galaxy from $L(8-1000)$ and $L(0.28)$, where
$L(0.28)=\nu L_\nu(\mathrm{0.28})$ is the monochromatic luminosity at
0.28\mic\, measured directly from the stellar population model for each
galaxy. The conversion factor is taken from
\citet{1998ARA&A..36..189K} for a \citet{1955ApJ...121..161S} IMF:

\begin{equation}
SFR=1.8\times10^{-10}[L(8-1000)+3.3
L(0.28)]/L_\sun\,\,\mathcal{M}_\sun\mathrm{yr}^{-1}
\end{equation}

In order to characterize the uncertainties of the SFRs derived with
our models, we also calculated IR-based SFRs by estimating
monochromatic luminosities at rest-frame wavelengths 6.7\mic, 12\mic,
and 15\mic. The integrated luminosity $L(8-1000)$ can be obtained from
these monochromatic luminosities by applying the empirical
relationships found in \citet{2001ApJ...556..562C}. Another estimation
of the SFR can be obtained from the rest-frame monochromatic luminosity
at 24\mic\, applying the equation given in
\citet{2006ApJ...650..835A}. We will discuss the uncertainties in the 
SFR estimations in Section~\ref{statsfr}.

\subsection{Evaluation of the modelling procedure and derived parameters}
\label{evaluation}

\subsubsection{Some examples of SED fits}

\begin{figure*}
\begin{center}
\includegraphics[angle=-90,width=6.8cm]{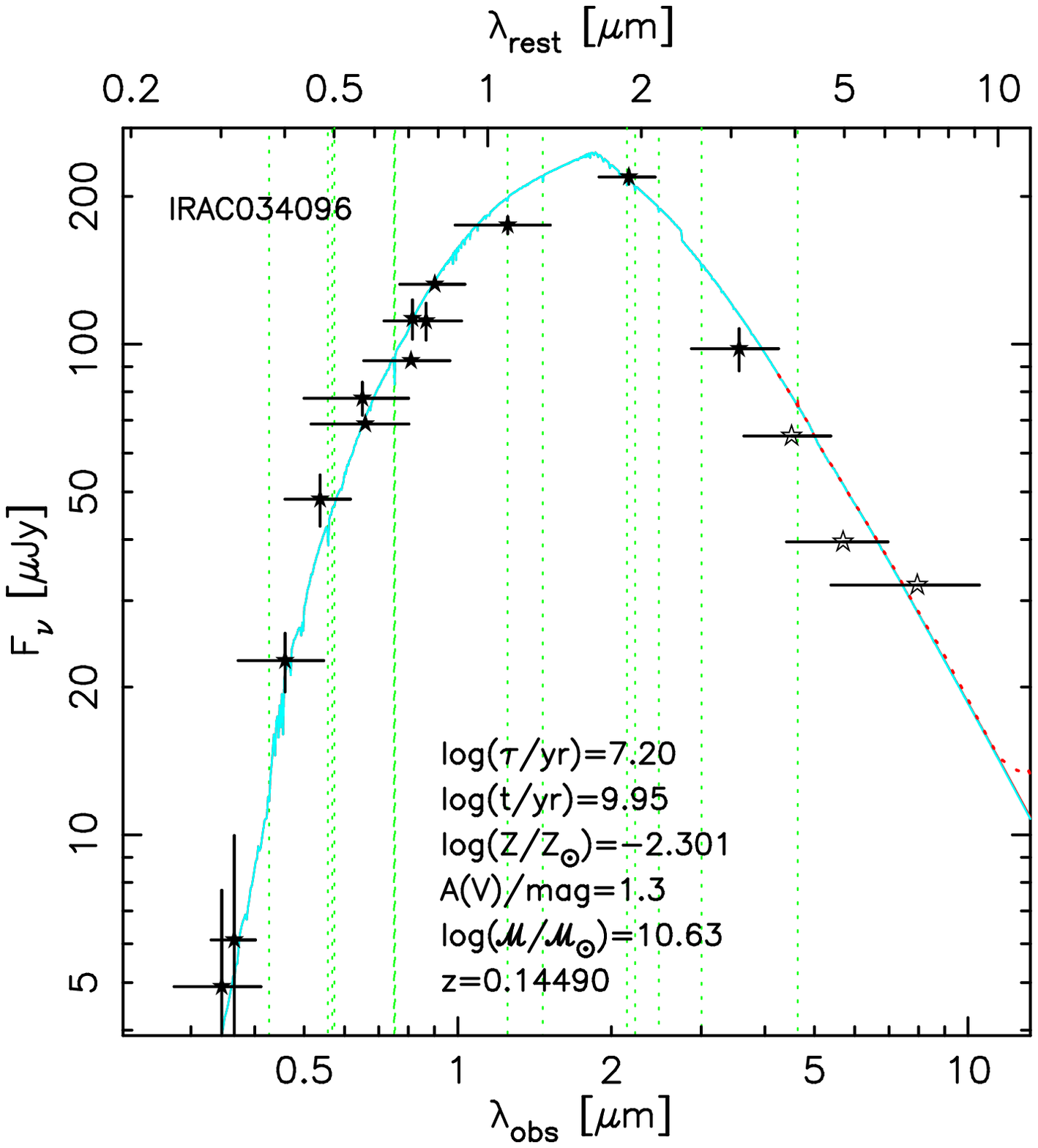}
\hspace{1cm}
\includegraphics[angle=-90,width=6.8cm]{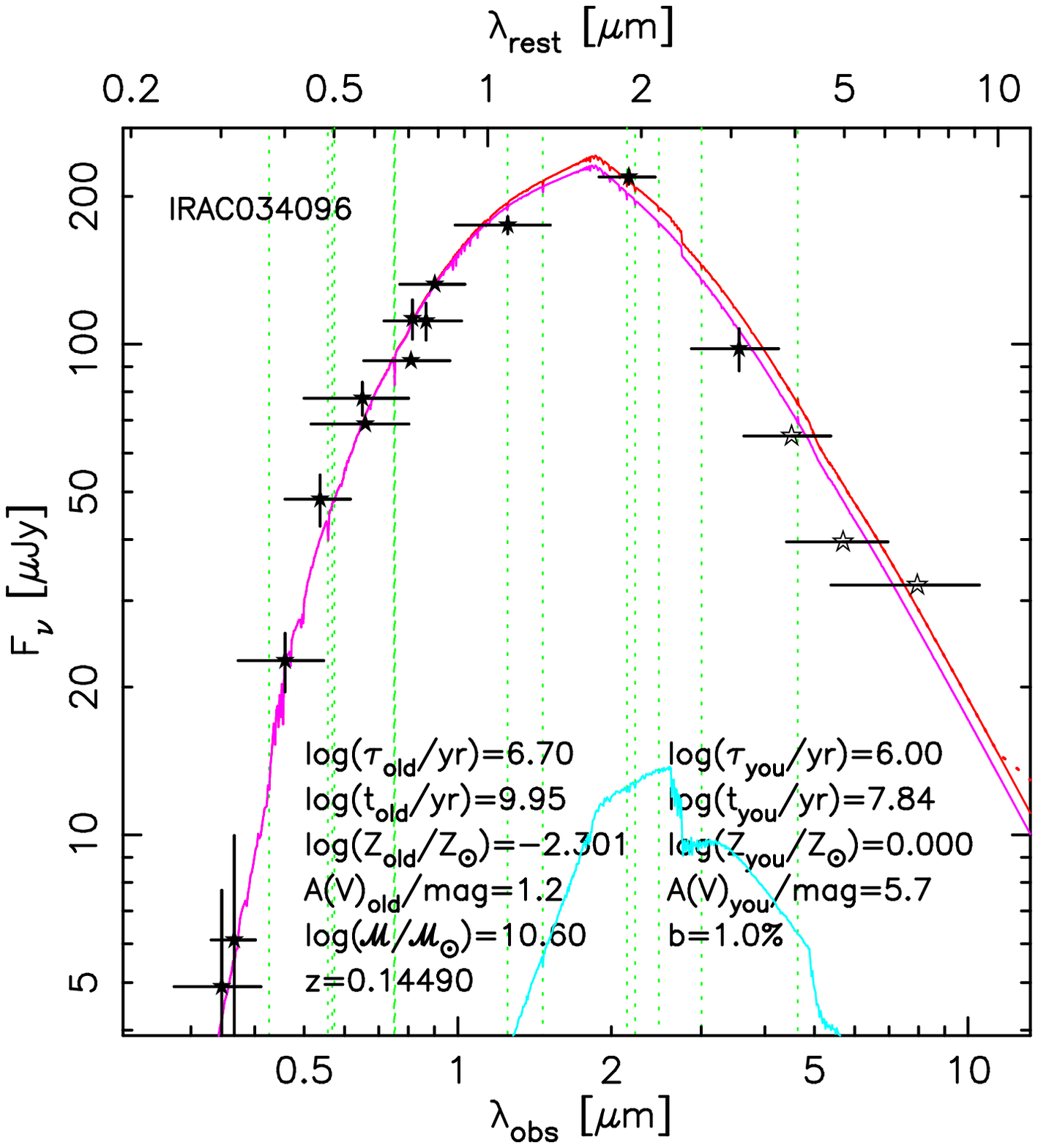}
\hspace{1cm}
\includegraphics[angle=-90,width=6.8cm]{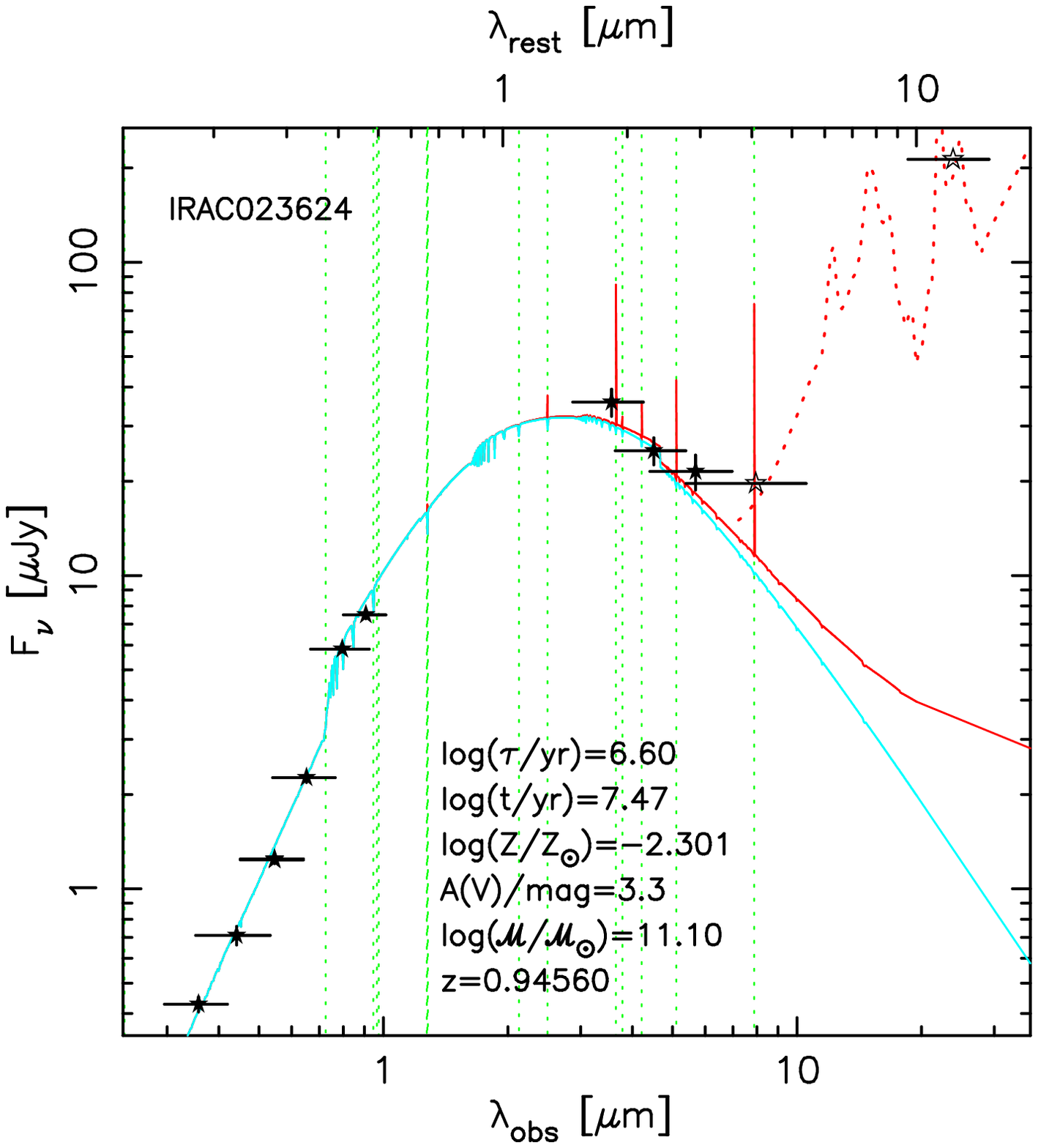}
\hspace{1cm}
\includegraphics[angle=-90,width=6.8cm]{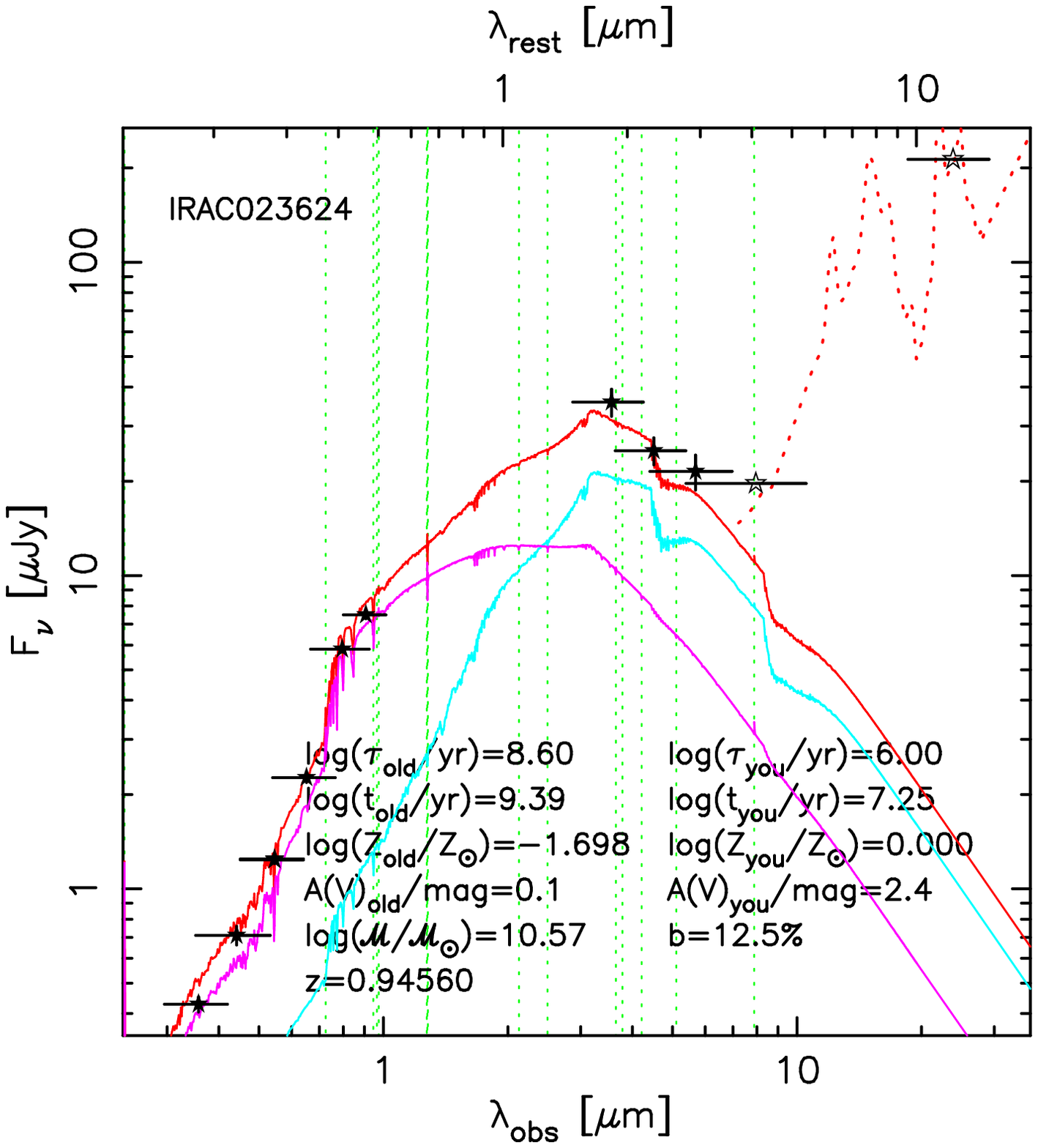}
\hspace{1cm}
\includegraphics[angle=-90,width=6.8cm]{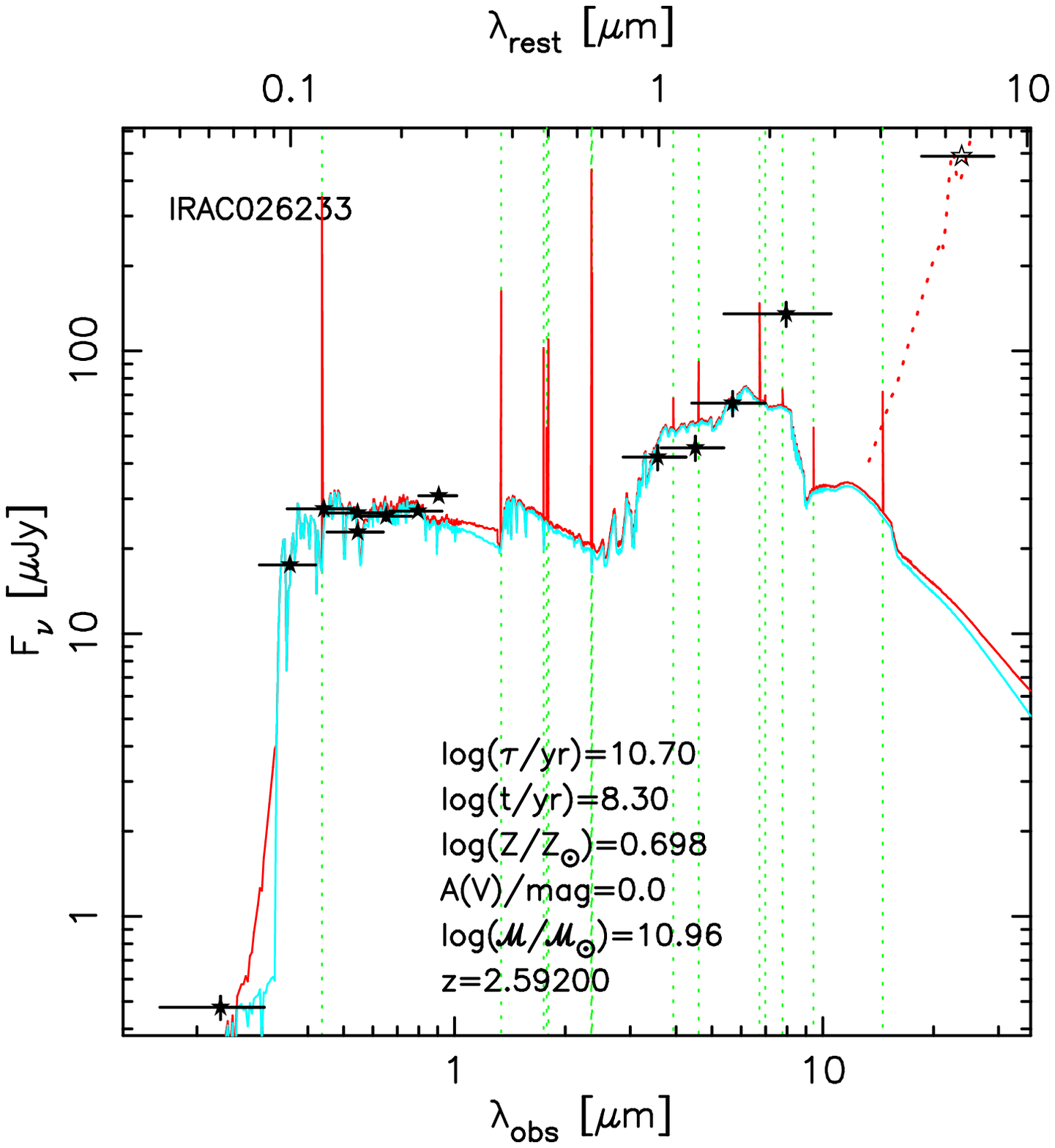}
\hspace{1cm}
\includegraphics[angle=-90,width=6.8cm]{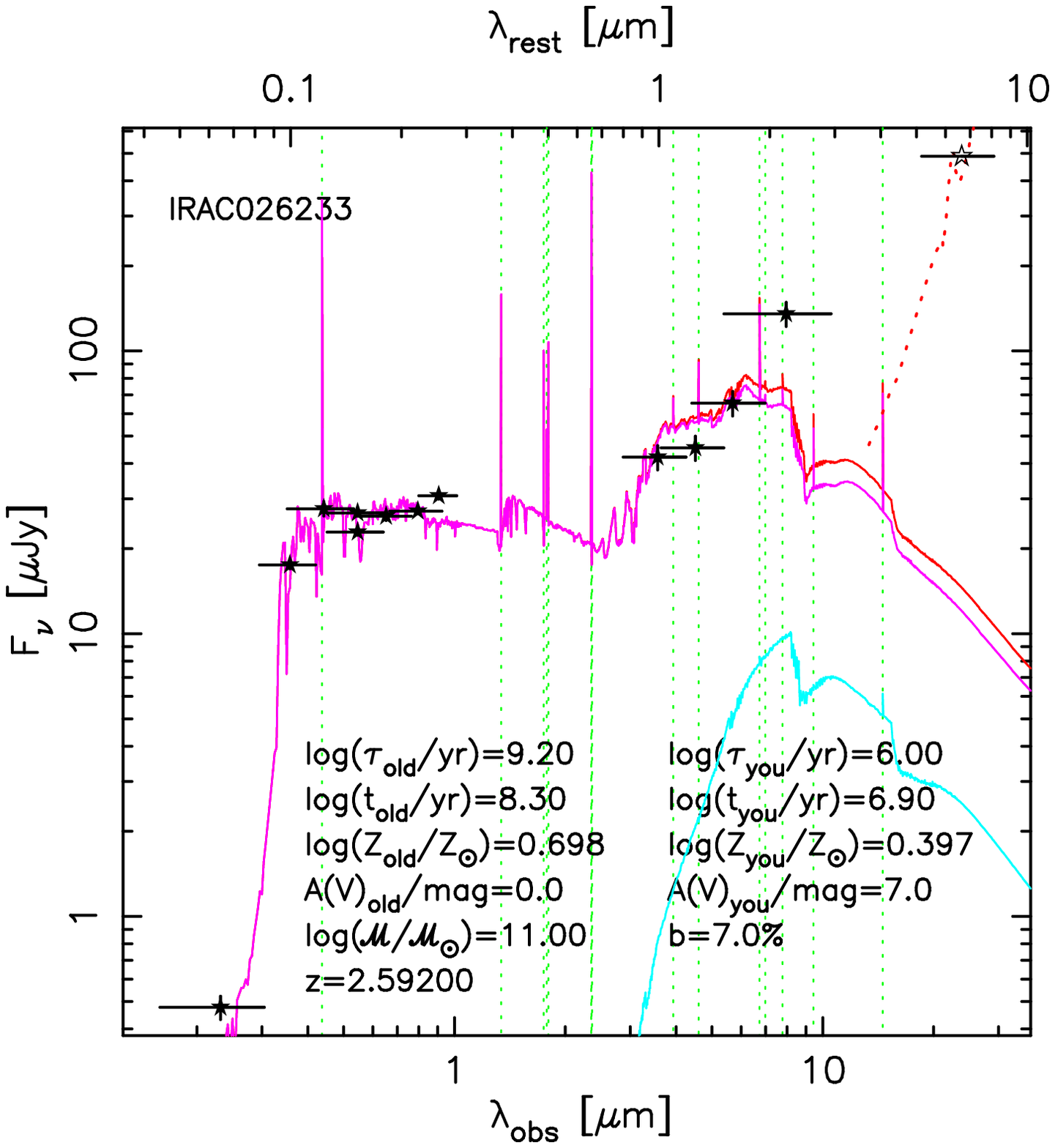}
\figcaption{\label{fig_template}Three examples of the stellar 
population and dust emission modelling of IRAC selected sources in the
spectroscopic sample. The spectroscopic redshift and main stellar
population parameters of the best fit are given in each panel. Filled
black stars and vertical error bars show the photometric points used
in the stellar population fits (wavelengths bluer than
4\mic). Horizontal error bars for each photometric point show the
width of the filter. Open black stars are the photometric data points
used in the modelling of the dust emission. The left panel of each row
shows the 1-POP stellar emission fits with a cyan line, and the final
fit (including nebular continuum and emission lines) with a red
continuous line. On the right panel of each row, the same SEDs have
been fitted with the 2-POP models, where one stellar population is
plotted with a cyan line, the other population with a magenta line,
and the addition of both with a red line (including nebular continuum
and emission lines). For all panels, the dust emission model taken
from \citet{2001ApJ...556..562C} which best reproduces the MIR
emission (if present) has been plotted with a dashed red line. Green
vertical lines show the positions of the most interesting
emission-lines in the optical and NIR spectral ranges.}
\end{center}
\end{figure*}

\slugcomment{Please, plot this figure with the width of two columns}
\placefigure{fig_template}

Figure~\ref{fig_template} shows three examples of the dust and stellar
population models for IRAC sources at different redshifts. The three
panels on the left show the fits for the 1-POP case, and the three
panels on the right show the fits for the 2-POP case for the same
sources. 

The upper two panels present a source nicely fitted by a single old
stellar population with intermediate extinction, no current star
formation (a bulge dominated galaxy), and no detection at 24\mic. In
this case, although the photometry at rest-frame wavelengths redder
than 4\mic\, was not used in the model fitting, those points are well
reproduced with just stellar emission. When fitting the same SED with
a 2-POP model, we recover very similar parameters to the 1-POP case,
with a minor contribution (just 1\% in mass) from a more recent
burst. Note that both types of models give very similar stellar mass
values.

The second example (middle row) shows an intermediate redshift galaxy
detected at 24\mic. This galaxy can be fitted either by an old single
stellar population with a large extinction or with a combination of an
old stellar population with very low attenuation and a more recent
burst contributing about 13\% to the total stellar mass of the
galaxy. This recent burst presents a relatively large dust attenuation
that could be responsible for the emission in the MIR/FIR (the galaxy
is detected at 24\mic). For this galaxy, dust emits a significant
fraction (about 50\%) of the 8\mic\, luminosity (rest-frame 4\mic) and
almost 100\% of the 24\mic\, luminosity (rest-frame 12\mic). The 1-POP
models give a larger stellar mass value than the 2-POP models (still,
the difference is a factor of 3, comparable to the typical uncertainty
in stellar population studies) because a lot of stars are necessary to
fit the high NIR photometric data points, while a lot of extinction is
necessary to simultaneously fit the UV/optical fluxes.

The third example (bottom row) is a high redshift galaxy with a very
blue spectrum. It can be fitted either with an almost continuous
unattenuated star formation (based on the high $\tau$ value) lasting
about 100~Myr, or with a similar primary burst (producing 93\% of all
the stellar mass) followed by a more recent (10~Myr) and much more
attenuated ($A(V)_\mathrm{you}$$=$7~mag and a strong MIR emission
detected at 24\mic) event of star formation. Note also that the IRAC
photometric point at 8.0\mic\, is too high in comparison with the
combined star$+$dust models. At wavelengths around
$\lambda$$\sim$4--10\mic\, or even at $\lambda$$\sim$2--10\mic\, for
very luminous IR sources (with very hot dust), the integrated emission
comes from both the dust and the stars in comparable fractions. In
this overlap region between the dust and stellar models, the spectrum
may show prominent emission-lines, PAH features, or emission from hot
dust (e.g., coming from a dust torus surrounding a nuclear massive
black hole) which are not found in the stellar and dust emission
models. For example, there is a PAH feature at rest-frame 3.3\mic\,
(very weak in all dust models in the
\citealt{2001ApJ...556..562C} or \citealt{2002ApJ...576..159D} 
libraries) which may have a non-negligible contribution to the global
emission in this spectral region for very luminous IR sources.

\subsubsection{Statistical evaluation of the photometric redshifts}

\begin{figure}
\begin{center}
\includegraphics[angle=-90,width=9.cm]{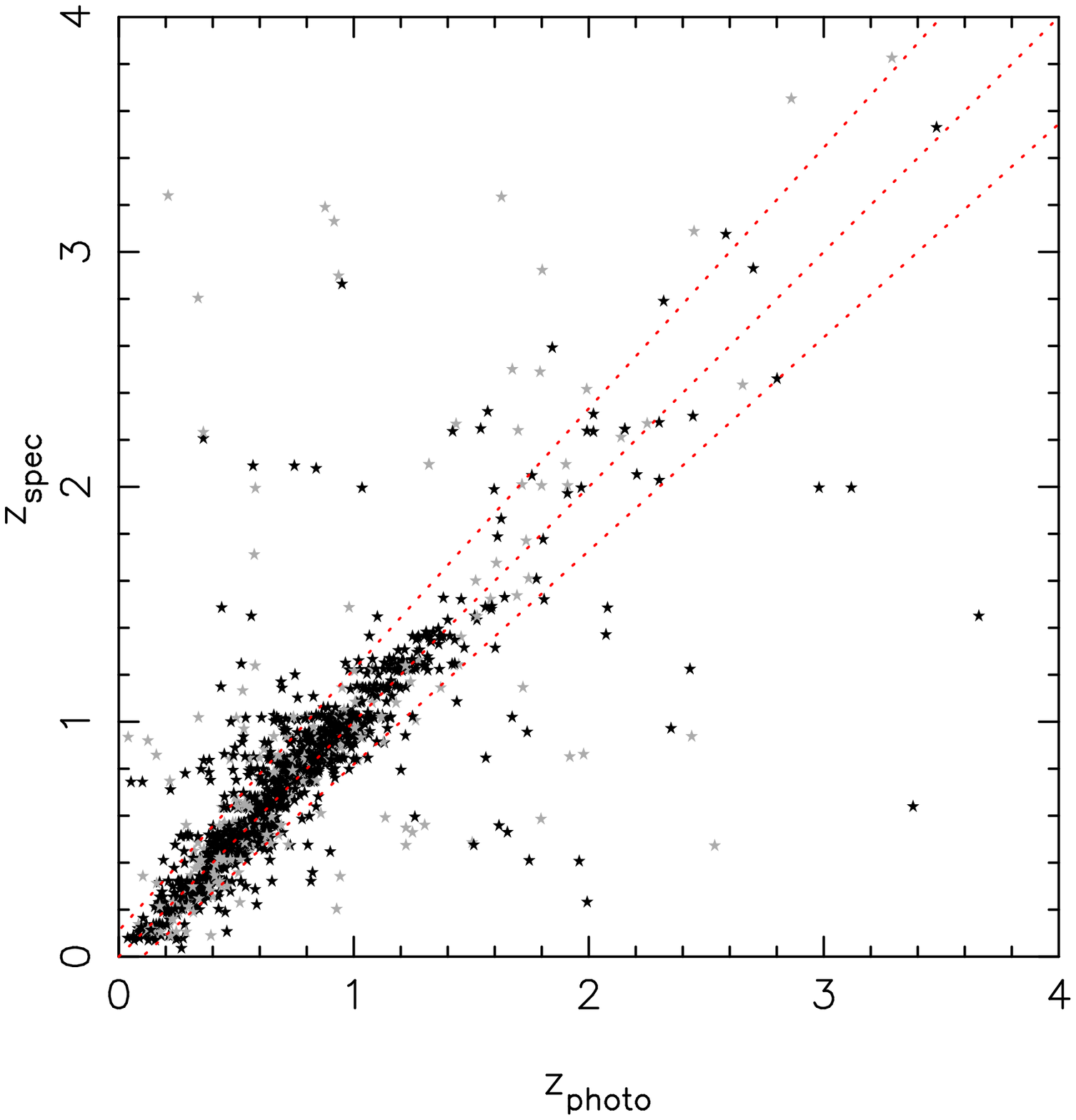}
\includegraphics[angle=-90,width=9.cm]{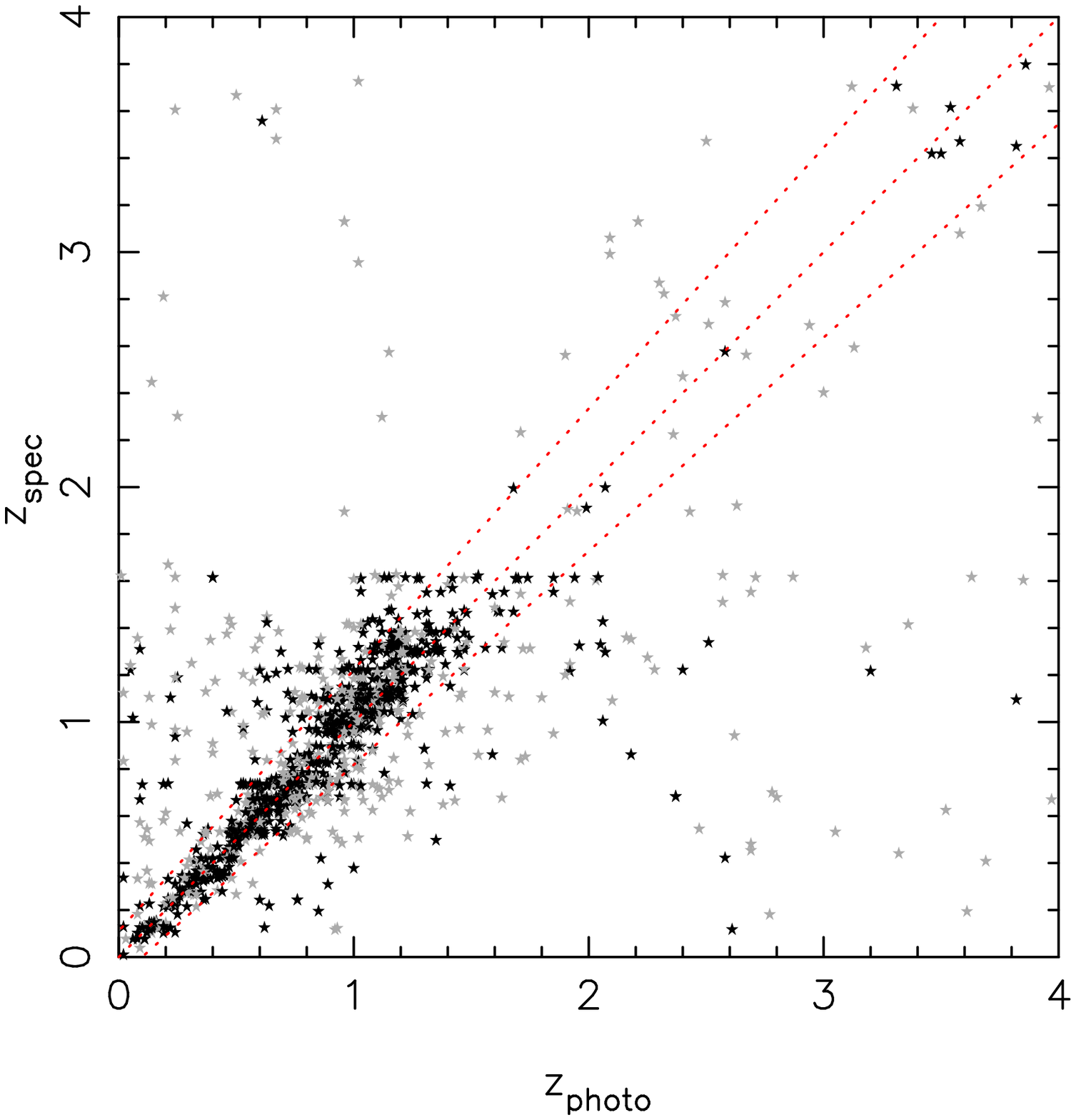}
\figcaption{\label{fig_specz-photoz}Comparison of the spectroscopic and 
photometric redshifts for IRAC selected sources in the HDF-N (top
panel) and the CDF-S (bottom panel). Gray symbols are sources with
spectroscopic redshifts which have a reliability probability lower
than 80\%. Open stars are sources detected in less than five
bands. The dashed lines show the equality line, and the
$\sigma_\mathrm{z}$/(1+z)$<$0.1 area. }
\end{center}
\end{figure}

\slugcomment{Please, plot this figure with the width of two columns}
\placefigure{fig_specz-photoz}

The main three parameters that we want to extract from the SED fits
are the photometric redshift, the stellar mass, and the SFR of each
galaxy. The quality of our photometric redshifts is checked in
Figure~\ref{fig_specz-photoz} for the fields with extensive
spectroscopic data: the HDF-N and the CDF-S. Unfortunately, given that
there is not a systematic public spectroscopic survey in the LHF, we
cannot check our photometric redshifts directly in this field. In
spite of this, the photometry in the LHF is as good or even better
(given that the optical images are ultra-deep observations taken with
Subaru) than in the other two fields, and the general redshift
distribution for the LHF sources is similar to that in the HDF-N and
the CDF-S. Consequently, we conclude that the quality of the
photometric redshifts in the LHF must be comparable to the other two
fields (see also the discussion about the redshift distribution of our
sample in Section~\ref{photozsect}).

The top panel of Figure~\ref{fig_specz-photoz} shows the comparison of
our photometric redshifts and spectroscopic redshifts for the IRAC
selected sample in the HDF-N (for the 1,702 sources with available
spectroscopy). The average (median) redshift difference
($\delta$z$=$z$_\mathrm{spec}$-z$_\mathrm{photo}$) is 0.014 (0.010),
comparable to the redshift step used in our photometric redshift
technique. This demonstrates that there are no systematic errors in
our redshifts. Almost all sources, 95\%, have values of
$\sigma_\mathrm{z}$/(1+z)$<$0.2 (where $\sigma_\mathrm{z}$ is the
absolute value of $\delta$z), 88\% of the objects present values of
$\sigma_\mathrm{z}$/(1+z)$<$0.1, and 70\% have
$\sigma_\mathrm{z}$/(1+z)$<$0.05. The average (median) value of
$\sigma_\mathrm{z}$/(1+z) is 0.055 (0.032). Very similar statistics
are obtained for the $I$-band selected sample: 94\% of these sources
present $\sigma_\mathrm{z}$/(1+z)$<$0.2, 86\%
$\sigma_\mathrm{z}$/(1+z)$<$0.1, and the average (median)
$\sigma_\mathrm{z}$/(1+z) is 0.060 (0.036).

The quality of the photometric redshifts in the CDF-S for IRAC
selected sources (for the 1,410 sources with available spectroscopy)
is shown in the bottom panel of Figure~\ref{fig_specz-photoz}.  The
average (median) value $\delta$z is 0.020 (0.015). In this field, 93\%
of the objects have values of $\sigma_\mathrm{z}$/(1+z)$<$0.2, 85\% of
the objects have values of $\sigma_\mathrm{z}$/(1+z)$<$0.1, and 67\%
have $\sigma_\mathrm{z}$/(1+z)$<$0.05. The average (median)
$\sigma_\mathrm{z}$/(1+z) is 0.060 (0.040). For the $I$-band selected
sources, the numbers are similar: 92\% of these sources present
$\sigma_\mathrm{z}$/(1+z)$<$0.2, 80\% $\sigma_\mathrm{z}$/(1+z)$<$0.1,
and the average (median) $\sigma_\mathrm{z}$/(1+z) is 0.080 (0.047).

Some of the sources used in the photometric redshift evaluation
depicted in Figure~\ref{fig_specz-photoz} were used in the building of
the templates (77\% of all sources plotted in the HDF-N, and 54\% in
the CDF-S). The validity of our procedure (including the merging of
photometric data) and the templates built with data from the HDF-N and
the CDF-S was also tested in completely different and independent
fields. For example, in the Extended Groth Strip (P\'erez-Gonz\'alez
et al. 2007, in preparation), we compared our photometric redshifts
(based on photometry measured in the same way as in this paper) with
spectroscopic values for 6,828 sources, obtaining that for 87\% of the
galaxies, our photometric redshifts were better than
$\sigma_\mathrm{z}$/(1+z)$<$0.1, and for 95\% were better than
$\sigma_\mathrm{z}$/(1+z)$<$0.2.

Figure~\ref{fig_specz-photoz} demonstrates the high quality of our
photometric redshifts at z$\lesssim$1.5. Beyond this redshift,
spectroscopic surveys have severe limitations due to the intrinsic
faintness of the sources (most of them are below the typical R$\sim$25
spectroscopic limit) and the absence of bright spectroscopic features
in the observed optical range for sources at
1.5$\lesssim$z$\lesssim$2.5 (the redshift desert). Therefore,
photometric redshifts cannot be extensively tested at high-z, given
that very few spectroscopic redshifts are available. To overcome this
problem as much as possible, we included up to 59 galaxies at z$>$1.5
in our template set, most of them extracted from spectroscopic surveys
carried out with spectrographs with enhanced sensitivity in the blue
\citep[e.g.,][]{2004ApJ...604..534S,2006ApJ...653.1004R}. Using the 
very few sources with reliable spectroscopic redshifts at z$>$1.5, our
photometric redshifts seem to degrade to some extent. In the HDF-N,
69\% of the sources at z$>$1.5 have photometric redshifts
$\sigma_\mathrm{z}$/(1+z)$<$0.2 and 50\% with
$\sigma_\mathrm{z}$/(1+z)$<$0.1. In the CDF-S, 86\% of the sources at
z$>$1.5 have photometric redshifts $\sigma_\mathrm{z}$/(1+z)$<$0.2 and
59\% with $\sigma_\mathrm{z}$/(1+z)$<$0.1. 

These statistics are highly biased against red objects, and for blue
sources, a very small number of sources is used in the comparison
(less than 30 in each field). To further test our results at high-z,
we analyzed the photometric redshift distribution of samples of
galaxies selected with the different color techniques described in
Section~\ref{comparison}. We considered all the IRAC sources
identified as LBGs with $R$$<$25.5, and DRGs or/and $BzK$ galaxies
with $K$$<$22.9 ($K$[Vega]$<$21). In Section~\ref{comparison}, we
demonstrate that our IRAC survey detects virtually all these sources
(given that the number densities in our sample are very similar to
those measured by other surveys focused on the detection of these
high-z populations). Here, we test the photometric redshifts derived
for these sources, a topic that will be discussed in more detail in a
future paper (Barro et al., 2007, in preparation).

The average redshift of the LBG-BM sources in our IRAC sample is
$<$z$>$$=$1.4$\pm$0.3. Both the average and standard deviation values
agree, within the typical photometric redshift uncertainties, with the
average spectroscopic value of $<$z$>$$=$1.7$\pm$0.3 given by
\citet{2004ApJ...604..534S}. In
\citet{2006ApJ...653.1004R}, they also find an average $<$z$>$$=$1.7$\pm$0.3 
for the LBG-BM sources in the HDF-N, where our average photometric
redshift is $<$z$>$$=$1.6$\pm$0.3. If we calculate the average
spectroscopic redshift for the LBG-BM sources in our IRAC sample with
available spectroscopy (9\% of the total), we obtain
$<$z$>$$=$1.3$\pm$0.3. Our average is also consistent with the average
photometric redshift published by \citet{2007AJ....134.1103Q} for
LBG-BM galaxies in MUSYC, $<$z$>$$=$1.4, and the first peak of the
photometric redshift distribution of LBGs in the GOODS-MUSIC sample
\citep{2007A&A...465..393G}, also at z$\sim$1.4.

The average redshift of the LBG-BX sources in our IRAC sample is
$<$z$>$$=$2.0$\pm$0.4, consistent with the spectroscopic values of
$<$z$>$$=$2.2$\pm$0.3 found by \citet{2004ApJ...604..534S} and
$<$z$>$$=$2.2$\pm$0.4 by \citet{2006ApJ...653.1004R} in the HDF-N
(where we obtain $<$z$>$$=$2.1$\pm$0.3). The LBG-BX sources in our
sample with available spectroscopy (5\% of this sub-sample) have an
average spectroscopic redshift of $<$z$>$$=$1.7$\pm$0.4.  Our average
is again in perfect agreement with the average photometric redshift
published by \citet{2007AJ....134.1103Q} for this population,
$<$z$>$$=$2.1, and the second peak of the photometric redshift
distribution of LBGs in
\citet{2007A&A...465..393G}, placed at z$\sim$2.2.

The sources in our sample identified as ``classical'' LBGs lie at an
average redshift of $<$z$>$$=$3.1$\pm$0.5, which compares well with
the spectroscopic values from \citet{2003ApJ...592..728S},
\citet{2004ApJ...604..534S} and \citet{2006ApJ...653.1004R}, 
all of them being $<$z$>$$=$3.0$\pm$0.3. Only 2\% of our sources
identified as ``classical'' LBGs have spectroscopy, and the average
spectroscopic redshift for them is $<$z$>$$=$2.5$\pm$1.0.

The average photometric redshift of the population of DRGs identified
in our IRAC survey is $<$z$>$$=$2.2$\pm$1.0, and the median is
z$=$2.5, in good agreement (taking into account photometric redshift
uncertainties) with the average spectroscopic redshift
$<$z$>$$=$2.5$\pm$0.4 in the HDF-N (we obtain $<$z$>$$=$2.4$\pm$0.9
just in this field) published by
\citet{2005ApJ...633..748R}, the median and rms photometric values 
z$=$2.6$\pm$0.7 published by
\citet{2003ApJ...587L..79F}, the median photometric redshift 
z$=$2.2 from \citet{2006ApJ...640...92P}, and the median photometric
redshift z$\sim$2.5 from \citet{2007AJ....134.1103Q}.  Spectroscopic
redshifts are available for just 4\% of the DRGs in our sample, with
an average of $<$z$>$$=$1.5$\pm$0.9, a lower value than the
photometric estimation, probably due to the bias of spectroscopic
surveys towards the optically brightest sources (whose probability of
being at lower redshifts is relatively larger).

$BzK$ (combining both PE and SF sub-types) sources in our IRAC sample
have an average photometric redshift $<$z$>$$=$2.1$\pm$0.6, which
compares nicely with the average spectroscopic value
$<$z$>$$=$2.1$\pm$0.4 from \citet{2005ApJ...633..748R}. Other
photometric redshift studies obtain similar redshift distribution,
e.g., \citet{2007AJ....134.1103Q} and
\citet{2007A&A...465..393G}. The average spectroscopic redshift for $BzK$ 
sources in our sample (available just for a 1\% of the total number
of $BzK$ galaxies) is $<$z$>$$=$1.7$\pm$0.3.

The previous statistics and the consistency with spectroscopic and
photometric values found in the literature demonstrate that our
photometric redshifts for the galaxies at z$>$1.5 are also
reliable. Still, a spectroscopic survey focused on high redshift
sources is necessary to increase the reliability (narrow the
uncertainties) of our results at z$>$1.5.

\subsubsection{Statistical evaluation of the stellar masses}


In this section, we discuss the quality of our stellar masses, and the
possible systematics introduced by our fitting algorithm and the {\it
a priori} assumptions of the models.

First, we checked how well the minimization algorithm of our SED
fitting technique recovered the stellar mass value corresponding to
the model best fitting the data. For that purpose, we used 1,000
randomly selected galaxies for which we probed all the nodes in the
solution grid for the 1-POP case. On average, the difference between
the stellar mass estimated with the minimization algorithm and the
stellar mass given by the model best fitting the SED is 0.002dex, the
median is 0.000dex, the standard deviation is 0.07dex, and there are
not any absolute differences larger than 0.20dex. For the 2-POP case,
the number of points in the solution grid is too large to attempt the
individual evaluation of all of them. To test this case, we only
considered 100 randomly selected galaxies and degraded the resolution
of the parameter space grid by one third for all the free parameters
(thus, we only considered $1\times10^{8}$ models). On average, the
difference between the stellar mass estimated with the minimization
algorithm and the stellar mass given by the 2-POP model (with a coarse
solution grid) best fitting the SED is 0.04dex, the median is 0.02dex,
the standard deviation is 0.15dex, and there are not any absolute
differences larger than 0.30dex. These statistics confirm that the
minimization algorithm is able to recover the best stellar mass
estimate within the typical uncertainties in stellar populations
synthesis analysis (a factor of 2--3).

We also compared the stellar masses obtained with the 1-POP and 2-POP
models. For about 70\% (55\%) of the galaxies, both estimates are
equal within a factor of 0.3dex (0.2dex). However, for the rest of
galaxies (virtually all of them with
$\mathcal{M}$$<$$10^{10.5}$~$\mathcal{M}_\sun$), the 2-POP estimates
are higher (with the most extreme cases showing a difference of up to
a factor of 10). On average, including all galaxies, stellar masses
derived with 2-POP models are 0.18dex higher than those derived with
1-POP models. For galaxies with
$\mathcal{M}$$>$$10^{10.5}$~$\mathcal{M}_\sun$, the average difference
is significantly smaller, just 0.02dex (with a scatter of
0.15dex). This can be explained by the fact that most of the
photometric data points in the modelling fits are found in UV/optical
wavelengths, where the emission of relatively young stars is
significant. Older stars, possibly much more numerous and dominating
the global stellar mass of a galaxy, may be hidden by the intensity of
more recent starbursts. This effect should be more noticeable in less
massive systems presenting bright recent bursts involving a relatively
high fraction (larger than what is normally observed in very massive
galaxies with old stellar populations) of the total stellar mass of
the galaxy. Only in the 2-POP models are we able to take this effect
into account, and that is why in this case we systematically obtain
larger stellar masses for some galaxies with relatively low masses. We
conclude that the choice of the 1-POP or 2-POP models does not change
the stellar masses significantly (more than the typical uncertainties
of a factor of 2--3) in a statistical sense, and the effect on the
masses for massive galaxies (which dominate the stellar mass density
at any redshift) is very small.

Uncertainties in the stellar emission models are known to introduce
systematic errors in the estimation of stellar masses from photometry
\citep[see, e.g.,][]{2006ApJ...636L..21V,2006ApJ...652...97V}. In order to 
check this effect, the stellar masses obtained with the PEGASE code
\citep{1997A&A...326..950F} were compared with the values estimated by 
using the BC03 models from \citet{2003MNRAS.344.1000B}. On average,
the BC03 models give stellar masses larger by 0.03dex (less than
10\%), with a scatter of 0.18dex. For 95\% of the galaxies, the
stellar mass difference is lower than a factor of 3. We also fitted
the SEDs with the M05 models developed by \citet[][see also
\citealt{2007astro.ph..2091B}]{2005MNRAS.362..799M}, which include a 
more refined treatment of the emission from thermally pulsating
asymptotic giant branch stars, and are claimed to obtain stellar
masses that can be lower by as much as 60\% (based on the prediction
of lower NIR mass-to-light ratios for some ages). On average, the M05
models give stellar masses smaller by 0.14dex (less than 30\%), with a
scatter of 0.22dex (and no clear dependence on redshift). For 96\% of
the galaxies, the stellar mass difference is lower than a factor of 3.

One important {\it a priori} assumption of any stellar population
modelling is the treatment of extinction by dust. We compared the
stellar masses obtained with the two different extinction recipes we
considered (CF00 and CALZ00). For about 80\% (65\%) of the galaxies,
both estimates are equal within a factor of 0.3dex (0.2dex). For the
rest of galaxies (again, most of them with
$\mathcal{M}$$<$$10^{10.5}$~$\mathcal{M}_\sun$), the estimates using
the CF00 recipe are higher up to a factor of 5. On average, stellar
masses derived with the CF00 recipe are 0.10dex higher than those
derived with CALZ00 law. As discussed in \citet{2003MNRAS.338..525P},
in the CF00 recipe the attenuation of the emission arising from the
stars is always (except for very young bursts) larger than the
attenuation of the gas emission. The CALZ00 recipe shows a opposite
behavior, given that the attenuation of the stellar emission is
roughly half of the attenuation of the gas emission. Moreover, the
attenuation wavelength dependence (from the UV to the NIR) proposed by
CF00 is shallower than the one in CALZ00. This leads to a need of more
stars to obtain the same observed luminosity for equal values of the
extinction in the CF00 case, which explains the larger stellar masses
derived for this case (on average). However, the final effect on the
masses is of the order of 0.1dex, which demonstrates that choice of an
extinction recipe does not change the stellar masses more than the
typical uncertainties.

Another important assumption of the stellar population models is the
IMF, which has a direct effect on the derived stellar
masses. Different IMFs produce stellar spectra with very similar
colors, but with less or more stars, which causes a systematic
uncertainty in the final stellar mass estimations. For example, a
\citet{1993MNRAS.262..545K} IMF (as the one used in
\citealt{2006A&A...453..869B}) predicts stellar masses smaller than
ours by a factor of $\sim$1.7, or a
\citet{2003ApJ...593..258B} IMF (used in, for example,
\citealt{2004Natur.430..181G}) predicts also smaller masses by a
factor of $\sim$1.8. All our results and those extracted from the
literature were normalized to a \citet{1955ApJ...121..161S} IMF with
0.1$<$$\mathcal{M}$$<$100~$\mathcal{M}_\sun$. If the IMF is universal
(the same at all redshifts), this choice should not affect our results
other than an overall normalization. A discussion about changes in the
IMF from galaxy to galaxy is far out of scope of this paper.

Finally, we performed another test of the goodness of our stellar mass
estimates by comparing the results obtained from direct comparison of
the SEDs with the entire grid of stellar population models (once the
redshift of a galaxy is known) with the results obtained with the
photometric redshift technique using the empirically built set of
models, from which we obtained stellar mass estimates for all
galaxies. We find a very good agreement between these two stellar mass
calculations: 90\% of galaxies present an average difference between
the two mass estimates of less than 0.01dex, and the scatter around
this value is 0.15dex.

Based on this discussion, the choices of 1-POP or 2-POP models,
distinct stellar population libraries, different IMFs, or different
extinction recipes produce changes in the derived stellar masses of
the same order or smaller than the typical error in any stellar
population synthesis analysis (a factor of 2--3), directly linked to
the degeneracies of the solutions to the problem. Thus, in the
Sections~\ref{photozsect} to \ref{sfrs}, we will only present the
results obtained with the stellar masses estimated with the 1-POP
models, the \citet{2000ApJ...533..682C} extinction law, and a
\citet{1955ApJ...121..161S} IMF. This choice will also allow us 
to compare directly with other previous works found in the literature,
that usually assume these characteristics in their modelling
procedures.


\subsubsection{Statistical evaluation of the SFRs}
\label{statsfr}

In order to understand the systematic and random uncertainties of our
estimations of the SFR for each galaxy, we carried out two
tests. 

First, we used 3 different dust emission template sets built by
\citet{2001ApJ...556..562C}, \citet{2002ApJ...576..159D}, and
Rieke et al. (2007, in preparation). The values of the IR SFR
[estimated from $L(8-1000)$ using the conversion factor found in
\citealt{1998ARA&A..36..189K}] derived with the
\citet{2001ApJ...556..562C} models were systematically smaller than
the SFRs derived with the
\citet{2002ApJ...576..159D} models (on average, 0.1dex) and Rieke et
al. (2007, in preparation) templates (on average, 0.2dex). To take
into account the systematic uncertainties introduced by the use of a
particular set of models, we finally considered an average value of
the estimations from the three template sets. The typical uncertainty
of this average value (based on the standard deviation of the 3
estimations) is about 50\%.

The second test consisted in obtaining IR-based SFRs with different
methods. Classically, IR-based SFRs are calculated from the integrated
IR luminosity $L(8-1000)$. The quantity $L(8-1000)$ can be estimated
for each galaxy by fitting the IR spectrum with models of the dust
emission. For our galaxies, this translates to a significant
extrapolation in the templates, since the reddest point in our SEDs
corresponds to the observed MIPS 24\mic\, emission, and we are
assuming that a color or a single photometric point in the MIR is
closely related to the emission in the FIR, which dominates the
integrated IR luminosity. However, one can also avoid this large
extrapolation by estimating monochromatic luminosities at specific
wavelengths which are not far from the reddest photometric point in
our SEDs. In this sense, we estimated monochromatic luminosities at
6.7\mic, 12\mic, and 15\mic, and then calculated the integrated
luminosities $L(8-1000)$ using the empirical relationships built by
\citet{2001ApJ...556..562C}. Since they are based in the same templates, these 
estimations of $L(8-1000)$ based on different monochromatic emissions
are not independent. However, another independent SFR estimation was
obtained by extrapolating in the models to measure the rest-frame MIPS
24\mic\, monochromatic luminosity. This luminosity was converted to a
SFR using the calibration given in
\citet{2006ApJ...650..835A}. The typical scatter of these different 
IR-based SFR estimations obtained from monochromatic emissions is
30\%.

From these tests, we conclude that our IR-based SFR estimations are
good within a factor of 2, which is consistent with other evaluations
of IR-based SFRs \citep[e.g., ][]{2002ApJ...579L...1P,
2005ApJ...632..169L,2006ApJ...637..727C}.


\subsection{Evaluation of parameters derived for galaxies with AGNs}
\label{agn_pars}

According to Section~\ref{agns}, a small fraction (less than 5\%) of
our galaxy sample probably harbor an AGN which emits strongly at X-ray
and/or IR wavelengths. Given that our goal is to estimate the total
stellar mass content of the Universe at any redshift up to z$\sim$4,
we must try to keep this type of sources in our sample. However, the
emission of the dust heated by the nuclear massive black hole can
extend to the NIR (if very hot dust, with a temperature of
T$\gtrsim$100~K, is present) and even to optical bands (e.g., in the
case of Type 1 QSOs), affecting the estimations of the photometric
redshifts, stellar masses, and SFRs.

Given that our photometric redshifts are mainly based on stellar
population synthesis models, we can expect that galaxies whose
UV-to-NIR SEDs are not stellar are not well represented by our
template set, and there is a large probability that the photometric
redshift estimation fails. However, only the most extreme and powerful
AGNs in our sample would affect the UV-to-NIR global SED of the host
galaxy. This is demonstrated by \citet{2007ApJ...660..167D}, who build
median rest-frame SEDs of X-ray-detected IRAC sources in the HDF-N,
finding that only galaxies with X-ray observed luminosities
(integrated from 0.5 to 8 keV) $L(X)$$>$10$^{44}$~erg~s$^{-1}$ present
non-stellar SEDs, and galaxies with $L(X)$$=$10$^{43-44}$~erg~s$^{-1}$
start to show significant emission from hot dust at
$\lambda_\mathrm{rest}$$\gtrsim$2\mic. Consequently, our stellar mass
estimates should only be affected by the presence of an AGN for bright
X-ray sources. To further test our stellar masses for AGN, we run a
set of stellar population models on 1,000 randomly selected galaxies
fitting the SED only up to $\lambda_\mathrm{rest}$$=$2.3\mic\, (the
$K$-band) instead of up to $\lambda_\mathrm{rest}$$=$4\mic, to exclude
the hot dust emission that can arise at
$\lambda_\mathrm{rest}$$\sim$2-4\mic\, (note that only one photometry
data point at most is removed from our SEDs in this case). The average
difference between the masses estimated in this way and our original
values is just 0.002dex, and the scatter is 0.10dex. Cutting the SEDs
at even bluer wavelengths, $\lambda_\mathrm{rest}$$=$1.5\mic\, has a
similar negligible effect: the average difference is 0.004dex, and the
scatter is 0.13dex. This demonstrates that the (possible) AGN emission
in the NIR does not bias our stellar mass estimates for X-ray sources
of moderate brightness. For the brightest sources, however, the
UV-to-MIR SED is significantly affected by the AGN emission, so we
decided to remove from our sample all the X-ray detected sources with
$L(X)$$>$10$^{44}$~erg~s$^{-1}$. These sources are just 0.4\% of the
entire IRAC sample (0.5\% of the sample in the HDF-N and the CDF-S,
and 0.1\% in the LHF), slightly more common at z$>$2 (1\%-2\% of all
IRAC selected galaxies at high redshift), so they should have a very
small additive contribution (not accounted in our results) on the
stellar mass functions and densities.

Concerning the estimation of photometric redshifts, their reliability
for X-ray sources is slightly lower than for the global sample: we
obtain redshifts with $\sigma_\mathrm{z}$/(1+z)$<$0.1 for 81\% of the
X-ray detections, and with $\sigma_\mathrm{z}$/(1+z)$<$0.2 for
92\%. As mentioned earlier, these sources are very few in comparison
with the entire IRAC sample, and they are accounted in the stellar
mass function estimate by using the photometric redshift uncertainties
(which are estimated without removing this type of sources).

In the case of the SFR estimations, dust obscured AGNs are expected to
radiate the absorbed emission in the MIR/FIR. Although it is common
that star formation coexists with AGN activity, it is not possible to
decompose the IR emission into the components coming from the two
different phenomena, mostly when very few photometric points are
available in this wavelength range. Therefore, all X-ray sources were
removed in the analysis of the (specific) SFRs carried out in
Section~\ref{sect_sfr}. Given that we were only interested in the
distribution of specific SFRs of our sample, our results are not
significantly affected by the exclusion of this type of sources.

\end{appendix}

\end{document}